\documentclass{ieeeaccess}
\usepackage{cite}
\usepackage{amsmath,amssymb,amsfonts}
\usepackage{algorithmic}
\usepackage{graphicx}
\usepackage{textcomp}
\def\BibTeX{{\rm B\kern-.05em{\sc i\kern-.025em b}\kern-.08em
    T\kern-.1667em\lower.7ex\hbox{E}\kern-.125emX}}

\usepackage[colorlinks,linkcolor=blue, breaklinks=true]{hyperref}

\usepackage[nogroupskip=true, style=super]{glossaries}

\newacronym{IVA}{IVA}{Intelligent Video Analytics}
\newacronym{RIVA}{RIVA}{Real-time IVA}
\newacronym{BIVA}{BIVA}{Batch IVA}
\newacronym{VAS}{VAS}{Video Analytics System}
\newacronym{EVAS}{EVAS}{Embedded Video Analytics System}
\newacronym{OVAS}{OVAS}{On-site Video Analytics System}
\newacronym{CVAS}{CVAS}{Cloud-based Video Analytics System}
\newacronym{FVAS}{FVAS}{Fog-based Video Analytics System}
\newacronym{IR}{IR}{Intermediate Results}
\newacronym{VSDS}{VSDS}{Video Stream Data Source}
\newacronym{IVAS}{IVAS}{IVA services}
\newacronym{SIAT}{L-CVAS}{Lambda CVAS}
\newacronym{aaS}{aaS}{as-a-Service}
\newacronym{IaaS}{IaaS}{Infrastructure-as-a-Service}
\newacronym{PaaS}{PaaS}{Platform-as-a-Service}
\newacronym{SaaS}{SaaS}{Software-as-a-Service}
\newacronym{IVAaaS}{IVAaaS}{IVA-as-a-Service}
\newacronym{IVAAaaS}{IVAAaaS}{IVA-Algorithm-as-a-Service}
\newacronym{CBVR}{CBVR}{Context-Based Video Retrieval}
\newacronym{ITS}{ITS}{Intelligent Transportation System}
\newacronym{ML}{ML}{Machine Learning}
\newacronym{SVM}{SVM}{Support Vector Machine}
\newacronym{VBDCL}{VBDCL}{Video Big Data Curation Layer}
\newacronym{VBDPL}{VBDPL}{Video Big Data Processing Layer}
\newacronym{VBDML}{VBDML}{Video Big Data Mining Layer}
\newacronym{KCL}{KCL}{Knowledge Curation Layer }
\newacronym{WSL}{WSL}{Web Service Layer}
\newacronym{RVSAS}{RVSAS}{Real-time Video Stream Acquisition and Synchronization}
\newacronym{DPDS}{DPDS}{Distributed Persistent Data Store}
\newacronym{DMBM}{DMBM}{Distributed Message Broker Manager}
\newacronym{VSAS}{VSAS}{Video Stream Acquisition Service}
\newacronym{IRM}{IRM}{Intermediate Results Manager}
\newacronym{VSCS}{VSCS}{Video Stream Consumer Service}
\newacronym{LVSM}{LVSM}{Lifelong Video Stream Monitor}
\newacronym{VSAC}{VSAC}{Video Stream Analytics Consumer}
\newacronym{ISDDS}{ISDDS}{Immediate Structured Distributed Data Store}
\newacronym{UPDDS}{UPDDS}{Unstructured Persistent Distributed Data Store}
\newacronym{DFS}{DFS}{Distributed File System}
\newacronym{DBDS}{DBDS}{Distributed Big Datastore}
\newacronym{HDFS}{HDFS}{Hadoop File System}
\newacronym{LSH}{LSH}{Locality-sensitive Hashing}
\newacronym{MCA}{MCA}{Multiple Correspondence Analysis}
\newacronym{AFS}{AFS}{Adaptive Feature Scaling}
\newacronym{SIFT}{SIFT}{Scale Invariant Feature Transform}
\newacronym{LBP}{LBP}{Local Binary Pattern}
\newacronym{HOG}{HOG}{Histogram of Oriented Gradients}
\newacronym{CNN}{CNN}{Convolutional Neural Network}
\newacronym{DAG}{DAG}{Directed Acyclic Graph}
\newacronym{CCDG}{CCDG}{Controlled Cyclic Dependency Graph}
\newacronym{RTSP}{RTSP}{Real-time Stream Protocol}
\newacronym{PCA}{PCA}{Principal Component Analysis}
\newacronym{SM}{SM}{Similarity Measures}
\newacronym{LSI}{LSI}{Latent Semantic Indexing}
\newacronym{RNN}{RNN}{Recurrent neural network}
\newacronym{LSTM}{LSTM}{Long Short-Term Memory}
\newacronym{FVSA}{FVSA}{Fused Video Surveillance Architecture}
\newacronym{C2C}{C2C}{Customer-to-Customer}
\newacronym{B2B}{B2B}{Business-to-Business}
\newacronym{B2C}{B2C}{Business-to-Customer}
\newacronym{QoS}{QoS}{Quality of Service}
\newacronym{ACID}{ACID}{Atomicity, Consistency, Isolation, Durability}
\newacronym{RDBMS}{RDBMS}{Relational Database Management System}
\newacronym{CAP}{CAP}{Consistency, Availability, Partition Tolerance}
\newacronym{CRUD}{CRUD}{create, read, update, and delete}
\newacronym{AI}{AI}{Artificial Intelligence}
\newacronym{IoT}{IoT}{Internet of Things}

\makeglossaries

\usepackage{multirow}
\usepackage{longtable}
\usepackage{makecell}

\newcommand{\cmark}{\ding{51}}%
\newcommand{\xmark}{\ding{55}}%

\usepackage{amssymb}
\usepackage{pifont}
\usepackage[margin=1in]{geometry}
\usepackage{lscape}

\hypersetup{draft}

\usepackage{import}

\usepackage{soul,color}
\usepackage{xcolor}
\usepackage{ulem}

\begin{document}
\history{Date of publication xxxx 00, 0000, date of current version xxxx 00, 0000.}
\doi{10.1109/ACCESS.2017.DOI}

\title{Video Big Data Analytics in the Cloud: A Reference Architecture, Survey, Opportunities, and Open Research Issues}
\author{
	\uppercase{Aftab~Alam}, 
	\uppercase{Irfan~Ullah, and 
		Young-Koo~Lee}}

\address{Department of Computer Science and Engineering, Kyung Hee University (Global Campus), Yongin 1732, South Korea}

\tfootnote{This work was supported by the Institute for Information and Communications Technology Promotion Grant through the Korea Government (MSIT) under Grant R7120-17-1007 (SIAT CCTV Cloud Platform).}

\markboth
{A Alam \headeretal: Video Big Data Analytics in the cloud}
{A Alam \headeretal: Video Big Data Analytics in the cloud}

\corresp{Corresponding author: Young-Koo~Lee (e-mail: yklee@khu.ac.kr)}

\begin{abstract}
The proliferation of multimedia devices over the \gls{IoT} generates an unprecedented amount of data. Consequently, the world has stepped into the era of big data. Recently, on the rise of distributed computing technologies, video big data analytics in the cloud has attracted the attention of researchers and practitioners. The current technology and market trends demand an efficient framework for video big data analytics. However, the current work is too limited to provide a complete survey of recent research work on video big data analytics in the cloud, including the management and analysis of a large amount of video data, the challenges, opportunities, and promising research directions. To serve this purpose, we present this study, which conducts a broad overview of the state-of-the-art literature on video big data analytics in the cloud. It also aims to bridge the gap among large-scale video analytics challenges, big data solutions, and cloud computing. In this study, we clarify the basic nomenclatures that govern the video analytics domain and the characteristics of video big data while establishing its relationship with cloud computing. We propose a service-oriented layered reference architecture for intelligent video big data analytics in the cloud. Then, a comprehensive and keen review has been conducted to examine cutting-edge research trends in video big data analytics. Finally, we identify and articulate several open research issues and challenges, which have been raised by the deployment of big data technologies in the cloud for video big data analytics. To the best of our knowledge, this is the first study that presents the generalized view of the video big data analytics in the cloud. This paper provides the research studies and technologies advancing the video analyses in the era of big data and cloud computing.
\end{abstract}


\titlepgskip=-15pt

\maketitle

\section{Introduction}
\IEEEPARstart{V}{ideos} are generated and uploaded regularly to the cloud. Many sources include CCTV, smartphones, drones, etc., are actively contributing to video generation leads to the evolution of \gls{IVA} and management systems. \gls{IVA} is a domain that uses advanced computer vision technologies to process and extract insights from streaming or stored videos automatically.  Over the past two decades, IVA is extensively rising from real-world needs driving by a broader range of application domains ranging from security and surveillance to transportation and healthcare. The \gls{IVA} market is anticipated to rise from 4.9 billion USD in 2020 to 11.7 billion USD by 2025 at a compound annual growth rate of 19.0\% \cite{ApacheRocketMQ2012}.

Video management and services providers such as Facebook \cite{Facebook2020}, YouTube \cite{YouTube2019}, and Netflix \cite{Netflix2020} are considered as valuable sources of large-scale video data. Along with these, various leading industrial organizations have successfully deployed video management and analytics platforms that provide more bandwidth and high-resolution cameras collecting videos at scale and has become one of the latest trends in the video surveillance industry. For example, more than 400 hours of videos are uploaded in a minute on Youtube \cite{pouyanfar2018multimedia}, and more than one hundred and seventy million video surveillance cameras have been installed in china only \cite{olatunji2018dynamic}. It has been reported that the data generated by various \gls{IoT} devices will see a growth rate of 28.7\% over the period 2018-2025, where surveillance videos are the majority shareholder \cite{idg2019}.

Such an enormous video data is considered as “big data” because a variety of sources generates a large volume of video data at high velocity that holds high Value. Even though 65\% of the big data shares hold by surveillance videos are monitored, but still, a significant portion of video data has been failed to notice\cite{huang2014surveillance}. That neglected data contain valuable information directly related to real-world situations. Video data provide information about interactions, behaviors, and patterns, whether its traffic or human patterns. However, handling such a large amount of complex video data is not worthwhile utilizing traditional data analytical approaches. Therefore, more comprehensive and sophisticated solutions are required to manage and analyses such large-scale unstructured video data.

Due to the data-intensive and resources hungry nature of large scale video data processing, extracting the insights from the video is a challenging task. A considerable size of video data poses significant challenges for video management and mining systems that require powerful machines to deal with large-scale video data. Moreover, a flexible solution is necessary to store and mine this large volume of video data for decision making. However, large-scale video analytics becomes a reality due to the popularity of big data and cloud computing technologies. 

Cloud computing is an infrastructure for providing convenient and ubiquitous remote access to a shared pool of configurable computing resources. These resources can be managed with minimal management effort or service \cite{mell2011nist}. Big data technologies, such as Hadoop or Spark echo system, are software platforms designed for distributed computing to process, analyze, and extract the valuable insights from large datasets in a scalable and reliable way. The cloud is preferably appropriate to offer the big data computation power required for the processing of these large datasets. \cite{zikopoulos2012harness}, Amazon web service \cite{amazon2015amazon}, Microsoft Azure \cite{copeland2015overview}, and Oracle Big Data Analytics \cite{dijcks2012oracle} are some examples of video big data analytics platforms. Large-scale video analytics in the cloud is a multi-disciplinary area, and the next big thing in big data, which opens new research avenues for researchers and practitioners.

This work aims to conduct a comprehensive study on the status of large scale video analytics in the cloud-computing environment while deploying video analytics techniques. First, this study builds the relationship between video big data and cloud computing and defines the terminologies that govern the study. Then service-oriented and a layered reference architecture have been proposed for large-scale video analytics in the cloud while focusing on architectural properties like reliability, scalability, fault-tolerance, extensibility, and intermediate results orchestration. Further, an intensive survey has been conducted to project the current research trends in video analytics that encompass the taxonomy of video analytics approaches, and cloud-based scholarly and industrial study. Finally, open research issues and challenges are discussed, with a focus on proposed architecture, i.e., the deployment of an array of computer vision algorithms for large-scale videos in the cloud. 

\subsection{Video Big Data, Cloud Computing, and their Relationship}
The term big data appeared and popularized by John R. Masey in the late 1990s \cite{mashey1997big}, which refers to a large volume of data that are impractical to be stored, processed and analyzed using traditional data management and processing technologies \cite{snijders2012big}. The data can be unstructured, semi-structured, and structured data, but mostly unstructured data is considered. The definition of big data evolved and has been described in terms of three, four, or five characteristics. In literature, among these characteristics, three are shared, i.e., Volume, Velocity, and Variety, while the others are Veracity and Value \cite{hashem2015rise, elgendy2014big, ozkose2015yesterday, fadiya2014advancing}. Various video stream sources generate a considerable amount of unstructured video data on a regular bases and becoming a new application field of big data. The data generated by such sources are further subject to contextual analysis and interpretation to uncover the hidden patterns for decision-making and business purposes. 

In the context of a large volume of video data, we specialize the generic big data characteristics. The size of data is referred to as Volume \cite{marr2015big}, but the majority of the shares, i.e., 65\%, are held only by surveillance videos. The type of data generated by various sources such as text, picture, video, voice, and logs are known as Variety \cite{marr2015big}. The video data are acquired from multimodal video stream sources, e.g., IP-Camera, depth camera, body-worn camera, etc., and from different geolocations, which augments the Variety property. The pace of data generation and transmission is known as Velocity \cite{o2013artificial}. The video data also possess the Velocity attribute, i.e., the \gls{VSDS} primarily produce video stream 24/7 and acquired by the data center storage servers. Veracity can be defined as the diversity of quality, accuracy, and trustworthiness of the data \cite{berman2013principles}. Video data are acquired directly from real-world domains and meet the Veracity characteristic. The Value refers to contextual analysis to extract the significant values for decision-making and business purpose \cite{john2014big, bernard2015bigdata}. Video data has high Value because of its direct relation with real-word. Automatic criminal investigation, illegal vehicle detection, and abnormal activity recognition are some of the examples of Value extraction. Almost all the big data properties are dominated by the video data, which encourage us to give birth to Video Big Data.

These five characteristics impose many challenges on the organizations when embracing video big data analytics. Storing, scaling, and analyzing are some apparent challenges associated with video big data.  To cope with these challenges, converged and hyper-converged infrastructure and software-defined storage are the most convenient solutions. Distributed databases, data processing engines, and machine learning libraries have been introduced to overcome video big data management, processing, and analysis issues, respectively.

These big data technologies are deployed over a computer cluster to process and manage a massive amount of video data in parallel. A computer cluster may consist of few to hundreds, and even thousands of nodes work together as a single integrated computing resource, on different parts of the same program \cite{pfister1998search, buyya1999high}. Deploying an indoor computer cluster is an option for big data technologies, but hardware cost and maintenance issues are associated with it.  An alternative solution can be cloud computing that elegantly reduces the costs associated with the management of hardware and software resources \cite{brian2008cloud}.

Typically, cloud services are provided on-demand in a “pay-as-you-go” manner for the conveniences of end-users and organizations \cite{aceto2013cloud}. Cloud computing follows the philosophy of the “as-a-Service” and offers its “services” according to different models, for example, \gls{IaaS}, \gls{PaaS}, \gls{SaaS} \cite{mell2011nist}.

Under \gls{IaaS} (e.g., Amazon’s EC2 ), the cloud service provider facilitates and allows the consumers to provision fundamental computing resources and deploy arbitrary software. In \gls{PaaS}, the service provider provides a convenient platform enabling customers to develop, run, and manage applications without considering the complexities of building and maintaining the infrastructure. The examples of \gls{PaaS} are Google's Apps Engine and Microsoft Azure. In \gls{SaaS}, applications (e.g., email, docs, etc.) are deployed on cloud infrastructure by service providers and allow the consumer to subscribe. These applications can be easily accessed from various client devices using a thin client or program interfaces \cite{mell2011nist}.

\subsection{Research Objectives and Contributions}
\label{ResearchObjectivesandContributions}
This paper presents a detailed survey and review of cloud-based large-scale \gls{IVA}. We also propose big data technological solutions for the challenges faced by \gls{IVA} researchers and practitioners. The contributions of this paper are listed below: 

\begin{itemize}
	\item We standardize the basic nomenclatures that govern the \gls{IVA} domain. The term video big data has been coined and clarified how it inherits the big data characteristics while establishing its relationship with cloud computing.  
	
	\item We propose a distributed, layered, service-oriented, and lambda style \cite{marz2015big} inspired reference architecture for large-scale \gls{IVA} in the cloud. Each layer of the proposed \gls{CVAS} has been elaborated technologically, i.e., layerwise available big data technological alternatives. The base layer of the proposed architecture, i.e., video big data curation layer, is based on the notion of \gls{IR} orchestration, which can play a significant role in the optimization of the \gls{IVA} pipeline. Under the proposed architecture, we perform a thorough investigation of scalable traditional video analytics and deep learning techniques and tools on distributed infrastructures along with popular computer vision benchmark datasets.
	
	\item To the best of our knowledge, this is a first surveying video big data analytics, our research targets the most recent approaches that encompass broad \gls {IVA} research domains like content-based video retrieval, video summarization, semantic-based approaches, and surveillance and security. 
	
	\item We also investigate how the researchers exploit big data technologies for large-scale video analytics and show how the industrial \gls{IVA} solutions are provided. 
	
	\item In this study, we further write real word \gls{IVA} application areas, which projects the significance of big data and cloud computing in \gls{IVA}. 
	
	\item Finally, we identify the research gap and list several open research issues and challenges. These research issues encompass orchestration and optimization of \gls{IVA} pipeline, big dimensionality, online learning on video big data, model management, parameter servers, and distributed learning, evaluation issues and opportunities, \gls{IVA} services statistics maintenance and ranking in the cloud, \gls{IVAaaS} and cost model, video big data management, and privacy, security and trust. 
\end{itemize}

The remaining paper is organized as follows. Section~\ref{sec:RecentStudies} and \ref{sec:ScopeAndNomenclature} is about recent studies, and scope and nomenclature, respectively. Section~\ref{sec:CVAS} discusses the proposed \gls{CVAS}. Literature review has been presented in Section~\ref{sec:IVAConstituentsPredominantTrends} and \ref{sec:StateOfTheArtCVAS}. The \gls{IVA} applications can be seen in Section~\ref{sec:IVAApplications}. In Section~\ref{sec:ResearchIssues} several video big data challenges and opportunities are discussed. Finally, Section~\ref{sec:conclusion} concludes this study.

\section{Recent Studies}
\label{sec:RecentStudies}
Various studies have been conducted that discus video analytics, which can be classified as \gls{IVA} and Big data, as shown in Table~\ref{tab:RecentStudies}. In the former class, the surveys overlook the big data characteristics and challenges of video analysis in the cloud. The majority of the work focuses on a specific domain of \gls{IVA} and the application of video analyses. For instance, a comprehensive survey was presented by Liu et al. \cite{liu2013intelligent}, and Olatunji et al. \cite{olatunji2019video}, where they have discussed \gls{IVA} and its applications in the context of surveillance conventionally. \gls{CBVR} was reviewed by Hu et al. \cite{hu2011survey}, Patel et al. \cite{patel2012content}, and Haseyama et al. \cite{haseyama2013survey}. Similarly, vehicle surveillance systems in \gls{ITS}, abnormal behavior analytics, in surveillance videos were studied by Tian et al. \cite{tian2014hier}, and \cite{mabrouk2018abnr}, respectively.  



\begin{table*}[!htb]
	\centering
	\caption{Recent Surveys Related to Video Big Data Analytics,}
	\label{tab:RecentStudies}
	\resizebox{\textwidth}{!}{%
		
		\begin{tabular}{|l|l|l|l|l|}
			\hline
			
			\multicolumn{1}{|l|}{Class} & References & Target Domain & Pros & Cons \\ \hline

			\multirow{4}{*}{\rotatebox[origin=c]{90}{Int. Video Analytics}} & 
			\makecell[l]{Liu et al. \cite{liu2013intelligent} \\ Olatunji et al. \cite{olatunji2019video}} & 
			\begin{tabular}[c]{@{}l@{}}
				Video analytics \\ based surveillance 
			\end{tabular}	
			& 
			\begin{tabular}[c]{@{}l@{}}
				Comprehensive theory \& applications of 
				\\ video analytics-based surveillance \end{tabular}&

			\begin{tabular}[c]{@{}l@{}}
				-Lacking big data \& cloud computing \\ and its research issues
				\\ - Not service oriented 
			\end{tabular}	
			
			\\ \cline{2-5} 
			
			& \makecell[l]{
				Hu et al. \cite{hu2011survey}	\\ 
				Patel et al. \cite{patel2012content} \\ 
				Haseyama et al. \cite{haseyama2013survey}}
			
			& \begin{tabular}[c]{@{}l@{}}\gls{CBVR}\end{tabular}	& 
			
			\begin{tabular}[c]{@{}l@{}}
				- \gls{CBVR} components, research trends \\ and issues
			\end{tabular}                                 & 
			
			\begin{tabular}[c]{@{}l@{}}
				-Lack of big data \& cloud computing, \\ \&	its research issues.
				\\-Limited to \gls{CBVR}
			\end{tabular}                                                                                        \\ \cline{2-5} 
			
			& \makecell[l]{Khan et al. \cite{khan2014big} \\ Tsai et al. \cite{tsai2015big}}
			
			& 
			
			\begin{tabular}[c]{@{}l@{}}
				Vehicle surveillance \\systems in \gls{ITS}
			\end{tabular}        & 
			
			\begin{tabular}[c]{@{}l@{}}
				Generic layered architecture for video \\ 
				analytics-based vehicle  surveillance
			\end{tabular} & 
			
			\begin{tabular}[c]{@{}l@{}}
				- Lacking big data and cloud computing \\ techniques and challenges\\ 
				- Abstract architecture
				- Consider \gls{IVA} for \gls{ITS}
			\end{tabular}                                                \\ \cline{2-5} 
			
			& 
			Mabrouk et al. \cite{mabrouk2018abnr}          
			& 
			\begin{tabular}[c]{@{}l@{}}
				Abnormal behavior in \\
				surveillance videos
			\end{tabular} & 
			
			\begin{tabular}[c]{@{}l@{}}
				A generic architecture for abnormal behaviors \\ 
				in video surveillance systems
			\end{tabular}  & 
			
			\begin{tabular}[c]{@{}l@{}}
				-Lacking big data and cloud computing \\ discussion.\\  
				-Limited to abnormal behavior
			\end{tabular}                                                                
			
			\\ \hline
			
			\multirow{7}{*}{\rotatebox[origin=c]{90}{Big data}}                    &
			\makecell[l]{
				Hashem et al. \cite{hashem2015rise} \\ 
				Agrawal et al. \cite{agrawal2011big}
			}
			& 
			
			\begin{tabular}[c]{@{}l@{}}
				Cloud computing
			\end{tabular}  &

			\begin{tabular}[c]{@{}l@{}}
				A thorough overview of big data techniques, \\
				challenges and open issues in cloud computing
			\end{tabular} & 
			
			\begin{tabular}[c]{@{}l@{}}
				-Big data processing over cloud computing only. \\ 
				-Lack of video big data analytics, \\ 
				service-oriented architecture, \\ 
				research issues, and challenges.
			\end{tabular} \\ \cline{2-5}

			& \makecell[l]{
				Khan et al. \cite{khan2014big} \\
				Tsai et al. \cite{tsai2015big}
			}
			
			& 
			
			\begin{tabular}[c]{@{}l@{}}
				Big data technologies
			\end{tabular}                           & 
			
			\begin{tabular}[c]{@{}l@{}}
				Overview of big data techniques,opportunities, \\ 
				and challenges
			\end{tabular}                         & 
			
			\begin{tabular}[c]{@{}l@{}}
				Limited discussion on video big data \\ analytics\end{tabular}                                                                                                  \\ \cline{2-5} 
			
			&	
			Zhou et al. \cite{zhou2017machine}
			& 
			
			\begin{tabular}[c]{@{}l@{}}
				\gls{ML}
			\end{tabular}                               & 
			
			\begin{tabular}[c]{@{}l@{}}
				A framework of \gls{ML} on big data, and open \\ 
				research issues
			\end{tabular}                               & 
			
			\begin{tabular}[c]{@{}l@{}}
				Lack of \gls{IVA} discussion
			\end{tabular}                                                                                                                               \\ \cline{2-5} 
			
			& 
			Che et al. \cite{che2013big} & 
			
			\begin{tabular}[c]{@{}l@{}}
				Big data mining
			\end{tabular}                                & 
			\begin{tabular}[c]{@{}l@{}}
				A comprehensive overview on big data \\ 
				mining and challenges
			\end{tabular}                        & 
			
			\begin{tabular}[c]{@{}l@{}}
				Lack of video analytics and its \\ 
				research issues
			\end{tabular}                                                                                                   \\ \cline{2-5} 
			
			& 
			Pouyanfar et al. \cite{pouyan2018mul} & 
			
			\begin{tabular}[c]{@{}l@{}}
				Multimedia 
			\end{tabular}                                & 
			
			\begin{tabular}[c]{@{}l@{}}
				A comprehensive overview of multimedia data	
			\end{tabular} 
			
			& 
			\begin{tabular}[c]{@{}l@{}}
				Limited discussion on video big data analytics
			\end{tabular}                                                                                                   \\ \cline{2-5} 	
			
			& 
			
			Zhu et al. \cite{zhu2018bigITS}
			
			& \gls{ITS}  & 
			\begin{tabular}[c]{@{}l@{}}
				-A study of big data analytics in \gls{ITS}\\
				-Consider multimedia for \gls{ITS}
			\end{tabular}                                      & 
			
			\begin{tabular}[c]{@{}l@{}}
				Limited discussion of video big data \\ 
				in the context of \gls{ITS} analytics
				
			\end{tabular}

		  \\ \cline{2-5} 	
	
		& 
		H. Zahid et al. \cite{zahid2019big}
		& 
		Telecommunications
		& 
		A study of big data analytics in telecommunication. 
		& 
		\begin{tabular}[c]{@{}l@{}}	
			Limited discussion of video big data \\ 
			in the context of telecommunication analytics
		\end{tabular}
		\\ \hline
		\end{tabular}%
	}
\end{table*}


Furthermore, big data studies were conducted from different perspectives, i.e., cloud computing, \gls{ML}, mining, multimedia, and \gls{ITS}. Hashem et al.  \cite{hashem2015rise} and Agrawal et al. \cite{agrawal2011big} presented research issues and challenges on big data analytics in the cloud computing. Tsai et al. \cite{tsai2015big} and Khan et al. \cite{khan2014big} focus on big techniques and general applications along with challenges. Zhou et al. \cite{zhou2017machine}, Che et al. \cite{che2013big}, and Pouyanfar et al. \cite{pouyan2018mul} studied and reported the research issues of big data practices concerning \gls{ML} techniques, data mining algorithms, and multimedia data respectively. Zhu et al. \cite{zhu2018bigITS} presented a comprehensive study on big data analytics in \gls{ITS}, analytical methods and platforms, and categories of big data video analytics. In the study of big data, \gls{IVA} has been overlooked, and minimal discussion can be found in the study of Khan et al. \cite{khan2014big}, Tsai et al. \cite{tsai2015big}, Pouyanfar et al. \cite{pouyan2018mul}, Zhu et al. \cite{zhu2018bigITS}, H. and Zahid et al. \cite{zahid2019big}.

From the literature, it is clear that the recent studies ignored large-scale unstructured video analytics in the cloud in the discussion of big data. Some surveys focus on big data management and its related tools, while others have limited investigation of \gls{IVA} in a particular context. They also do not consider the growing unstructured videos as video big data. Unlike existing work, this paper provides a comprehensive assessment of state-of-the-art literature and proposes an in-depth distributed cloud computing-based \gls{IVA} reference architecture.
\section{Scope and Nomenclature}
\label{sec:ScopeAndNomenclature}
We clarify some nomenclature being used and scoping this study. The first one is \gls{IVA}, which is “any surveillance solution that utilizes video technologies to automatically manipulate and/or perform actions on live or stored video images \cite{elliott2010intelligent}”. The \gls{IVA} services are implemented through hardware called \gls{VAS} \cite{liu2013intelligent}. \gls{VAS} assist acquires videos continuously and monitors unblinkingly. The \gls{VAS} falls into four categories, i.e., \gls{EVAS}, \gls{OVAS}, \gls{FVAS}, and \gls{CVAS}. 

\gls{EVAS} embeds \gls{IVA} solutions and performs video analysis directly on the edge device, e.g., camera or encoder, and can produce alerts in case of abnormality. \gls{EVAS} provides very plain video analytics solutions and can simultaneously perform two or three rules on its stream and cannot accomplish complex algorithms such as fire detection, facial recognition, or cross video stream analytics. Under \gls{OVAS}, small and middle-sized companies consist of networked or wireless cameras, a network router, a system running the video analytics and management software (e.g., IBM smart surveillance system \cite{tian2008ibm}, and Zoneminder \cite{sujana2004zoneminder}), and a storage device. All the cameras send the video data for analytics against contextual video analytics algorithms and warn the operator if anomaly detected. \gls{OVAS} has many limitations, e.g., maintenance, software up-gradation, expensive hardware, scalability, and unable to deal with large-scale video data.

When \gls{IVA} solutions are provided in fog and cloud computing environment, then it is called \gls{FVAS}, and \gls{CVAS}, respectively. In such environments the \gls{IVA} solutions are made available under the \gls{aaS} paradigm, i.e., \gls{IVAaaS}.  In \gls{FVAS}, the \gls{IVA} solutions are geographically distributed and configured near the edge devices, i.e., video stream sources, to meet the strict real-time \gls{IVA} requirements of large-scale video analytics, which must address latency, bandwidth, and provisioning challenges. Whereas the \gls{CVAS} is more suitable for offline \gls{IVA} because of the relatively high response time and latency. The hierarchy and relation among \gls{CVAS}, \gls{FVAS}, and \gls{EVAS} is shown in Fig.~\ref{fig:CloudFogEdge}. The batch video data analytics are performed in the cloud, while the real-time \gls{IVA} is performed in the fog. The \gls{EVAS} can play a passive role in the proposed architecture, e.g., feed the video streams to the \gls{CVAS} if motion is detected. 

The scope of this paper is \gls{FVAS}, and \gls{CVAS}, i.e., the real-time and batch \gls{IVA} solutions are deployed in the fog and cloud computing environment, respectively, while utilizing big data computing technology. However, for the ease of understandability, throughout this paper, we use the notion of \gls{CVAS}.

\Figure[!t]()[width=0.47\textwidth]{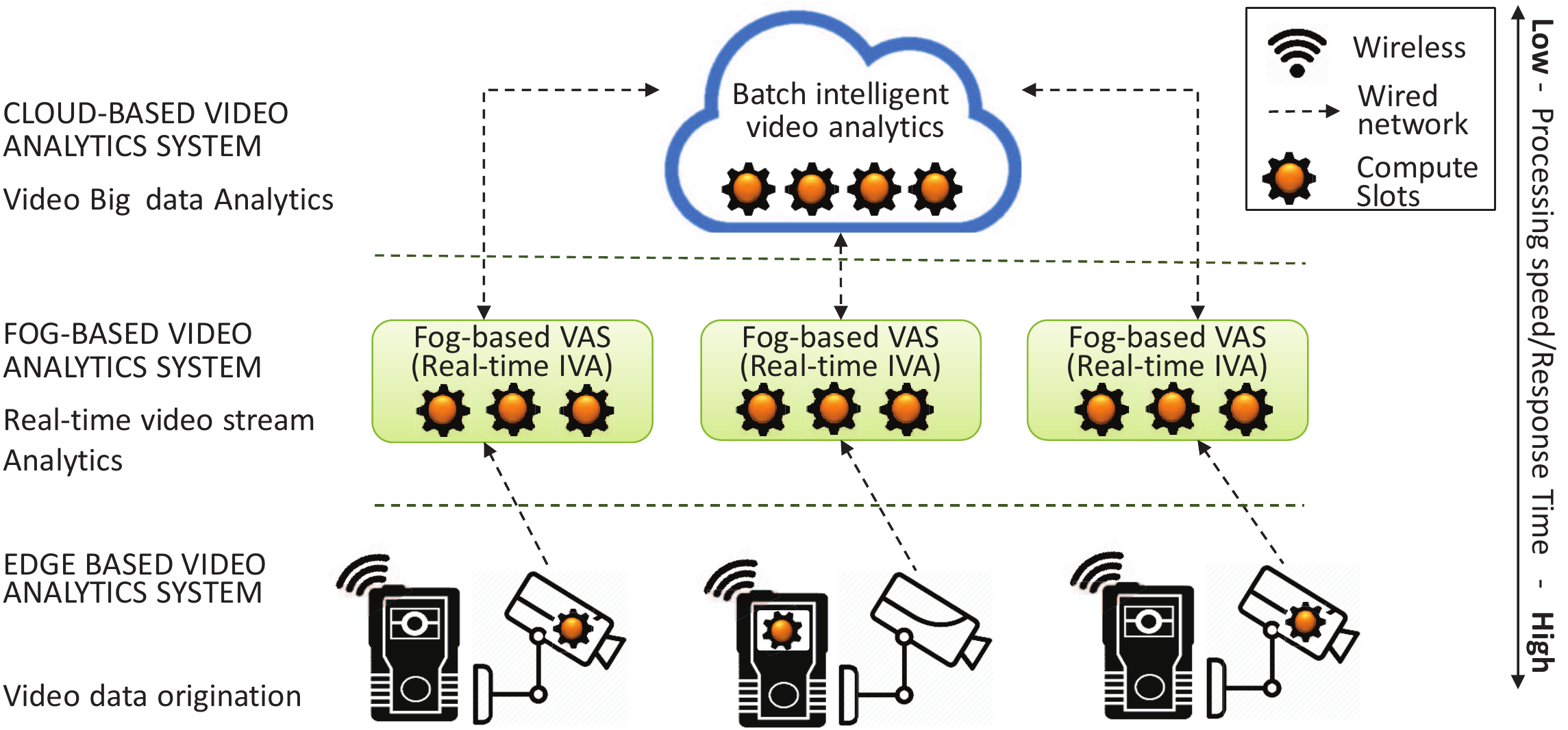}
{Geo-distributed video analytics infrastructure.
\label{fig:CloudFogEdge}}

\begin{figure*}[htbp]
	\centering
	\includegraphics[width=15cm]{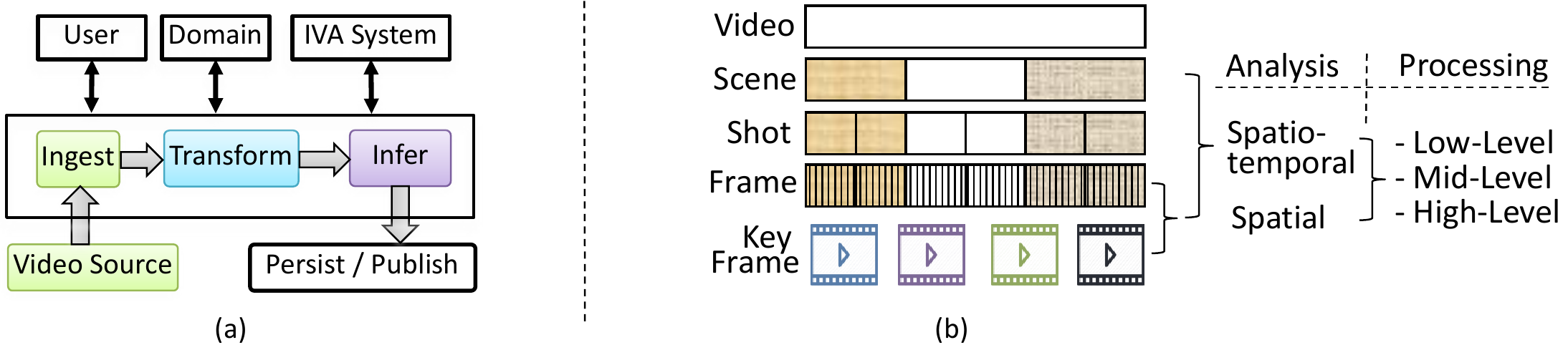}
	\caption{(a) A generic IVA service pipeline. (b) Hierarchal representation of video units.}
	\label{fig:BasicIVAPipeline}
\end{figure*}

To analyze a video in \gls{CVAS}, a video undergoes through different phases, as shown in Fig.~\ref{fig:BasicIVAPipeline}a. In Fig.~\ref{fig:BasicIVAPipeline}a, \texttt{Video Source} are the sources which either generate videos streams from sources connected directly to real-world domains such as IP-camera or can be already acquired videos in the form of datasets residing in a file system. If \gls{IVA} in the cloud are performed on real-time video streams, then we called it \gls{RIVA} and \gls{BIVA} if performed on batch videos. 

The \texttt{Ingest} phase implement interfaces to acquire videos from the \texttt{Video Source}. In the context of \gls{IVA}, the acquired videos can be represented as a hierarchy, as shown in Fig.~\ref{fig:BasicIVAPipeline}b. An acquired video from a \texttt{Video Source} may be decomposed into its constituent units either in the temporal or spatial domain.  In a video, a frame represents a single image, whereas a shot denotes a consecutive sequence of frames recorded by a single camera. A scene is semantically related shots in a sequence that depicts a high-level story. A collection of scenes composes a sequence/story. Frames and shots are low-level temporal features suitable for machines, while scenes and sequence/story are considered to be the high-level features that are suitable for human perceptions. Such constituent units are further subject to low, mid, or/and high-level processing. In low-level processing, primitive operations (in \texttt{Transformations} phase) are performed e.g., noise reduction, histogram equalizer. The \texttt{Infer} phase encompasses mid and high-level processing. The mid-level processing extracts features from the sequence of frames, e.g., segmentation, description, classification, etc. The high-level processing, make sense of an ensemble of recognized objects; perform the cognitive functions normally associated with vision. Finally, the extracted information can be persisted to the data store and/or published to the end-user.

Furthermore, the basic unit of an \gls{IVA} service pipeline is an algorithm, e.g., encoder, feature extractor, classifier, etc.  The input of an algorithm can be a \texttt{Video Source}, keyframes, or features. Similarly, the output of an \gls{IVA} algorithm can be high, mid, or low-features. Throughout this study, all possible outputs of \gls{IVA} algorithms are termed as \gls{IR}. Multiple algorithms can be pipelined to build a domain-specific \gls{IVA} service. The input and output of an \gls{IVA} service are restricted to the \texttt{Video Source} and \gls{IR}, respectively. The \texttt{User} represents the stakeholder of the \gls{CVAS}, such as administrator, consumer, \gls{IVA} researchers, and practitioners. A \texttt{Domain} is a specific real-word environment, e.g., street, shop, road traffic, etc., for which an \gls{IVA} service needs to be built for automatic monitoring. Domain knowledge facilities \gls{IVA} in discovering interesting patterns from domain video streams. The combination of software and hardware constitutes s distributed \texttt{System} (cloud environment) where \gls{IVA} service and algorithms can run fast. Nevertheless, upgrading existing \gls{IVA} algorithms to distributed architecture requires customization of how \gls{IVA} algorithms should be implemented and deployed. Moreover, the specific needs of \gls{IVA} may require the design and development of new system architecture. All the symbols being used in this paper are listed below.

\printglossary[type=\acronymtype, title= List of abbreviations, nonumberlist]
\section{Lambda CVAS: A Reference Architecture}
\label{sec:CVAS}

In this section, we briefly presents the technical details of the proposed \gls{CVAS} (called \gls{SIAT}) and the technical detail of each layer in the consecutive sub-sections. Fig.~\ref{fig:IVAaaSRefArchitecture} presents the proposed reference cloud-based layered architecture for distributed \gls{RIVA} and \gls{BIVA}. The proposed \gls{SIAT} architecture consists of five layers i.e., \gls{VBDCL}, \gls{VBDPL}, \gls{VBDML}, \gls{KCL}, and \gls{WSL}.

\gls{VBDCL} is the foundation layer and is responsible for large-scale big data management throughout the life-cycle of \gls{IVA}, i.e., from data acquisition to early persistence to archival and deletion \cite{alam2020tornado}. \gls{VBDPL} is responsible for distributed video pre-processing, feature extraction, etc. The \gls{VBDML} deploys \gls{IVA} algorithms on the top of distributed processing engines intending to produce high-level semantics from the processed sequence of frames. On top of the \gls{VBDML} layer, the KCL layer has been designed to link the low-level features in spatial and temporal relation across videos in a multi-stream environment. \gls{KCL} deploys a generic video ontology. The \gls{KCL} layer maps the extracted IR to the video ontology to bridge the semantic gap between the low-level features in Euclidean space and temporal relation across videos while utilizing semantic rich queries. The proposed architecture incorporates top-notch functionalities of the above four layers into a simple unified role base \gls{WSL}, which enables the \gls{SIAT} users to manage, built, and deploy a wide array of domain-specific near \gls{RIVA} and \gls{BIVA} services. 

All the layers are made available as \gls{aaS}. These \gls{IVAaaS} are provided to the domain experts and allow them to pipelined in a specific context to built an \gls{IVA} service. These \gls{IVA} services are made available as \gls{IVAaaS} to which users can subscribe Video Sources.

Functionalities like security, scalability, load-balancing, fault-tolerance, and performance are mandatory and common to all the layers, which are shown as a cross-cutting in Fig.~\ref{fig:IVAaaSRefArchitecture}. The cloud infrastructure provides the underlying hardware and software under \gls{IaaS}, on which the \gls{SIAT} can be deployed. The cloud infrastructure is out of the scope of this paper. It has already been studied in detail in the context of big data by \cite{hashem2015rise,agrawal2011big,mabrouk2018abnr}. However, in the discussion of \gls{IVA} in the cloud, some resources like CPU, GPU, FPGA, HDD, and SSD can be considered.

\Figure[!t]()[width=0.98\textwidth]{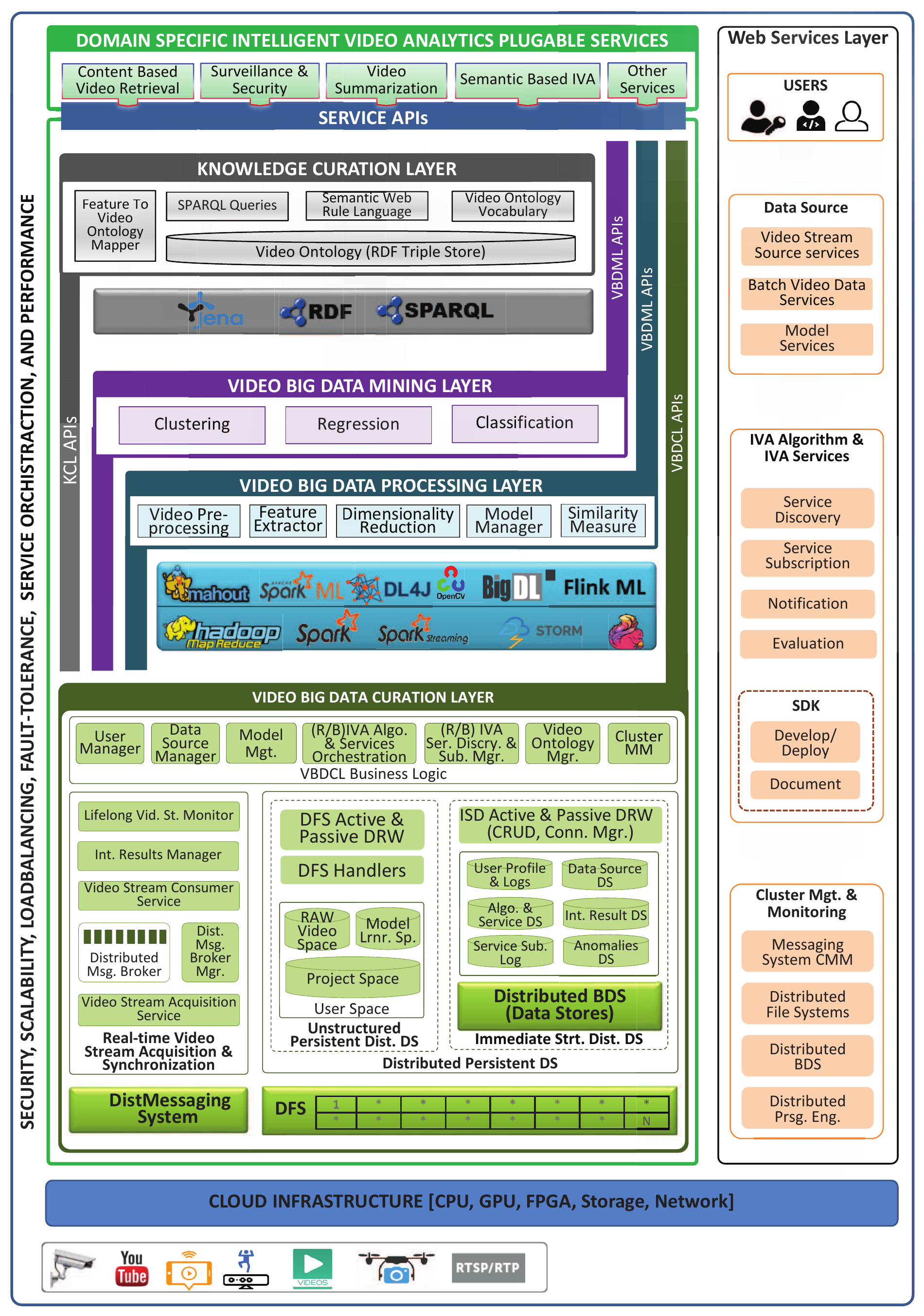}
{Lambda CVAS: A reference architecture for real-time and batch intelligent video analytics in the cloud.
\label{fig:IVAaaSRefArchitecture}}

\subsection{Video Big Data Curation Layer}
\label{Sec:VBDCL}
Effective data management is key to extract insights from the data. It is a  petascale storage architecture that can be accessed in a highly efficient and convenient manner. We design the \gls{VBDCL} for \gls{SIAT} to efficiently manage video big data. \gls{SIAT}’s data storage stack consists of three main components: \gls{RVSAS}, \gls{DPDS}, and \gls{VBDCL} Business Logic.


\begin{table*}[!htb]
	\centering
	\caption{Popular message-oriented middlewares.}
	\label{tab:MessageBrokers}
	\resizebox{\textwidth}{!}{%
		\begin{tabular}{|l|l|l|l|l|}

			\hline
			\textbf{} & 
			\textbf{Kafka} \cite{kreps2011kafka}& 
			\textbf{RabbitMQ} \cite{videla2012rabbitmq} & 
			\textbf{RocketMQ} \cite{ApacheRocketMQ2012} & 
			\textbf{ActiveMQ} \cite{snyder2011activemq}
			\\ \hline
			
			Format &
			String, Byte &
			Map, String &
			String, Byte &
			String, Map, Byte, Object 	
			
			\\ \hline
			Protocol &
			Kafka &
			\begin{tabular}[c]{@{}l@{}}AMQP, MQTT,\\ STOMP, XMPP\end{tabular} &
			Rocket, MQ &
			\begin{tabular}[c]{@{}l@{}}
				STOMP, REST, \\ XMPP, AMQP
			\end{tabular}
			\\ \hline
			Pub/Sub & 
			\cmark  & 
			\cmark  & 
			\cmark  & 
			\cmark                  
			\\ \hline
			
			\begin{tabular}[c]{@{}l@{}}Point-to-Point\end{tabular}  & 
			-              & 
			\cmark                  & 
			\cmark                  & 
			\cmark                  
			\\ \hline
			
			Cluster &
			\begin{tabular}[c]{@{}l@{}}
				Stateless cluster
			\end{tabular} &
			\begin{tabular}[c]{@{}l@{}}
				Simple clusters
			\end{tabular} &
			\begin{tabular}[c]{@{}l@{}}
				Stateless cluster
			\end{tabular} &
			\begin{tabular}[c]{@{}l@{}}
				Simple clusters
			\end{tabular} 
			\\ \hline
			
			\begin{tabular}[c]{@{}l@{}}Load balancing\end{tabular}  & 
			\cmark               & 
			\cmark                  & 
			\cmark                  & 
			\cmark                  
			\\ \hline
			
			Persistence                                                & 
			\cmark               & 
			\cmark                  & 
			\cmark                  & 
			\cmark                  \\ 
			\hline
			Encryption                                                 & 
			SSL     & 
			TLS     & 
			TLS     & 
			TLS     
			\\ \hline
			\begin{tabular}[c]{@{}l@{}}
				Authentication
			\end{tabular} & 
			SASL & 
			SASL & -                 & 
			JAAS 
			\\ \hline
			
			APIs &
			
			\begin{tabular}[c]{@{}l@{}}
				Java, .NET,	PHP, Go, JS\\
				Python, Ruby, Perl, Erlang 
			\end{tabular} &
			
			\begin{tabular}[c]{@{}l@{}}
				Java, .NET, PHP, Ruby, \\
				Python, Go, JavaScript
			\end{tabular} &
			
			\begin{tabular}[c]{@{}l@{}}
				Java, C++, Go
			\end{tabular} &
			
			\begin{tabular}[c]{@{}l@{}}
				Java, PHP, C++, Ruby, \\ Perl, C\#, Python, C
			\end{tabular} \\ \hline

		\end{tabular}%
	}
\end{table*}


\subsubsection{Real-time Video Stream Acquisition and Synchronization}
\label{sec:RVSAS}
The real-time video stream needs to be collected from the source device and forwarded to the executors for on-the-fly processing against the subscribed \gls{IVA} service. Handling a tremendous amount of video streams, both processing and storage are subject to lose \cite{zhang2015video}. To handle, large-scale video stream acquisition in real-time, to manage the \gls{IR}, anomalies, and the communication among \gls{RIVA} services, we design the \gls{RVSAS} component while assuming a distributed messaging system.  

Distributed Message Broker, also known as message-oriented-middleware \cite{10.1007/3-540-48169-9_1}, is an independent application that is responsible for buffering, queuing, routing, and delivering the messages to the consumers being received from the message producer \cite{ejsmont2015web}. Message broker should be able to handle permission control and failure recovery. A message broker generally supports routing methods like direct worker queue, and/or publish-subscribe \cite{2015:WSS:2935460}. Similarly, the message consumer component receives the messages from the message broker either periodically (cron-like consumer) or continuously (daemon-like consumer). Generally and for the sake of scalability, message consumers are deployed on separate servers independently of message producers \cite{2015:WSS:2935460}. Some popular distributed messaging systems are shown in table~\ref{tab:MessageBrokers}.  

\gls{RVSAS} provides client APIs on the top of a distributed messaging system for the proposed framework. The \gls{RVSAS} component is responsible for handling and collecting real-time video streams from device-independent video data sources. Once the video stream is acquired, then it is sent temporarily to the distributed broker server. The worker system, on which an \gls{IVA} service is configured, e.g., activity recognition, reads the data from the distributed broker and process. The \gls{RVSAS} component is composed of five sub-modules, i.e., \gls{DMBM}, \gls{VSAS}, \gls{VSCS}, \gls{IRM}, and \gls{LVSM}.

\begin{figure}[h]
	\centering
	\includegraphics[width=7.4 cm]{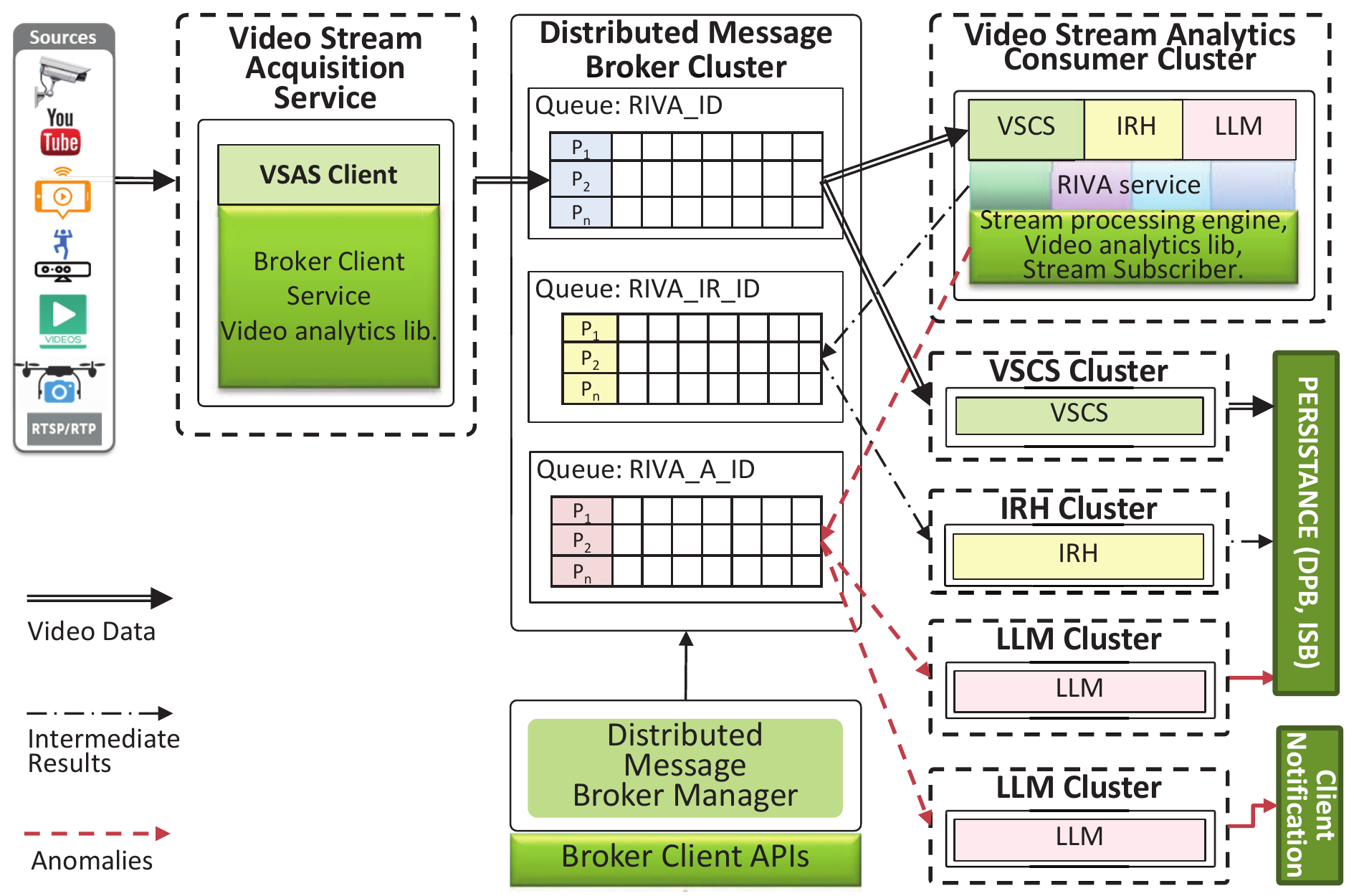}
	\caption{Real-time Video Stream Acquisition and Sync.}
	\label{fig:RVSAS}
\end{figure}

\paragraph{Distributed Message Broker Manager} 
\gls{DMBM} are used to manage the queues in the distributed message broker cluster considering \gls{RIVA} services. Three types of queues, \texttt{RIVA\_ID}, \texttt{RIVA\_IR\_ID}, and \texttt{RIVA\_A\_ID} as shown in Fig.~\ref{fig:RVSAS}, are automatically generated by the \gls{DMBM} module  on the distributed message broker when a new \gls{RIVA} service is created. Here \texttt{RIVA}, \texttt{ID}, \texttt{IR}, and \texttt{A} stands for \gls{RIVA} service, unique identifier of the service, \gls{IR}, and Anomalies, respectively. These queues are used to hold the actual video stream being acquired by \gls{VSAS}, \gls{IR} produced by an algorithm, and anomalies detected by the video analytics services.

\paragraph{Video Stream Acquisition Service} 
\gls{VSAS} module is used to provide interfaces to \gls{VSDS} and acquires large-scale streams from device-independent video data sources for on-the-fly processing. If a particular video stream source is subscribed against an \gls{RIVA} service, then the \gls{VSAS} gets the configuration metadata from the Data Source DS in \gls{ISDDS} and configure the source device for video streaming. After successful configuring the source device, \gls{VSAS} decodes the video stream, detects the frames, and then performs some necessary operations on each frame such as meta-data extraction and frame resizing, which is then converted to a formal message. These messages are then serialized in the form of mini-batches, compressed, and sent to the Distributed Broker. If a video-stream source “$C_1$” is subscribed to the \gls{RIVA} service “$S_1$” then the \gls{VSAS} will rout the mini-batch of the video stream to queue \texttt{RIVA\_1} in the Broker Cluster as shown in Fig.~\ref{fig:RVSAS}.

\paragraph{Video Stream Consumer Service} 
As the acquired video streams are now residing in the distributed broker in different queues in the form of mini-batches. To process these mini-batches of the video stream, we have different groups of computer cluster know as \gls{VSAC} Cluster. On each cluster, three types of client APIs are configured, i.e., \gls{RIVA} services, \gls{VSCS}, and \gls{LVSM}. Each \gls{VSAC} cluster has different domain-specific \gls{RIVA} services where the \gls{VSCS} are common for all.  The \gls{VSCS} assists the \gls{RIVA} service to read the mini-batches of the video stream from the respective queue in the distributed broker for analytics, as shown in Fig.~\ref{fig:RVSAS}. The \gls{VSCS} module has two main functions. First, this module allows \gls{RIVA} service to read the mini-batches of the video stream from the respective queue in the distributed broker. The second task is to save the consumed unstructured video streams and its meta-data to the row video space in the \gls{DPDS} and to the Data Source DS meta-store, respectively.

\paragraph{Intermediate Results Manager} 
During the \gls{IVA} service life-cycle, a sequence of algorithms are executed. Thus the output of an algorithm can be the input of another algorithm. The \gls{IR} demands proper management in the distributed environment because one algorithm may be on one computer while the other may be on another computer. Thus we design \gls{IRM} that sends and gets the \gls{IR} to and from the topic \texttt{RIVA\_IR\_ID} in the distributed broker cluster. Similarly, this module is also responsible for reading the \gls{IR} from the respective queue and persists to the \gls{IR} data store for future use so that to avoid recomputation.

\paragraph{Lifelong Video Stream Monitor} 
The domain-specific \gls{RIVA} service process the video stream for anomalies or abnormal activities. If \sout{an} anomalies are detected, then it is sent to the distributed broker queue (i.e., \texttt{RIVA\_A\_ID}) by using the \gls{LVSM} instance. To generate notification base response, \gls{LVSM} follow standard observer-based concept \cite{gamma1995design}. Based on this approach, the \gls{LVSM} module reads the anomalies from the respective distributed broker queue, i.e., \texttt{RIVA\_A\_ID} and notify the clients in near real-time and simultaneously persisted to the \gls{ISDDS}.

\subsubsection{Distributed Persistence Data Store}
The second component of the \gls{VBDCL} is \gls{DPDS}. The \gls{DPDS} component provides the facilities of permanent and distributed big-data persistence storage of both structured and unstructured data. The \gls{DPDS} provides abstraction in two levels on the acquired video data ie. \gls{ISDDS} and \gls{UPDDS}. The philosophy behind \gls{DPDS} and two levels of abstraction in the context of \gls{SIAT} is many folds. From the users’ perspective, the \gls{CVAS} demands geo-based, real-time, low latency, and random read-write access to the data in the cloud. Similarly, the \gls{DPDS} should also provide high-performance locality-based access to the data when the other layers deploy data-intensive \gls{IVA} services. To meet such a diverse amount of requirements of the \gls{DPDS} component, technologically, a \gls{DFS} and \gls{DBDS} can be leveraged. 

\paragraph{Immediate Structured Distributed Datastore} 


\begin{table*}[!htb]
	\centering
	\caption{Popular Distributed File Systems.}
	\label{tab:dfs}
	\resizebox{\textwidth}{!}{%
		
		\begin{tabular}{|l|l|l|l|l|l|l|}
			
			\hline
			
			\textbf{Features} &
			\textbf{CEPH} \cite{Weil:2006:CSH:1298455.1298485} &
			\textbf{GlusterFS} \cite{boyer2012glusterfs} &
			\textbf{HDFS} \cite{5496972} &
			\textbf{Lustre} \cite{braam2002lustre} &
			\textbf{MooseFS} \cite{GASTON20091768} &
			\textbf{Quantcast File} \cite{Ovsiannikov:2013:QFS:2536222.2536234} 
			
			\\ \hline

			Initial release &
			2007 &
			2007 &
			2006 &
			2003 &
			2008 &
			2011 
			\\ \hline

			License &
			LGPL 2.1 &
			GPLv3+ &
			Apache 2.0 &
			GPLv2 &
			GPLv2 &
			Apache 2.0 
			\\ \hline
			
			Availability &
			\cmark &
			\cmark &
			\begin{tabular}[c]{@{}l@{}}Transparent \\ master failover\end{tabular} &
			\cmark &
			Master &
			Master 
			\\ \hline
			
			Tolerate disk failures &
			\cmark &
			\cmark &
			\cmark &
			\cmark &
			\cmark &
			\cmark 
			\\ \hline
			Scale &
			PBs &
			PBs &
			PBs &
			PBs &
			PBs &
			PBs 
			\\ \hline
			Shards &
			\cmark &
			\cmark &
			\xmark &
			\cmark &
			\xmark &
			\xmark 
			\\ \hline
			Commodity Hardware &
			\cmark &
			\cmark &
			\cmark &
			\cmark &
			\cmark &
			\cmark 
			\\ \hline
			Language &
			C++ &
			C &
			Java &
			C &
			C &
			C 
			\\ \hline
			POSIX Complaint &
			\cmark &
			\cmark &
			\xmark &
			\cmark &
			\cmark &
			\cmark 
			\\ \hline
			APIs &
			\begin{tabular}[c]{@{}l@{}}
				Librados (C, \\
				Python, C++, \\
				Ruby), S3, \\
				Swift, FUSE \\
			\end{tabular} &
			
			\begin{tabular}[c]{@{}l@{}}
				NFS, FUSE, SMB, \\
				Swift, libgfapi, \\
				libglusterfs\end{tabular} &
			
			\begin{tabular}[c]{@{}l@{}}
				Java and \\
				C client, HTTP
			\end{tabular} &
			
			\begin{tabular}[c]{@{}l@{}}
				NFS-Ganesha, \\
				NFS, SMB
			\end{tabular} &
			
			FUSE &
			\begin{tabular}[c]{@{}l@{}}
				C++ client, \\
				FUSE
			\end{tabular} 
			\\ \hline
			
		\end{tabular}%
	}
\end{table*}


The \gls{ISDDS} is provided to manage large-scale structured data in the distributed environment over \gls{DBDS}. Because of the data-intensive operation and according to the requirements of the other layer, technologically, a distributed big data store, can be deployed. The \gls{ISDDS} hosts five types of data. The detailed description of each type of data has been described in this section.

\gls{SIAT} provides role-based access to its user. \gls{SIAT} user logs and the respective role information are maintained through the User Profile and Logs meta-store. The proposed framework manages two types of video data sources through the Data Source meta-store. These are video data sources, for example, IP-cameras, Kinect, body-worn cameras, etc., and batch video datasets. The former one can be subscribed to RIVA service while the later one is eligible for BIVA services. The meta-information of these sources, along with access rights, are managed through the Data Source meta-store. Administrator and developer roles can develop, create, and deploy video analytics algorithms through the \gls{SIAT}. Similarly, different \gls{IVA} algorithms can be pipelined into an \gls{IVA} service. The management of video analytics algorithms and services is managed through Video Analytics Algorithm and Service meta-store, respectively. As stated that in \gls{IVA} pipelining environment, the output of one \gls{IVA} algorithm can be the output of another algorithm. In this context, we design a general container called \gls{IR} datastore to persist and index the output of an \gls{RIVA} algorithm, and services. This datastore is significant and can play a vital role in \gls{IVA} pipeline optimization, and fast content-based searching and retrieval. Finally, the \gls{SIAT} users are allowed to subscribe to the data sources to the \gls{IVA} services. The subscription information is maintained through the Subscription meta-store, and the anomalies are maintained through the Anomalies meta-store. 


The \gls{ISDDS}  Data Model of the \gls{SIAT} demands an efficient distributed data store. The distributed data store should have the ability of horizontal scalability, high availability, partition tolerance, consistency, and durability. Furthermore, the data store should fulfill the read/write access demands of the \gls{BIVA} operations, and \gls{RIVA}, interaction, and visualization. It is a fact that traditional relational databases have little or no ability to scale-out to accommodate the growing demands of the big data, and resultantly new distributed data stores have emerged. The distributed data stores can be grouped into two major categories, i.e., NoSQL (Not Only SQL) and NewSQL (excluding graph data stores).

NoSQL is a schema-free data store designed to support massive data storage across distributed servers \cite{GANESHCHANDRA201513, moniruzzaman2013nosql}. The features of NoSQL data stores include horizontal scalability, data replication, distributed indexing, simple API, flexibility, and consistency \cite{Cattell:2011:SSN:1978915.1978919}. NoSQL lacks true \gls{ACID} transactions, unlike \gls{RDBMS}. In the context of \gls{CAP} theorem \cite{Brewer:2012:PCS:2360751.2360957}, it has to compromise on either consistency or availability while choosing partition tolerance. NoSQL can further be categorized as Document, Key-value, and Extensible stores. A key-value data store is responsible for storing values and indexes for searching. Document datastore is used for document storage, indexing, and retrieval. Extensible data store stores extensible records that can be partitioned vertically and horizontally across the nodes. 

NewSQL data stores provide the characteristics of both NoSQL and RDBMS:  \gls{ACID} transactional consistency of relational databases with facilities of SQL; and the scalability and performance of NoSQL.  MySQL Cluster, VoldDB, ClustrixDB are examples of NewSQL. Some of the popular NewSQL and NoSQL datastores are shown in Table~\ref{tab:DDS} along with the respective properties. The ‘A’ and ‘C’ in \gls{CAP} is not equal to the ‘A’ and ‘C’ in \gls{ACID} \cite{6133253}.


\begin{table*}[!htb]
	\caption{Popular Distributed Data Stores.}
	\label{tab:DDS}
	\resizebox{\textwidth}{!}{%
		\begin{tabular}{|l|l|l|l|l|l|l|l|l|l|}
			\hline
			\textbf{Type}              & \textbf{DS Name}       & \textbf{Release} & \textbf{Developer} & \textbf{OS} & \textbf{Language}  & \textbf{D} & \textbf{HS} & \textbf{SL} & \textbf{Properties}  \\ \hline

			\multirow{3}{*}{\rotatebox[origin=c]{90}{Key-value }} 
			& Voldemort     & 2009    & LinkedIn & \cmark  & Java                              & \cmark & \cmark  & \cmark & C\underline{\textbf{AP}} \\ \cline{2-10} 
			& Riak          & 2009    & \begin{tabular}[c]{@{}l@{}}Basho\\ Tech.\end{tabular}     & \cmark  & Erlang    & \cmark & \cmark & \cmark &  C\underline{\textbf{AP}} \\ \cline{2-10} 
			& Redis \cite{Carlson:2013:RA:2505464}         & 2009    & \begin{tabular}[c]{@{}l@{}}Salvatore\\ Sanfilippo\end{tabular} & \cmark & C & \cmark & \cmark  & \cmark & \underline{\textbf{C}}A\underline{\textbf{P}} \\ \hline

			\multirow{4}{*}{\rotatebox[origin=c]{90}{Document }} 
			& SimpleDB      & 2007    & Amazon & \xmark  & Erlang    & \cmark & \cmark  & \cmark & C\underline{\textbf{AP}} \\ \cline{2-10} 
			& CouchDB       & 2005    & ASF & \cmark  & Erlang       & \cmark & \cmark  & \cmark & C\underline{\textbf{AP}} \\ \cline{2-10} 
			& MongoDB \cite{Banker:2011:MA:2207997}      & 2009    & 10gen & \cmark  & C, C++     & \cmark & \cmark  & \cmark & \underline{\textbf{C}}A\underline{\textbf{P}} \\ \cline{2-10} 
			& DynamoDB      & 2012    & Amazon  & \xmark  & Java     & \cmark & \cmark  & \cmark & C\underline{\textbf{AP}} \\ \hline

			\multirow{4}{*}{\rotatebox[origin=c]{90}{ Extensible }} 
			& HBase \cite{vora2011hadoop} & 2010  & ASF     & \cmark    & Java      & \cmark    & \cmark  & \cmark  & \underline{\textbf{C}}A\underline{\textbf{P}} \\ \cline{2-10} 
			& HyperTable    & 2008    & Zvents    & \cmark    & C++       & \cmark  & \cmark  & \cmark & \underline{\textbf{C}}A\underline{\textbf{P}} \\ \cline{2-10} 
			& Cassandra     & 2008    & ASF       & \cmark    & Java      & \cmark  & \cmark  & \cmark & C\underline{\textbf{AP}} \\ \cline{2-10} 
			& BigTable    \cite{Chang:2008:BDS:1365815.1365816}  & 2005    & Google    & \xmark    & C, C++    & \cmark  & \cmark  & \cmark & \underline{\textbf{C}}A\underline{\textbf{P}} \\ \hline

			\multirow{4}{*}{\rotatebox[origin=c]{90}{NewSQL}} 
			& MySQL Cluster & 2004    & Oracle                                                         & \cmark  & C++       & \cmark & \cmark  & \xmark & ACID \\ \cline{2-10} 
			& VoldDB  \cite{stonebraker2013voltdb}      & 2009    & VoltDB Inc                                                     & \cmark  & Java, C++ & \cmark & \cmark  & \xmark & ACID \\ \cline{2-10} 
			& ClustrixDB    & 2007    & \begin{tabular}[c]{@{}l@{}}Clustrix,\\ Inc.\end{tabular}       & \xmark  & C         & \cmark & \cmark  & \xmark & ACID \\ \cline{2-10} 
			& NuoDB         & 2008    & NuoDB                                                          & \xmark  & C++       & \cmark & \cmark  & \xmark & ACID \\ \hline
			
			\multicolumn{10}{l}{\textit{
					\begin{tabular}[c]{@{}l@{}}\\ 
						P=Persistence, D=Durability, HS=High Scalability, SL=Schema-less \\
						CAP=Consistency, Availability, Partition Tolerance \\
						ACID= Atomicity, Consistency, Isolation, Durability \\
						In the properties column, the underline character represents the supported properties in the CAP theorem. \\
			\end{tabular}}}
		\end{tabular}
	}
\end{table*}

\begin{figure}[h]
	\centering
	\includegraphics[width=7.4 cm]{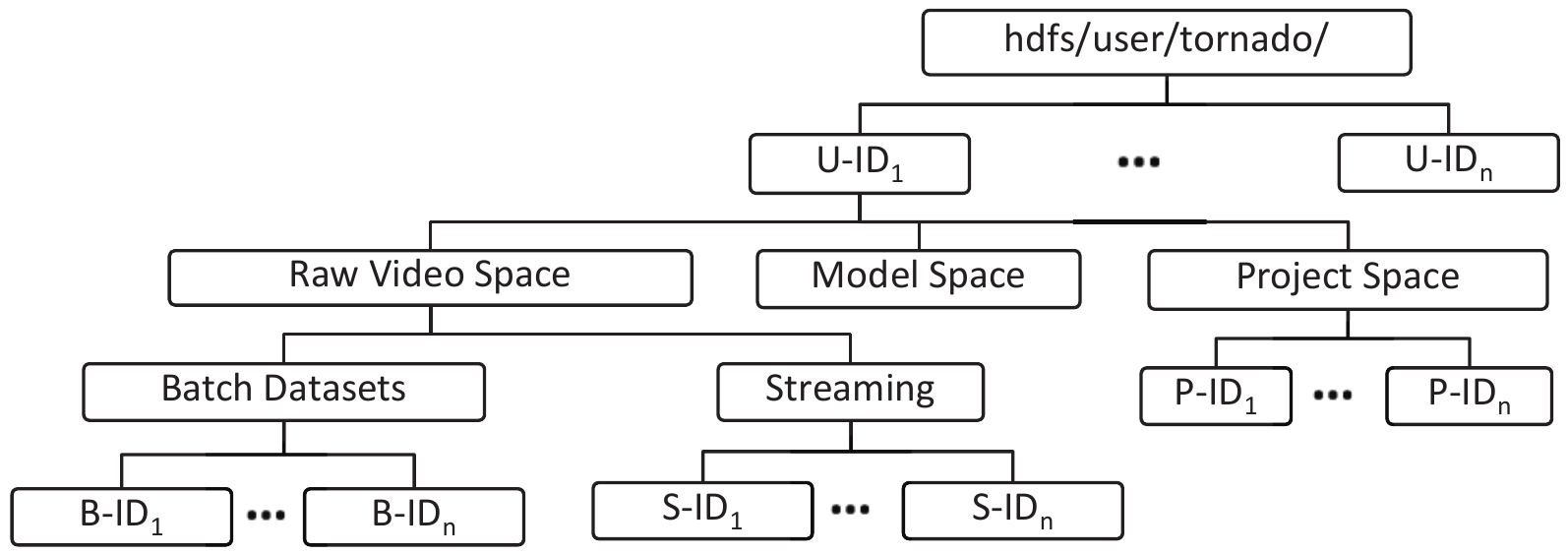}
	\caption{Hierarchical structure of the User Space in DFS.}
	\label{fig:DirectorySpace}
\end{figure}

\paragraph{Unstructured Persistent Distributed Datastore} The \gls{UPDDS} component built on the top of the \gls{DFS} such as \gls{HDFS} that facilitates permanent and distributed big-data storage. The data are stored and mapped systematically according to the business logic of the proposed system. The \gls{UPDDS} component is designed to effectively and dynamically manage the data workload in the life-cycle of an \gls{IVA} service. Upon new registration with the \gls{SIAT}, a formal User Space is created on the top of \gls{DFS}. The User Space is managed through the proper hierarchical directory structure and special read and writes access are granted to the owner. All the directories are synchronized and mapped in the corresponding user profile logs. Under the User Space, three types of distributed directories are created, i.e., Raw Video Space, Model Space, and Project Space. The hierarchical structure of the User Space in \gls{DFS} is shown in Fig.~\ref{fig:DirectorySpace}.

Raw Video Space is used for the management of the video data. Raw Video Space is further divided into two types of video spaces. The first type is a batch video which has been uploaded to the \gls{SIAT} for batch analytics, where the second type is acquired and persisted from the \gls{VSDS}.  The entire acquired stream is time-stamped on persistence. The granularity level of raw streaming videos is maintained through video data sources. \gls{IVA} life-cycle may need different models for training and testing purposes. The Model Space is provided to facilitate the developers to manage the training and testing model according to the deployed \gls{IVA} algorithm. Similarly, a developer can develop a new algorithm or a video analytics service. The Project Space is provided to manage the source code of the respective developer and practitioners.

The \gls{UPDDS} is supposed to be designed on the top of \gls{DFS}. In this context, diverse types of open source \gls{DFS} have been developed. The implementation quality of big data applications is relative to the storage tier’s file system. From an architectural perspective, working and handling large volumes and throughput of data is challenging. Commonly, big data solutions exploit a cluster of computers, ranging from few to hundreds of computers, connected through high-speed networks while deploying specialized distributed data and management software. In distributed data-intensive applications, large-scale data is always moving across the cluster and thus demands a distributed, scalable, reliable, and fault-tolerant file system \cite{Ghemawat:2003:GFS:945445.945450} known as \gls{DFS}. \gls{DFS} is a file system that provides access to replicated files across multiple hosts on a computer network, ensuring performance, data locality, high availability, scalability, reliability, security, uniform access, and fault-tolerance. Currently, various \gls{DFS} are available and may differ in terms of performance, fault-tolerance, content mutability, and read/write policy. Some state-of-the-art popular \gls{DFS}, which can be used for \gls{UPDDS}, along with a short description, are shown in Fig~\ref{tab:dfs}.

\paragraph{Active and Passive Data Readers and Writer} 
This module gives read-write access to the underlying data securely according to the business logic of the \gls{VBDCL} Business Logic and according to the registered user access rights. This sub-module is composed of four types of readers and writers, i.e., \gls{ISDDS} Active Data Reader, \gls{ISDDS} Passive Data Reader, \gls{UPDDS} Active Data Reader, and \gls{UPDDS} Passive Writer. For real-time, read-write operation over the data residing in the \gls{ISDDS} and \gls{UPDDS}, such as CRUD operation, file creation, video stream writing to \gls{DFS}, etc. the active data reader and writer are used. In the context of offline analytics over the bulk of videos while using distributed processing engines. The Passive Data Reader and Writer (PDRW) is provided to allow processing engines to load the bulk of data and persist the same to the \gls{ISDDS} and \gls{UPDDS}. 

\subsubsection{VBDCL Business Logic}
\gls{VBDCL} Business Logic provides the actual business logic. It implements six different modules, i.e., User Manager, Data Source Manager, Model Manager, (R/B)\gls{IVA} Algorithm and Service Manager, (R/B)\gls{IVA} Service Discovery and Subscription Manager, Video Ontology Manager, Cluster Management and Monitoring.

The \texttt{User Manager} module encapsulates all the user-related operations such as new user account creation, access role assignment, and session management. Through the \texttt {Data Source Manager} \texttt{Model Manager} modules, the user can manage the \gls{VSDS}, video data uploading, and model management. The \texttt{(R/B)IVA Algorithm and Service Manager} are built to manage, develop, and deploy new IVA algorithms and services, respectively. The former one is provided \gls{aaS} to the \gls{SIAT} developers, while the latter one is provided \gls{aaS} to the consumers. The developer role can create and publish a new video analytics algorithm. The algorithm is then made available \gls{aaS} to other developers and can use it. Once \gls{IVA} services are created, then the \gls{SIAT} users are allowed to subscribe to the streaming video data sources and batch data against the provided \gls{RIVA} and \gls{BIVA} services, respectively. 

Similarly, the \texttt{Ontology Manager} allows the developer to get the \gls{IR} for decision making. The ontology manager provides a secure way of getting the \gls{IR} and maps it according to ontology. This module also allows the user to manage the functionalities of the \texttt{KCL}. Finally, \texttt{Cluster Management and Monitoring} allows the administrator to monitor the health of the cluster.
\subsection{Distributed Intelligent Video Analytics}
\label{sec:DistributedVideoAnalytics}
\gls{IVA} performs the complex tasks of extracting significant knowledge and information of interest from the video data, i.e., structural patterns, behavior patterns, content characteristics, event patterns \cite{xie2008event, shyu2008video} and their relationship in the form of classification and clustering. Extracting knowledge and information from video big data is a processing-intensive task. Value extraction from video big data is behind the capabilities of traditional tools and demands for technological solutions to meet the processing requirements.

Therefore, video big data analytics is preferably performed in the distributed environment in a parallel manner while utilizing distributed scale-out computing technologies \cite{regazzoni2010video}, such as MapReduce and Apache Spark. Distributed \gls{IVA} not only significantly improves the performance, but also reduces the analytics cost. Based on the generic \gls{IVA} life-cycle and motivated by scikit-learn \cite{buitinck2013api}, the distributed \gls{IVA} are divided into two layers, i.e., \gls{VBDPL}, and \gls{VBDML}. These two layers are further elaborated in the following sub-sections.

\subsubsection{Video Big Data Processing Layer}
\label{sec:VBDPL}

\gls{IVA} requires video data pruning and strong feature extraction. With such intentions, the \gls{VBDPL} layer consists of three components, i.e., Video Preprocessing, Feature Extractor, and Dimensionality Reduction is designed.

\paragraph{Video Preprocessing} The quality of data plays an active and significant role in solving a problem with \gls{ML}. Raw videos have an unstructured format and contain noise/uncertainties, making it unsuitable for knowledge and information mining. \texttt{Video Preprocessing} component is designed with the same objectives and is supposed to deploy several distributed video preprocessing operations including frame extraction \cite{schultz1996extraction}, frame-resizing, frame-conversion from RGB to grayscale \cite{saravanan2010color}, shot boundary detection \cite{boreczky1996comparison}, segmentation \cite{hampapur1994digital}, transcoding \cite{vetro2003video}, and many more. In the first step, frames are extracted from a video for processing. Several frame selection algorithms are available, for example, keyframe extraction. The number of frames to be extracted is dependent on the user objective and task-dependent. Candidate frames can be all frames, step frames (every second frame, fifth frame, etc.) or keyframes \cite{zhuang1998adaptive}. The spatial operations highly depend on the scenario and objective. Spatial operations include frame resizing (for reducing computational complexity), corrections (brightness, contrast, histogram equalization, cropping, keyframes), mode (RGB, Grayscale, etc.), and many other operations. Segmentation is used for various purposes, such as partitioning video into semantically related chunks. 

\paragraph{Feature Extractor} 
The \texttt{Feature Extractor} component implements distributed feature extraction algorithms. The performance of \gls{ML} is highly dependent upon the type of data representation or features \cite{bengio2013representation}. Features represent the characteristics of classes in the dataset and have a heavy impact on the \gls{ML} algorithm's generalizability and performance. The data features that used to train \gls{ML} models have a huge influence on the performance of the algorithm. Inappropriate or irrelevant features affect the performance of the algorithm. Thus, feature extraction extracts the features from the raw videos that can be interpreted by the \gls{ML} algorithm \cite{baumann2014computation, baumann2016recognizing}. In this context, several feature extraction algorithms have been introduced for video data. These feature extraction approaches can be categorized into static features of keyframes \cite{amir2003ibm, adcock2004fxpal, yan2007review}, object features \cite{sivic2005person, visser2002object}, dynamic/motion feature extraction \cite{baumann2014computation, mattivi2009human, zhao2007dynamic}, trajectory-based features extraction \cite{baumann2014computation, baumann2016recognizing, nanni2011local, yi2017realistic}, and deep learning-based feature extraction \cite{uddin2019feature, simonyan2014two, karpathy2014large, wang2015action, lu2019action, yao2019learning, wang2018action, girdhar2017actionvlad, zhao2017pooling}.  

\paragraph{Feature Selection and Dimensionality Reduction}  
Feature selection and dimensionality reduction reduce the size of the features. Large sizes of feature sets are expensive in terms of time for training and/or performing classification acquired by trained classifiers. For example, \gls{PCA} and its variants are used to reduce the size of features.  During feature selection, most relevant features are selected by discarding irrelevant and weak features. The performance of \gls{ML} classifiers is also directly related to the quality of features; GIGO (garbage in garbage out). Inappropriate or partially relevant features can negatively affect model performance. Therefore, only a limited set of features should be selected and used for training classifiers. This is what precisely the purpose of this component is and deploy different algorithms in this context. Similarly, some feature reduction techniques available that selects the specific set of limited features in real-time. For example, Online Feature Selection selects and inputs a specific number of small features to the classifiers in real-time \cite{hoi2012online}. In order to accelerate the training process, \cite{tan2014approach} used non-linear and group-based feature selection techniques, based on \gls{AFS}, to process the data with substantial dimension sizes.  \cite{cong2016udsfs} proposed an unsupervised feature reduction technique that selects extremely relevant features and indicates suitable weights to the distinctive feature dimensions.


\subsubsection{Video Big Data Mining Layer}
\label{sec:VBDML}
The \gls{VBDML} utilizes diverse types of machine-learning algorithms, i.e., supervised, semi-supervised, and unsupervised algorithms to find different type of information from the videos \cite{xie2008event, shyu2008video}. In this context, \gls{VBDML} layer hosts three types of components, i.e., Classification, Regression, Clustering. 

Classification component provides various \gls{ML} algorithms, e.g., \gls{SVM}, Nearest Neighbors, Random Forest, Decision Tree, Naïve Bayes, etc., that identifies that a particular object in a video frame belongs to which category while using predefined classes. The Regression component includes different algorithms, e.g., Linear Regression, Decision Tree Regression, Logistic Regression, and many more, predicting a continuous-valued attribute associated with objects rather than discrete values. The Clustering component encapsulates algorithms, e.g., K-Mean, spectral clustering, etc., that produces groups of data depending upon the similarity of data items. 

\gls{SIAT} has the ability of \gls{BIVA} and \gls{RIVA}. In this context, the \gls{VBDML} should support batch learning and online learning. The former case considers the complete training data to learn and generates models. The batch-learning algorithm is expected to generalize, that usually does not perform well in the real environment. Unlike batch learning, online learning continuously learns from new input without making any statistical assumptions about the data \cite{dekel2009online}. In the context of model generalization, online learning is expected to work well by accurately predicting the predefined set of inputs  \cite{dekel2009online}. Online learning is used in the environment when continuous learning from the data is required to learn new patterns instead of batch learning.


\begin{table*}[!htb]
	\centering
	\caption{Comparision of popular deep learning architectures.}
	\label{tab:PopularDeepLearning}
	\resizebox{\textwidth}{!}{%
		\begin{tabular}{|l|l|l|l|l|l|l|}
			\hline
			\textbf{Architecture} &
			\textbf{Year} &
			\textbf{Error} &
			\textbf{Parameters} &
			\textbf{Depth} &
			\textbf{Description and Pros} &
			\textbf{Cons} \\ \hline
			LeNet &
			1998 &
			- &
			60 K &
			5 &
			\begin{tabular}[c]{@{}l@{}}-- First popular CNN architecture\\ -Utilize spatial correlation to decrease the \\ parameters and computation \\ -Automatic learning of feature hierarchies\end{tabular} &
			\begin{tabular}[c]{@{}l@{}}-Poor scaling to diverse classes \\ of images\\ -Large size filters\\ -Extraction of low-level features\end{tabular} \\ \hline
			AlexNet &
			2012 &
			15.3\% &
			60 M &
			8 &
			\begin{tabular}[c]{@{}l@{}}-Low, mid and high-level feature extraction \\ using large and small size filters on initial\\ (5x5 and 11x11) and last layers (3x3)\\ -The notion of deep and wide CNN architecture\\ -Introduced regularization in CNN\\ -Initiated GPUs parallelization\\ - Deeper and wider than the LeNet\\ - Uses Relu ,dropout and overlap Pooling\\ - GPUs NVIDIA GTX 580\end{tabular} &
			\begin{tabular}[c]{@{}l@{}}-Inactive neurons in the first and \\ second layers\\ -Aliasing artifacts in the learned \\ feature-maps due to large filter size\end{tabular} \\ \hline
			ZFNet &
			2013 &
			14.8\% &
			- &
			8 &
			\begin{tabular}[c]{@{}l@{}}-The notion of parameter tuning by visualizing \\ the output of intermediate layers\\ -Reduced both the filter size and stride in the \\ first two layers of AlexNet\end{tabular} &
			\begin{tabular}[c]{@{}l@{}}-Extra information processing is \\ required for visualization\end{tabular} \\ \hline
			GoogleNet &
			2014 &
			6.67\% &
			4 M &
			22 &
			\begin{tabular}[c]{@{}l@{}}- Introduces block concept\\ -The notion of utilizing Multiscale Filters within \\ the layers\\ -The notion of a split, transform and merge\\ -Parameters reduction by utilizing the bottleneck \\ layer, global average pooling at the last layer and \\ Sparse Connections\\ -Utilization of auxiliary classifiers to enhance the \\ convergence rate\end{tabular} &
			\begin{tabular}[c]{@{}l@{}}-Heterogeneous topology  led to \\ tedious parameter customization\\ -Chances of losing the\\ valuable information because of \\ representational bottleneck\end{tabular} \\ \hline
			VGGNet &
			2014 &
			7.3\% &
			138 M &
			19 &
			\begin{tabular}[c]{@{}l@{}}- Small kernel size\\ -The notion of effective receptive field\\ -The notion of simple and homogenous topology\end{tabular} &
			\begin{tabular}[c]{@{}l@{}}-Use of computationally expensive \\ fully connected layers\end{tabular} \\ \hline
			ResNet &
			2015 &
			3.6\% &
			- &
			152 &
			\begin{tabular}[c]{@{}l@{}}-The error rate was decreased for deeper networks\\ -Introduced the notion of residual learning\\ -Improves the effect of vanishing gradient issue\\ - Identity mapping based skip connection\end{tabular} &
			\begin{tabular}[c]{@{}l@{}}-A bit complex architecture\\ -Lowers information of feature-map \\ in feed forwarding\\ -Over adaption of hyperparameters \\ for a specific task, due to the \\ stacking of the same modules\end{tabular} \\ \hline
		\end{tabular}%
	}
\end{table*}


\subsubsection{Distributed Deep Learning for IVA}
\label{sec:DLDVA}
Handcrafted features, e.g., \gls{SIFT} \cite{abdel2006csift}, \gls{LBP} \cite{fang2008improving}, \gls{HOG} \cite{deniz2011face}, etc., generates high dimensional features vectors and resultantly facing the issue of scalability. Recently, \gls{CNN} based approaches have shown performance superiority in tasks like optical character recognition \cite{borisyuk2018rosetta}, and object detection \cite{lecun2015deep}. The motive of the deep learning is to scale the training in three dimensions, i.e., size and complexity of the models \cite{dean2012large}, proportionality of the accuracy to the amount of training data \cite{hestness2017deep}, and the hardware infrastructure scalability \cite{zhang2017poseidon}. Results of deep learning are so promising that soon the deep learning will give equivalent or higher performance compared to humans when trained over large data sets \cite{lecun2015deep}. A \gls{CNN} or ConvNet is a type of neural network that can recognize visual patterns directly from the pixels of images with less preprocessing. \gls{CNN} based video classification methods have been proposed in the literature to learn features from raw pixels from both short video and still images \cite{karpathy2014large, simonyan2014two, yue2015beyond, krizhevsky2012imagenet}. In the proposed \gls{SIAT} framework, both the \gls{VBDPL}, and \gls{VBDML} are capable to deploy deep-learning approaches for distributed \gls{IVA}. 

Since on the dawn of deep learning, various open-source architecture have been developed. Some of the well-known and state-of-the-art \gls{CNN} architectures are LeNet-5 \cite{lecun1998gradient}, AlexNet \cite{krizhevsky2012imagenet}, ZFNet \cite{szegedy2015going}, GoogleNet \cite{43022GoogleNet2015}, VGGNet \cite{simonyan2014very}, and ResNet \cite{he2016deep}. The comparison of these architectures can be found in Table~\ref{tab:PopularDeepLearning}. Similarly, several frameworks have been developed to eliminate the need for a manual definition of gradient propagation. Table~\ref{tab:DeepLearningLibs} summarizes these libraries along with the comparisons.

TensorFlow is a popular deep learning library designed for \gls{ML} and deep learning. It supports the deployment of computation on both CPUs and GPUs. TensorFlow allows the fast implementation of deep neural networks on the cloud. TensorFlow is also suitable for other data-driven research purposes and is equipped with TensorBoard (a visualization tool).  Higher-level programming interfaces such as Luminoth, Kera, and TensorLayer were built on the top of TensorFlow. Caffe2, developed by Berkeley AI Research, is another library to build their deep learning models efficiently along with GPUs' support in a distributed environment. PyTorch, maintained by Facebook, is a scientific computing framework with wide support for machine learning models and algorithms. PyTorch offers rich pre-trained models that can be easily reused. MXNET is a  deep learning library suitable for fast numerical computation for both single and distributed ecosystems. Likewise, some more deep learning libraries have been developed, such as CNTK, Deeplearning4j, Blocks, Gluon, and Lasagne, which can also be employed cloud environment. 


\begin{table*}[!htb]
	\centering
	\caption{Summary of popular deep learning libraries.}
	\label{tab:DeepLearningLibs}
	\resizebox{\textwidth}{!}{%
		\begin{tabular}{|l|l|l|l|l|l|l|}
			\hline
			\textbf{Features} &
			\textbf{TensorFlow \cite{abadi2016tensorflow}} &
			\textbf{Caffe2  \cite{jia2014caffe}} &
			\textbf{PyTorch \cite{ketkar2017introduction}} &
			\textbf{MXNet \cite{chen2015mxnet}}  &
			\textbf{CNTK \cite{seide2016cntk}} &
			\textbf{Matlab DL Toolbox \cite{mathworkdeep}}
			
			\\ \hline
			
			Release &
			2015 &
			2013 &
			2016 &
			2015 &
			2016 &
			2017
			\\ \hline
			
			Licence &
			Apache 2.0 &
			BSD &
			BSD &
			Apache 2.0 &
			MIT &
			Proprietary
			\\ \hline
			
			Written in &
			C++, Python, CUDA &
			C++ &
			Python, C, C++, CUDA &
			C++ &
			C++ &
			C, C++, Java, MATLAB
			\\ \hline
			
			Interface &
			\begin{tabular}[c]{@{}l@{}}
				Python (Keras), C/C++, \\
				Java, Go, JavaScript, \\
				R, Julia, Swift
			\end{tabular}
			&
			\begin{tabular}[c]{@{}l@{}}
				Python, \\
				MATLAB, C++ 
			\end{tabular}
			
			&
			
			Python, C++ &
			\begin{tabular}[c]{@{}l@{}}
				C++, Python, Julia, \\
				Matlab, JavaScript, \\
				R Go, Scala, Perl, Clojure 
			\end{tabular}	
			&
			\begin{tabular}[c]{@{}l@{}}
				Python, C++, \\
				C\#, and Java. 
			\end{tabular} 
		&
		MATLAB
			\\ \hline
			
			GPU &
			\cmark &
			\cmark &
			\cmark &
			\xmark &
			\cmark &
			\cmark
			\\ \hline
			
			DNN Support &
			RNN, CNN, LSTM, etc. &
			CNN, RNN, LSTM &
			RNN, CNN, LSTM, etc. &
			RNN, CNN, LSTM, etc. &
			CNN, RNN &
			ConvNets, CNN, LSTM, etc.
			\\ \hline
			\begin{tabular}[c]{@{}l@{}}
				Top-Level \\ Libraries 
			\end{tabular}
			&
			TensorLayer, Keras, &
			- &
			- &
			Gluon &
			Keras &
			-
			\\ \hline

			Pros &
			\begin{tabular}[c]{@{}l@{}}
				Tensorboard  for \\ 
				visualization, well \\
				documented, large \\
				user community 
			\end{tabular}
			
			&
			
			\begin{tabular}[c]{@{}l@{}}		
				Fast, scalable, \\
				lightweight, \\
				cross-platform \\
				support, server \\
				optimized interface \\
			\end{tabular}	
			&
			\begin{tabular}[c]{@{}l@{}}		
				Easy to use and debug, \\
				flexible, well documented, \\ 
				declarative data parallelism \\
			\end{tabular}
			&
			\begin{tabular}[c]{@{}l@{}}	
				Lightweight and fast, \\
				memory-efficient, highly \\ 
				scalable 
			\end{tabular} 		
			&
			\begin{tabular}[c]{@{}l@{}}	
				Good performance \\
				and scalability, \\
				highly optimized, \\
				support for Spark 
			\end{tabular}	
			&
			\begin{tabular}[c]{@{}l@{}}	
				easy for scientists, \\
				model visualization, \\
				high-performance \\ CUDA code generation,
			\end{tabular}
			
			\\ \hline
			
			Cons &
			\begin{tabular}[c]{@{}l@{}}	
				Debugging issues, heavy, \\
				difficult for novice 
			\end{tabular}
			&
			\begin{tabular}[c]{@{}l@{}}	
				Small community, \\
				modest documentation 
			\end{tabular} 	
			&
			\begin{tabular}[c]{@{}l@{}}		
				Lacking model serving, \\
				visualization tools
			\end{tabular} 
			&
			\begin{tabular}[c]{@{}l@{}}	
				Small community, \\
				difficult to learn
			\end{tabular} 
			&
			
			Limited community 
			&
			\begin{tabular}[c]{@{}l@{}}	
				Closed-source, \\
				expensive in terms of \\ 
				execution and price
			\end{tabular} 
			
			\\ \hline
			
		\end{tabular}%
	}
\end{table*}

\begin{figure}[h!]
	\centering
	\includegraphics[width=6 cm]{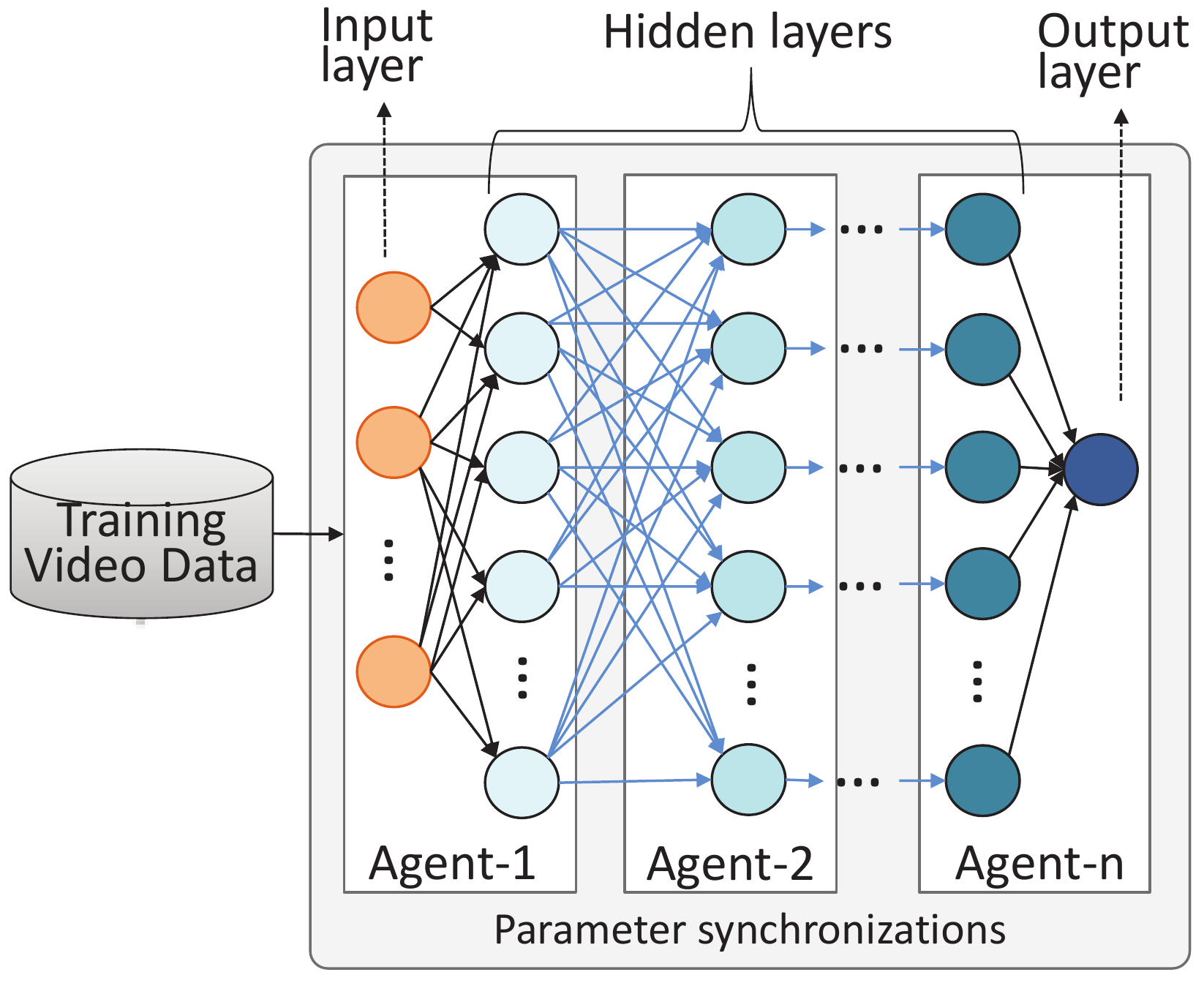}
	\caption{Scalable deep learning utilizing model distribution.}
	\label{fig:DLModelDistribution}
\end{figure}

Big DL models training with large-scale training data is a challenging task. For example, S Gao at al. \cite{gao2018dendritic} utilized six learning algorithms, i.e.,  biogeography-based optimization, particle swarm optimization, genetic algorithm, ant colony optimization, evolutionary strategy, and population-based incremental learning, for the best combination of neural network user-defined parameters during training. It is a hectic job for a single system when the training datasets are large. Distributed infrastructure with multiple computing nodes (equipped with powerful GPUs) is the best option, but it leads to numerous challenges. First is the effective utilization of resources (costly GPU stalling preventions). Second, the resources are shared among different users in the cloud for cost reduction and elasticity. Such challenges are attracting the attention of researchers. There are three approaches, i.e., model, data, and pipeline distribution for leveraging distributed computing in distributed computing \cite{mayer2019scalable, deshpande2018artificial}. In the former, the DL model is partitioned in logical fragments, and loaded to different worker agents for training, as shown in Fig.~\ref{fig:DLModelDistribution}.

\begin{figure}[h!]
	\centering
	\includegraphics[width=7.3 cm]{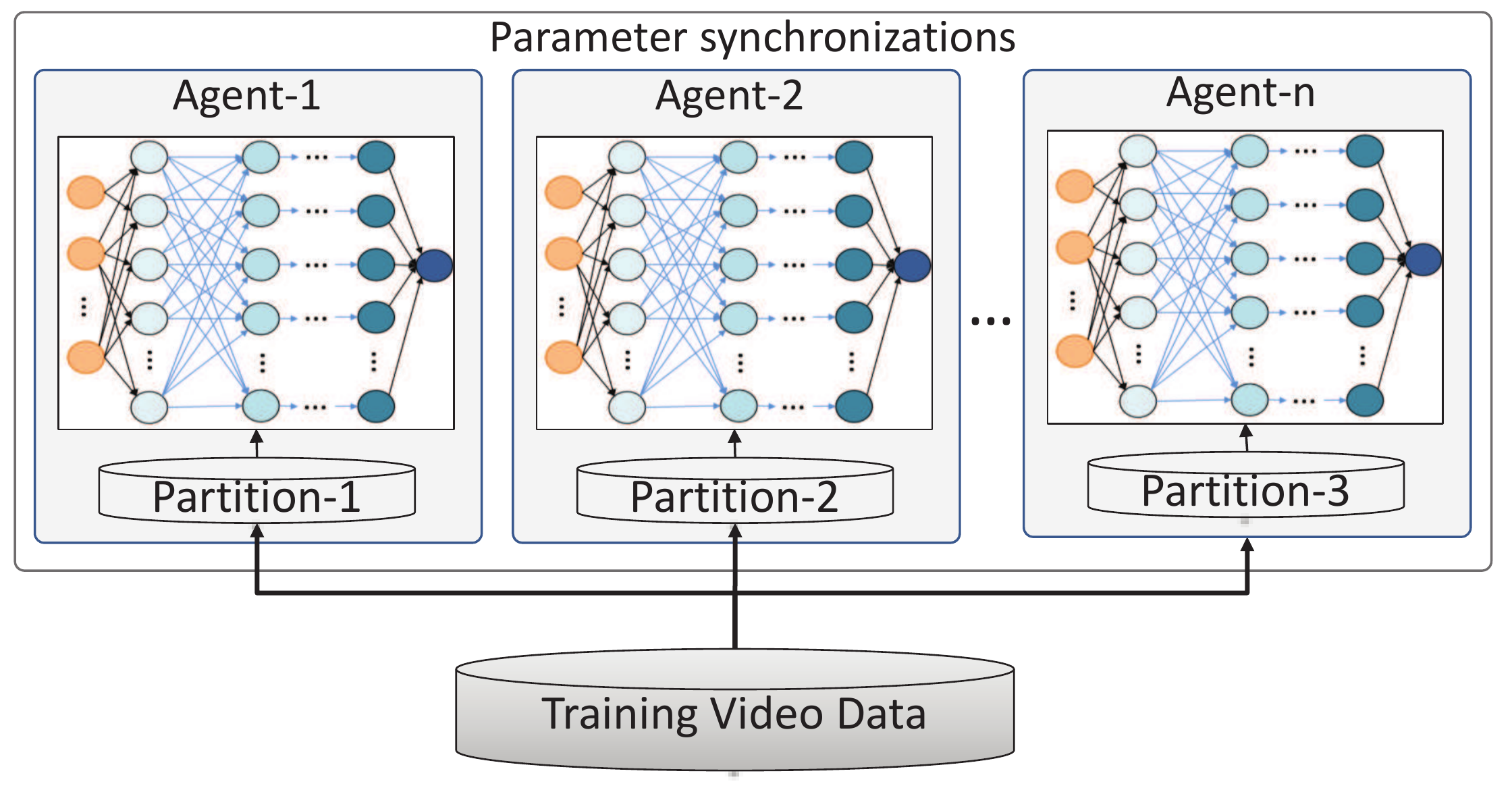}
	\caption{Scalable deep learning utilizing data distribution.}
	\label{fig:DLDataDistribution}
\end{figure}

In the second approach, the deep learning model is replicated to all the cluster's worker-agents, as shown in Fig.~\ref{fig:DLDataDistribution}. The training dataset is partitioned into non-overlapping sub-dataset, and each sub-dataset are loaded to the different worker-agents of the cluster.  Each worker-agent executes the training on its sub-dataset of training data. The model parameters are synchronized among the cluster worker-agents to updates the model parameters. The data distribution approached naturally fits in the distribute computing MapReduce paradigm \cite{deshpande2018artificial}. The MapReduce splits the input based on some predefined parameters. The map tasks, then, process these chunks in parallel a manner. After processing, the output is shuffled for relevance and is directed to map tasks for generating intermediate results. The output from the map tasks is shuffled for relevance and is given as input to the reduce tasks for generating intermediate results. The intermediate results are combined to produce the complete result. Hadoop and Spark require and process data naturally distributed in the manner and popular research trend nowadays.   

\begin{figure}[h!]
	\centering
	\includegraphics[width=7.3 cm]{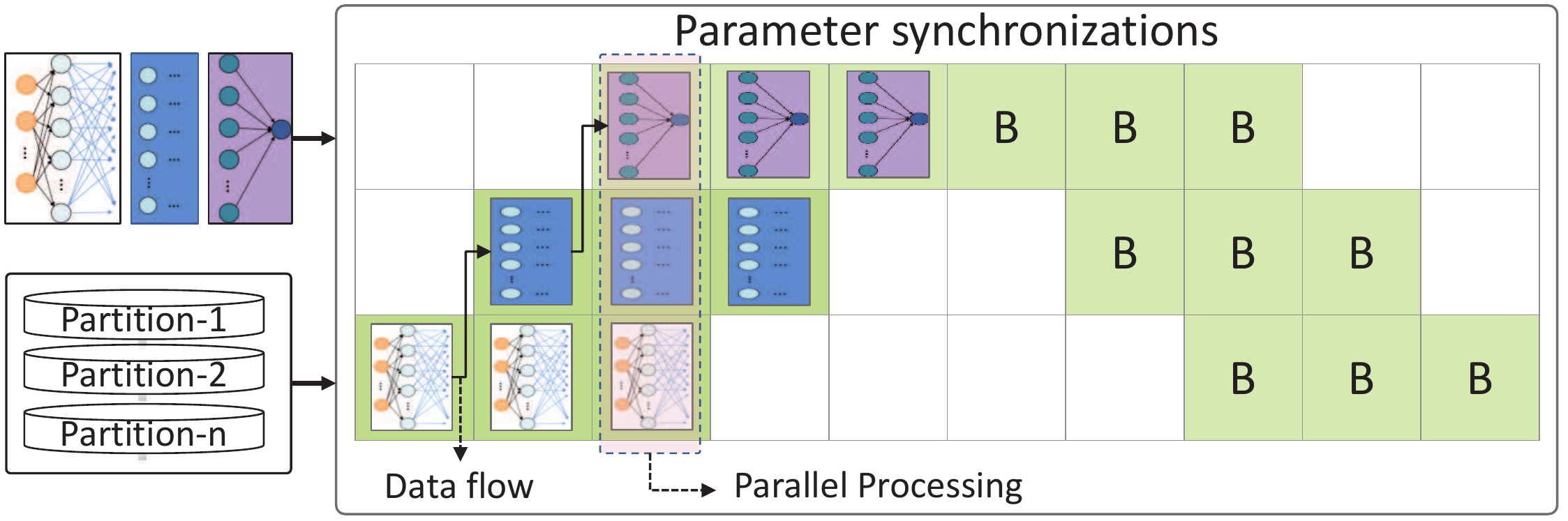}
	\caption{Deep learning with pipelining (Figure motivated by \cite{huang2019gpipe}).}
	\label{fig:DLPipelineDistribution}
\end{figure}

In the third case, the DL model is partitioned, and each worker-agent loads a different segment of the DL model for training, as shown in Fig.~\ref{fig:DLPipelineDistribution}. The training data are given to the worker-agents that carry the input layer of the DL model. In the forward pass, the output signal is computed, which is transmitted to the worker-agents that hold the next layer of the DL model.  In the backpropagation pass, gradients are calculated starting at the workers that carry the DL model's output layer, propagating to the workers that hold the input layers of the DL model \cite{huang2019gpipe}.

\begin{table*}[]
	\centering
	\caption{Comparison of big data engines.}
	\label{tab:ComparisionOfDistributedComputingEngine}
	\resizebox{\textwidth}{!}{%
		\begin{tabular}{|l|l|l|l|l|l|}
			\hline
			\textbf{Features} &	
			\textbf{Hadoop} \cite{dean2008mapreduce} &	
			\textbf{Spark} \cite{zaharia2016apache} &	
			\textbf{Flink} \cite{carbone2015apache} &	
			\textbf{Storm} \cite{toshniwal2014storm} & 
			\textbf{Samza} \cite{wang2015building} 		
			\\ \hline
			
			Batch & \cmark & \cmark & \cmark &\xmark & \xmark	
			\\ \hline
			
			Stream	& 
			\xmark &	
			near real-time	& 
			Real-time &	
			Real-time event based &	
			Event based 
			\\ \hline
			
			Data Flow &	
			\begin{tabular}[c]{@{}l@{}}
				No loops, Chain of stages	
			\end{tabular}	
			
			& 
			\begin{tabular}[c]{@{}l@{}}
				Direct Acyclic Graph
			\end{tabular}	
			
			& 
			\begin{tabular}[c]{@{}l@{}}
				Controlled Cyclic \\ Dependency Graph 
			\end{tabular}	
			
			& 
			\begin{tabular}[c]{@{}l@{}}
				Direct Acyclic Graph 
			\end{tabular}
			
			& 
			\begin{tabular}[c]{@{}l@{}}
				Kafka Dependent 
			\end{tabular}
			\\ \hline
			\begin{tabular}[c]{@{}l@{}}
				Fault-\\tolerance	
			\end{tabular}				
			& 
			
			\begin{tabular}[c]{@{}l@{}}
				High and application \\ 
				restart is not required \\
				in case of failure. 
			\end{tabular}
			& 
			\begin{tabular}[c]{@{}l@{}}
				Recovers lost work \\
				in case of failure. 
			\end{tabular}
			
			& 
			\begin{tabular}[c]{@{}l@{}}
				Based on distributed \\
				snapshots 
			\end{tabular}	
			& 
			\begin{tabular}[c]{@{}l@{}}
				Automatically restart dead \\
				workers or migrate in case \\
				of node failure. 
			\end{tabular}
			& 
			\begin{tabular}[c]{@{}l@{}}
				Migrates the task into \\
				another node in case of \\
				node failure. 
			\end{tabular}
			\\ \hline
			
			Scalability	& \cmark & \cmark & \cmark	& \cmark	& \cmark 	
			\\  \hline
			
			Latency & Higher & Low & Low & Low & Low 
			\\ \hline	
			
			Cost &	Less expensive	& Expensive & 	Expensive &	Expensive &	Expensive 
			\\ \hline	
			
			Security &	
			Kerberos authentication	&	
			\begin{tabular}[c]{@{}l@{}}
				Shared secret, HDFS \\
				ACLS and Kerberos 
			\end{tabular}	
			&	
			\begin{tabular}[c]{@{}l@{}}					
				Kerberos authentication, \\ SSL
			\end{tabular}
			&	
			Kerberos authentication	&	
			Kerberos authentication
			\\ \hline
			
			Scheduler & 
			\begin{tabular}[c]{@{}l@{}}
				By default Fair Scheduler, \\
				Capacity Scheduler, \\
				and is pluggable. 
			\end{tabular}
			& 
			\begin{tabular}[c]{@{}l@{}}
				Acts as its own flow \\
				scheduler because of \\
				in-memory computation.
			\end{tabular}
			
			& 
			\begin{tabular}[c]{@{}l@{}}
				Own scheduler and can \\
				use YARN scheduler. 
			\end{tabular}
			
			& 
			\begin{tabular}[c]{@{}l@{}}
				Default, isolation, \\
				multi-tenant and resource \\
				aware scheduler.	
			\end{tabular}
			&
			Use YARN scheduler. 
			\\ \hline
			
			\begin{tabular}[c]{@{}l@{}}
				Language \\ Support	
			\end{tabular}									
			& 
			\begin{tabular}[c]{@{}l@{}}
				Java, C/C++, Ruby, \\
				Groovy, Perl and Python	
			\end{tabular}					
			& 
			\begin{tabular}[c]{@{}l@{}}
				Java, Scala, Python, \\ and R.	
			\end{tabular}	
			& 
			\begin{tabular}[c]{@{}l@{}}			
				Java, Scala, \\
				Python, and R.	
			\end{tabular}	
			& 
			\begin{tabular}[c]{@{}l@{}}			
				Java, Clojure, and \\
				can support non-JVM \\ languages.	
			\end{tabular}	
			& 
			\begin{tabular}[c]{@{}l@{}}			
				Java, scala and can \\ 
				support JVM languages.
			\end{tabular}	
			\\ \hline
			
			Windowing	
			& \xmark	
			& Time-based
			& Time-based, Count based	
			& Time-based, Count based
			& Time-based
			\\ \hline
			
			\begin{tabular}[c]{@{}l@{}}	
				\gls{ML} Library	
			\end{tabular}					
			& 
			\begin{tabular}[c]{@{}l@{}}	
				Mahout \cite{owen2012mahout}, 
				DL4J \cite{team2016deeplearning4j}, \\
				Keras \cite{chollet2015keras}, Caffe \cite{jia2014caffe}, \\
				TensorFlow \cite{abadi2016tensorflow}, \\
				BigDL \cite{dai2018bigdl}	
			\end{tabular}				
			& 
			\begin{tabular}[c]{@{}l@{}}				
				MLLib (Native) \cite{meng2016mllib}, \\
				DL4J, Mahout, DL4J, \\
				TensorFlow, 
				Keras, 
				Caffe, \\
				H20, 
				PyTorch \cite{ketkar2017introduction}, \\
				BigDL \\
			\end{tabular}	
			& 
			\begin{tabular}[c]{@{}l@{}}				
				FlinkML (Native), \\
				Mahout, TensorFlow, \\
				Keras, Caffe, H20	
			\end{tabular}	
			& 
			\begin{tabular}[c]{@{}l@{}}			
				Don’t have native \\ \gls{ML} Lib, SAMOA 3rd \\ Party \cite{morales2015samoa}
			\end{tabular}	
			& 
			\begin{tabular}[c]{@{}l@{}}	
				Don’t have native. \\
				SAMOA 3rd Party 
			\end{tabular}					
			\\ \hline

		\end{tabular}%
	}
\end{table*}


\subsubsection{Big Data Engines, ML Libraries, and IVA}
\label{sec:DistrustedProcessingEnginesAndMLLibraries}
The \gls{VBDPL}, and \gls{VBDML} are assumed to be built on the top of distributed computing engines. This section overview some latest big data engines that can be utilized for scale-out \gls{IVA}.

Hadoop MapReduce \cite{dean2010mapreduce} is a distributed programming model, developed based on GFS \cite{Ghemawat:2003:GFS:945445.945450}, for data-intensive tasks. Apache Spark follows a similar programming model like MapReduce but extends it with Resilient Distributed Datasets (RDDs), data sharing abstraction \cite{zaharia2016apache}. Hadoop’s MapReduce operations are heavily dependent on the hard disk while Spark is based on in-memory computation, which makes Spark a hundred times faster than Hadoop \cite{zaharia2016apache, zaharia2010spark}. Spark support interactive operations, \gls{DAG} processing, and process streaming data in the form of mini-batches in near real-time \cite{zaharia2013discretized}. Apache Spark is batch centric and treats stream processing as a special case, lacking support for cyclic operations, memory management, and windows operators. Such issues of Spark has been elegantly addressed by Apache Flink \cite{carbone2015apache}. Apache Flink treats batch processing as a special and does not use micro-batching. Similarly, Apache Storm and  Samza is another prominent solution focused on working with large data flow in real-time. A brief description and comparison of the open-source big data frameworks are listed in Table~\ref{tab:ComparisionOfDistributedComputingEngine}.

To achieve scalability, big data techniques can be exploited by existing video analytics modules. The \gls{VBDPL} is not provided by default and needs its implementation on the top of these big data engines. However, The \gls{ML} approaches can be categorized into two classes in the context of \gls{VBDML}. One class re-implements the existing \gls{ML} task by providing a middleware layer to run them on a big data platform. This general type of middleware layer provides general primitives/operations that assists in various learning tasks. Users who want to try different \gls{ML} algorithms in the same framework can take benefits from it. In the second class, the individual video analytics and \gls{ML} algorithm are executed on a big data platform that is directly built on top of a big data engine for better scalability. 

Spark MLlib \cite{meng2016mllib}, Mahout \cite{owen2012mahout}, FlinKML \cite{carbone2015apache} are list of some open-source \gls{ML} packages built on the top of Hadoop, Spark and Flink, respectively, that support many scalable learning algorithms. For deep learning, various open-source libraries have been develop including TensorFlow \cite{abadi2016tensorflow}, DeepLearning4J \cite{team2016deeplearning4j}, Keras \cite{chollet2015keras},  Caffe \cite{jia2014caffe}, H20 \cite{cook2016practical}, BigDL \cite{dai2018bigdl}, and PyTorch \cite{ketkar2017introduction}. All these libraries provide support for various \gls{ML} algorithms and feature engineering. These libraries introduce an independent layer between front-end algorithms and a back-end engine to facilitate the migration from one big data engine to another, as shown in Table~\ref{tab:ComparisionOfDistributedComputingEngine}. These algorithms can be used to process large datasets, just like processing it on a single machine by providing the distributed environment abstraction and optimization.


\begin{table*}[]
	\centering
	\caption{Popular computer vision benchmark datasets.}
	\label{tab:VisualDatasets}
	\resizebox{\textwidth}{!}{%
		\begin{tabular}{|l|l|l|l|l|}
			\hline
			Dataset         & 
			Instances       & 
			Classes    & 
			Ground Truth & 
			Description                                      
			\\ \hline
			
			YFCC100M \cite{deng2009imagenet}             & 
			100M            & 
			8M         & 
			Partially    & 
			Video and image, understanding                   
			\\ \hline
			
			YouTube-8M \cite{thomee2015yfcc100m}         & 
			8M              & 
			4,716      & 
			Automatic    & 
			Video classification                             
			\\ \hline
			
			Trecvid 2016 \cite{trecvid2019}       & 
			Different       & 
			Different  & 
			Partially    & 
			Human action detection                           
			\\ \hline
			
			UCF-101 \cite{soomro2012ucf101}           
			& 13K             
			& 101        
			& Yes          
			& Human action detection                           
			
			\\ \hline
			Kinetics \cite{kay2017kinetics}         
			& 306K            
			& 400        
			& Yes          
			& Human action detection                           
			
			\\ \hline
			MediaEval \cite{MediaEval2015} 
			& -              
			& -          
			& -            
			& Videos from BBC                                  
			\\ \hline
			
			YACVID \cite{nambiar2014multi}             
			& 75207 Frames    
			& -          
			& Yes          
			& A multi-camera high-resolution dataset           
			
			\\ \hline
			ActivityNet \cite{caba2015activitynet}   
			& 849 video hours 
			& 203       
			& Yes          
			& Human activity                                   
			\\ \hline
			
			HMDB \cite{kuehne2011hmdb}                
			& 7K Videos       
			& 51 Classes 
			& Yes          
			& Human Motion Recognition                         
			\\ \hline
			
			Sports-1M \cite{karpathy2014large}          
			& IM              
			& 487        
			& Yes          
			& The YouTube Sports-1M Dataset                    
			\\ \hline
			
			AVA \cite{gu2018ava}                
			& 57K             
			& Different  
			& Yes          
			& 
			\begin{tabular}[c]{@{}l@{}}
				Spatiotemporally Localized Atomic Visual Actions 
			\end{tabular}	
			
			\\ \hline
			
			HowTo100M \cite{miech2019howto100m} 
			& 100M            
			& 23k        
			& -            
			& Narrated instructional web videos                
			\\ \hline
			
		\end{tabular}%
	}
\end{table*}


\subsubsection{Computer vision benchmark datasets}
In the advancement of \gls{IVA}, public datasets always play a vital and active role. Over the year, server benchmark datasets have emerged. Some of the recent popular datasets are listed in Table~\ref{tab:VisualDatasets}. ImageNet \cite{deng2009imagenet} is one of the significant datasets in deep learning and is utilized for training neural networks such as ResNet, AlexNet, and GoogleNet.  Some more datasets have been developed aiming human action and motion recognition, including \cite{trecvid2019, trecvid2019, kay2017kinetics, caba2015activitynet, kuehne2011hmdb}. Google released YouTube-8M \cite{abu2016youtube} and consisting of eight million diverse types of automatically labeled videos. Deng A et al. \cite{miech2019howto100m} proposed the HowTo100M dataset comprising of web videos with narrated instructions. Dataset like MediaEval2015  \cite{MediaEval2015}, and Trecvid2016 \cite {trecvid2019} are designed to support \gls{CBVR} related research. For sports \gls{IVA}, the Sports-1M dataset {karpathy2014large} is proposed and composed of 487 classes along with ground truth. YACVID \cite{nambiar2014multi} is a labeled image sequence dataset for benchmarking video surveillance algorithms. All these benchmark datasets are used for different \gls{IVA}, such as action recognition, event detection, and classification.

\subsection{Knowledge Curation Layer}
\label{sec:KCL}
Videos low-level processing produces feature descriptors that summarize characteristics of data quantitatively. The high-level analytics is more associated with the visual data understanding and reasoning. The features descriptors work as input to the high-level analytics and generate abstract descriptions about contents. The difficult problem is to bridge the semantic gap between the low-level features and high-level concepts suitable for human perception \cite{ballan2010video}. 

With the same indentations, the \gls{KCL} layer has been proposed under \gls{SIAT} architecture, on the top of \gls{VBDML}, which map the \gls{IR} (both online and offline) into the video ontology in order to allow domain-specific semantic video and complex event analysis. The \gls{KCL} is composed of five components, i.e., Video Ontology Vocabulary, Video Ontology, Semantic Web Rules, FeatureOnto Mapper, and SPARQL queries.

\texttt{Video Ontology Vocabulary} standardizes the basic terminology that governs the video ontology, such as concept, attributes, objects, relations, video temporal relation, video spatial relation, and events. \texttt{Video Ontology} is a generic semantic-based model for the representation and organization of video resources that allow the \gls{SIAT} users for contextual complex, event analysis, reasoning, search, and retrieval. \texttt{Semantic Web Rules} express domain-specific rules and logic for reasoning.  When videos are classified and tagged by the \gls{VBDML} then the respective \gls{IR} are persistent to \gls{VBDCL} and also mapped to the \texttt{Video Ontology} while using the \texttt{FeatureOnto Mapper}. Finally, SPARQL based semantic rich queries are allowed for knowledge graph, complex event reasoning, analysis, and retrieval.

\subsection{Web Services Layer}
\label{sec:WSL}
\begin{figure}[h!]
	\centering
	\includegraphics[scale=.4]{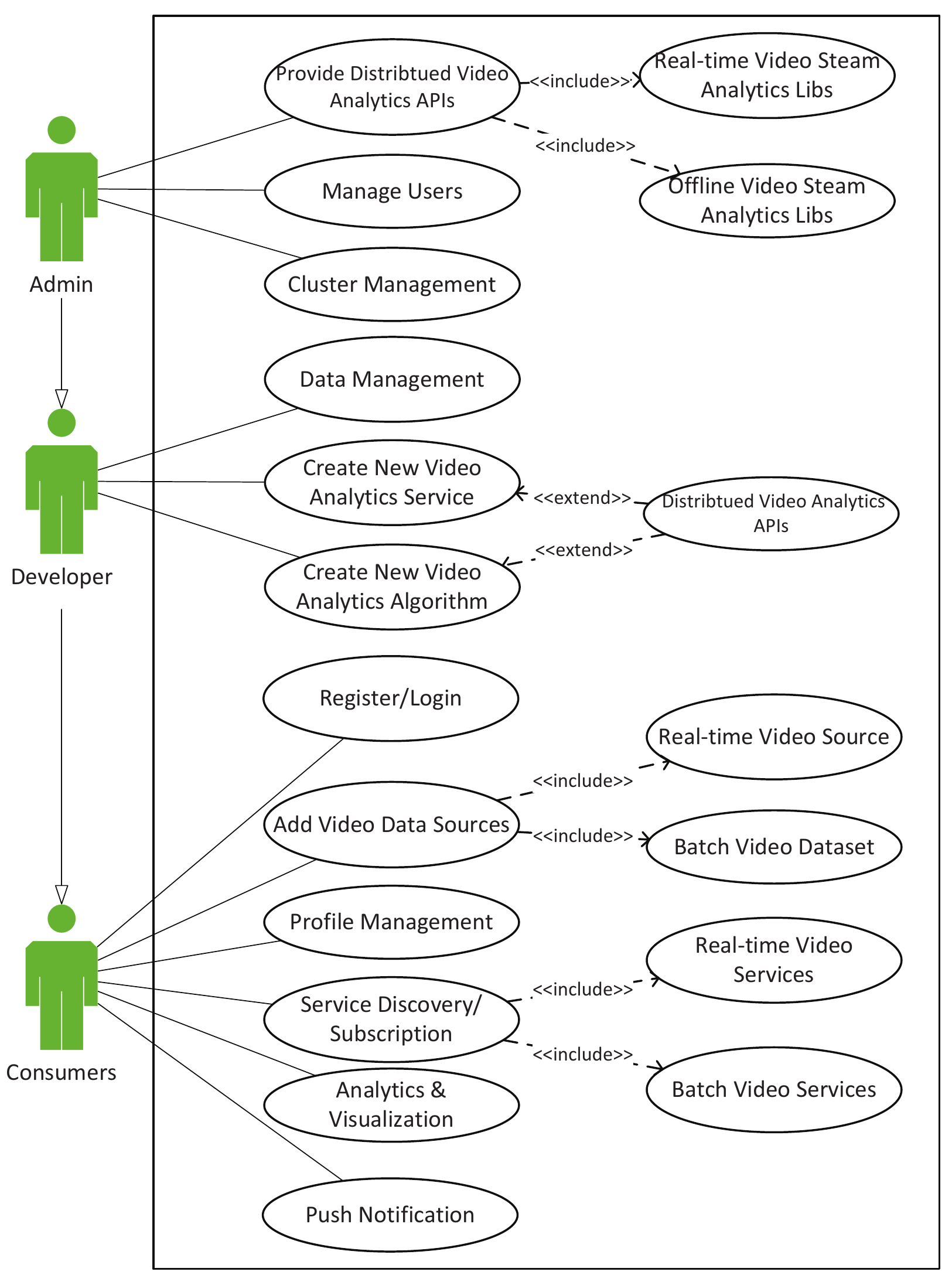}
	\caption{Lambda CVAS user roles and use case diagram.}
	\label{fig:UserUseCaseDiagram}
\end{figure}

Finally, to provide the functionality of the proposed \gls{SIAT} over the web, it incorporates top-notch functionality into simple unified role-based web services. The \texttt{Web Service Layer} is built on the top of \texttt{VBDCL Business Logic}. Sequence diagrams for \gls{IVA} algorithm and service creation is shown in Fig.~\ref{fig:AlgoServiceSequence}. Whereas, role-based use case diagram of the proposed platform is shown in Fig.~\ref{fig:UserUseCaseDiagram}.

\Figure[!t]()[width=0.46\textwidth]{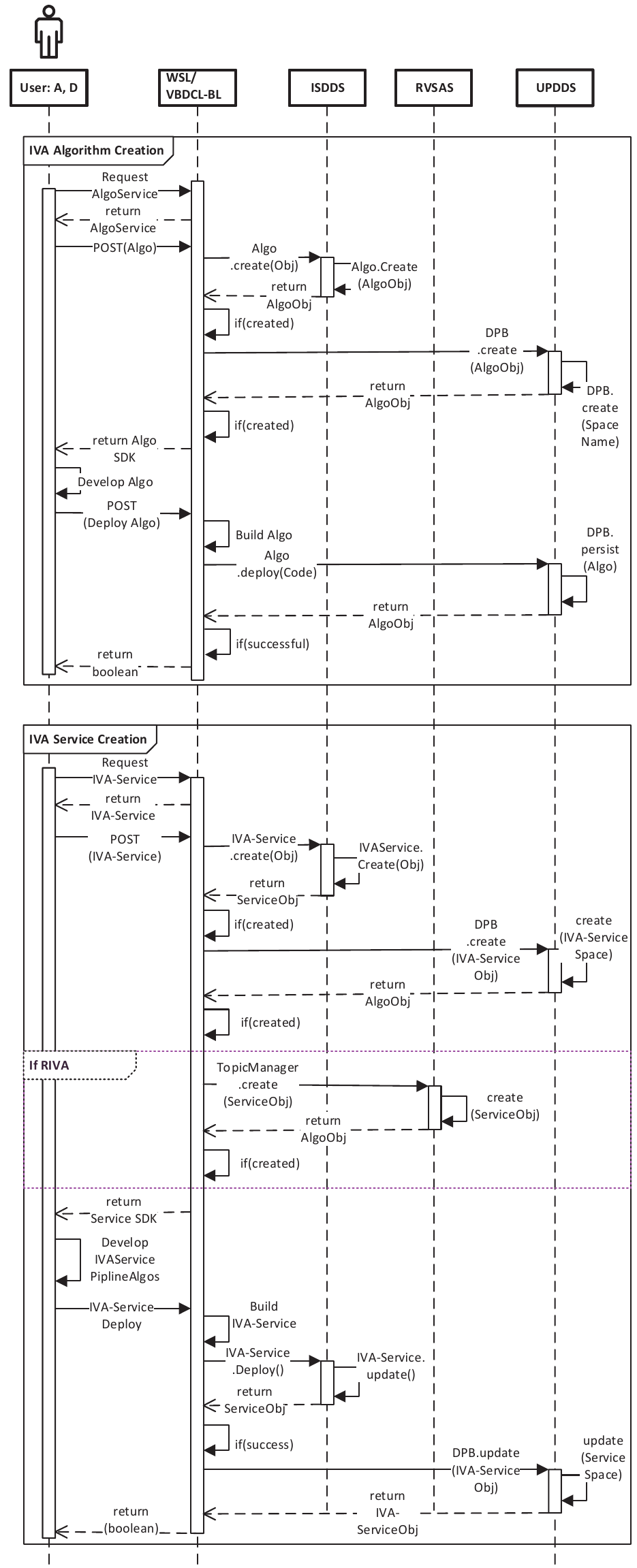}
{The sequence diagram for the IVA algorithm and service creation. In the Actor lifeline, \texttt{A} and \texttt{D} represent \texttt{Admin} and \texttt{Developer}, respectively. These two roles are allowed to create a new video analytics algorithm and service.
	\label{fig:AlgoServiceSequence}}

\subsection{IVA service example scenarios}
\label{sec:IVAServiceUseCase}
\begin{figure}[h!]
	\centering
	\includegraphics[width=7.3 cm]{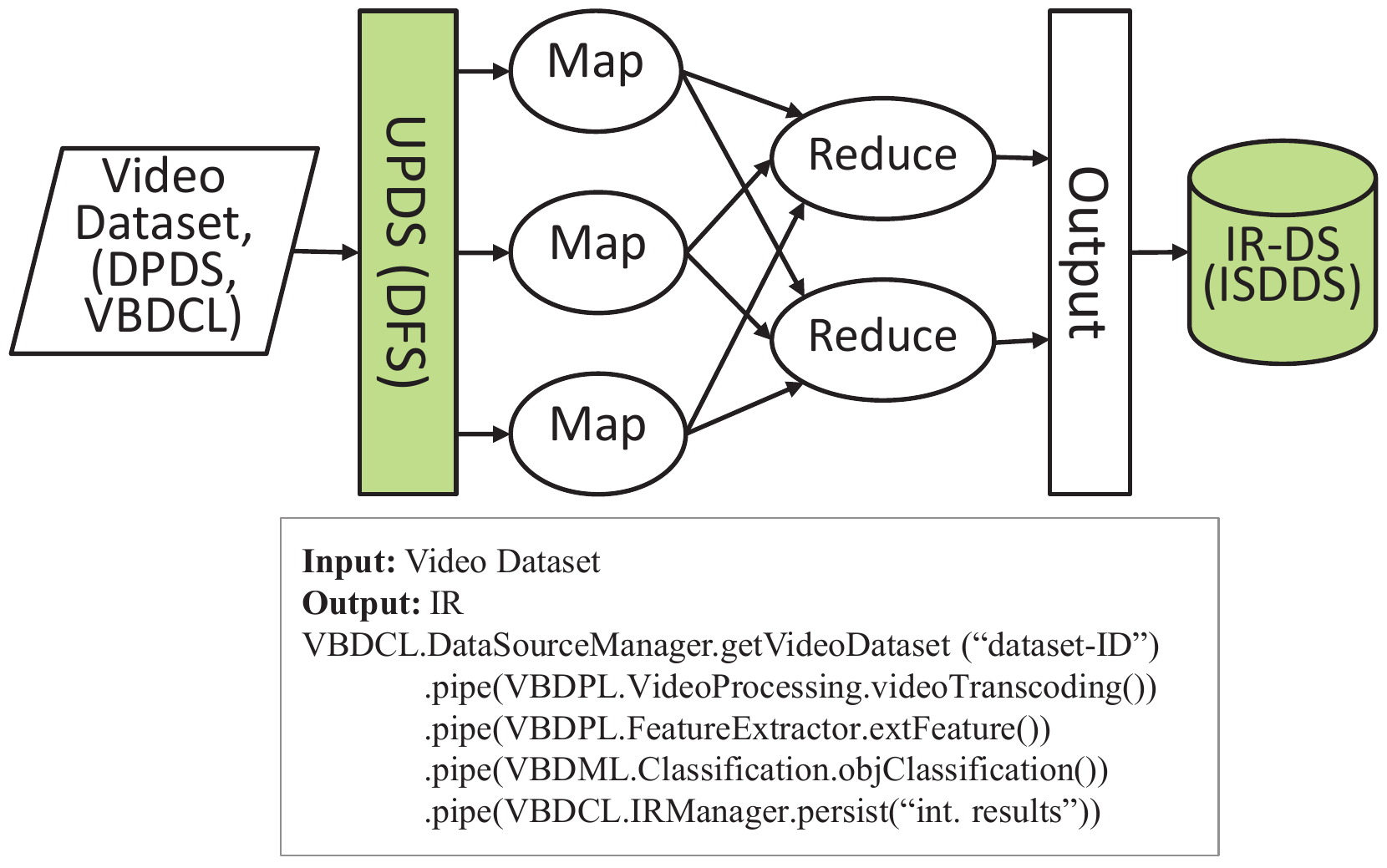}
	\caption{BIVA service example pipeline under Lambda CVAS.}
	\label{fig:MapReduceBatchPipeline}
\end{figure}

We show two example scenarios, i.e., how to develop \gls{BIVA} and \gls{RIVA} services under \gls{SIAT}. Hadoop and Spark MapReduce type operations naturally fit in a \gls{BIVA}.  Fig.~\ref{fig:MapReduceBatchPipeline}, shows the block diagram along with sample script for distributed \gls{BIVA}, i.e., object classification, where a \gls{DFS} is configured to read and store video files, e.g., in a standard video file format such as MPEG, AVI, H.264, etc. First, the videos are loaded, the distributed video transcoding (a preprocessing algorithm under \gls{VBDPL}) are performed. For example, a user uploads a MPEG or other video files to the \gls{DFS}. The transcoder algorithm first split the file into image frames, which may be throttled to key-value frames, and converts them into a sequence file format. These frames are then mapped, and features are extracted (utilizing some feature extractor from \gls{VBDPL}). The features are then classified (using a classification algorithm of \gls{VBDPL}). For each detected object, a key-value and coordinates of where the object is located within the image are computed. For each detected object in each frame, the map element process the frame. The map operation generates and provides output as composite visual value pair that includes the visual key, a time-stamp that identifies the frame, and the coordinates. The map stages then send the output in the form of a composite key-value pair to the respective reduce stages. The reduce stages provide the output to an output stage, which is persisted in the \gls{IR}-DS of \gls{ISDDS}. 

\begin{figure}[h!]
	\centering
	\includegraphics[width=7.3 cm]{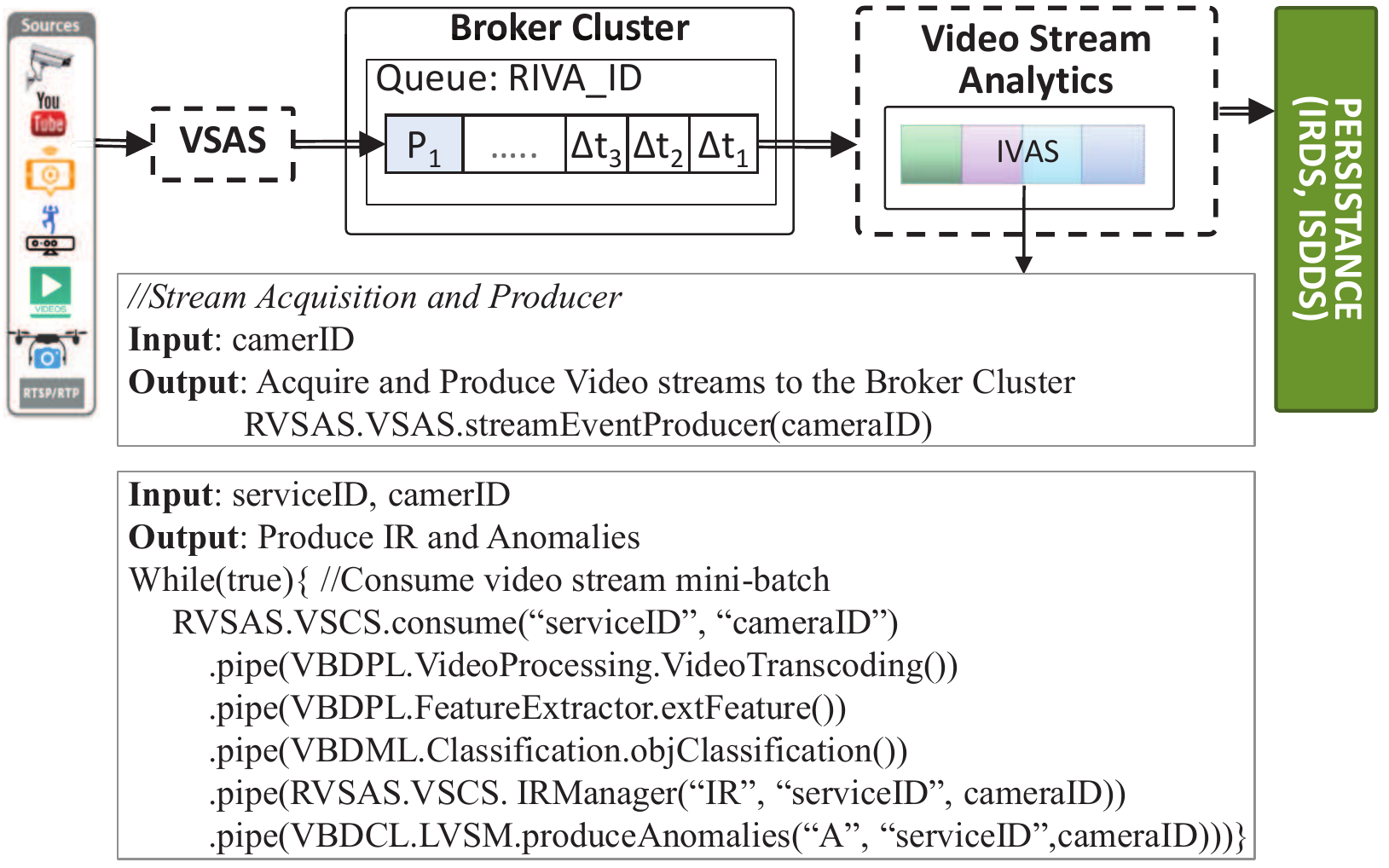}
	\caption{RIVA service example pipeline under Lambda CVAS.}
	\label{fig:RIVAUseCasePipeline}
\end{figure}

The flow for a single \gls{RIVA} along with an example pseudo-code is shown in Fig.~\ref{fig:RIVAUseCasePipeline}.  The \gls{VSAS} component sends the video stream to the queue in the broker cluster. Then the consumer service (VSCS) is used to extract the mini-batch of video streams from the queue and process it. Once the mini-batch is consumed, then it is transcoded, features are extracted for classification, and finally, the classification results are persisted to the required destination (\gls{IR}-DS and/or dashboard). 

\subsection{Execution Scenarios}
\label{sec:ExecutionScenarios}

\begin{figure*}[htbp]
	\centering
	\includegraphics[width=14 cm]{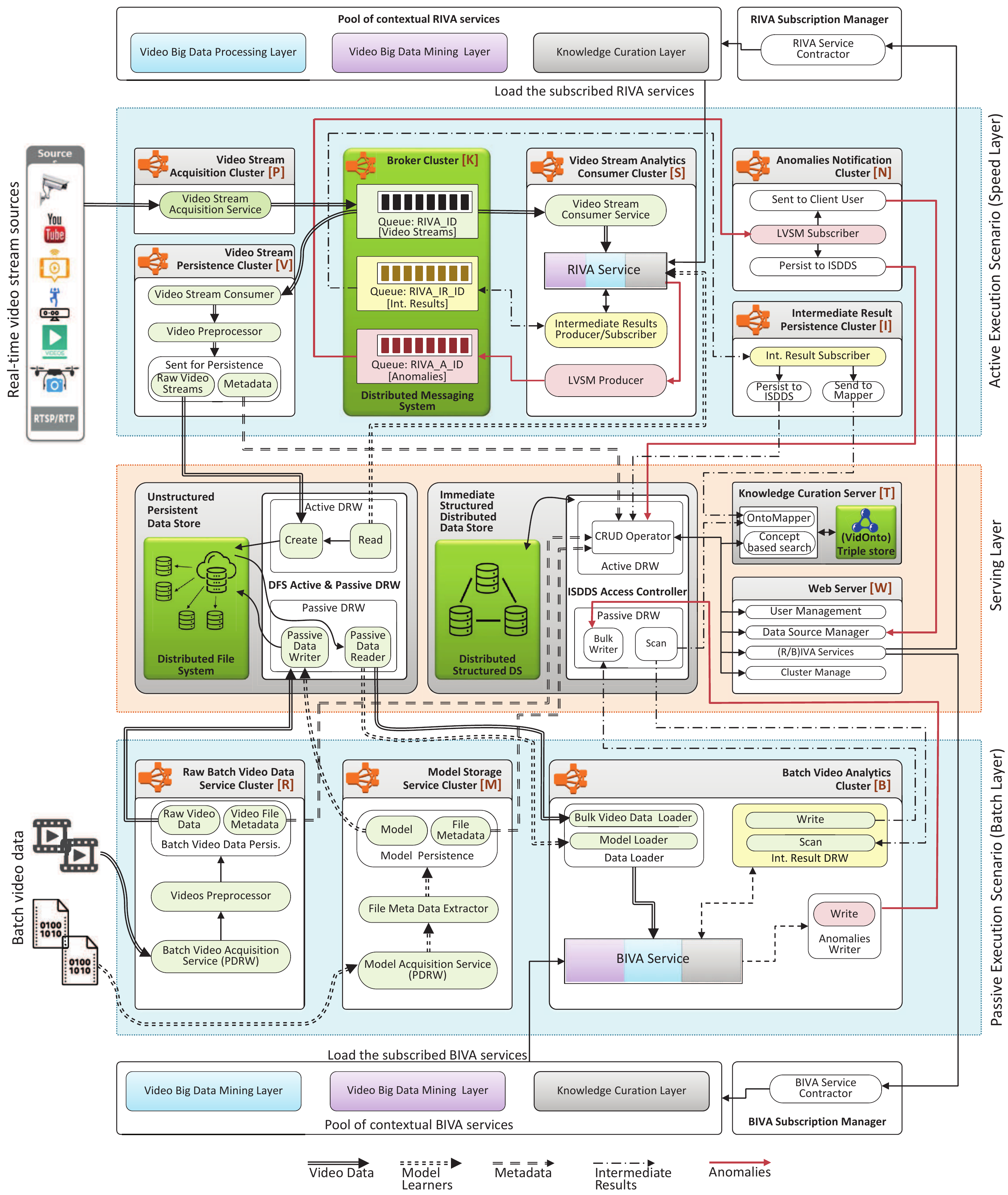}
	\caption{Lambda CVAS: Streaming and Batch execution scenarios.}
	\label{fig:ExecutionScenarios}
\end{figure*}

\gls{SIAT} follows the lambda architecture style \cite{marz2015big}, and the execution scenarios undergo through two types of execution scenarios, i.e., Streaming Execution Scenario, and Batch Execution Scenario. These two scenarios aim to execute a massive amount of real-time video stream and batch videos against the subscribed \gls{IVA} services. In literature, these execution scenarios are referred to as Speed Layer, and Batch Layer respectively \cite{marz2015big}. The data of both the scenarios are managed through a common layer called Serving Layer. \gls{SIAT} components are deployed on various types of clusters in the cloud, and each cluster is subject to scale-out on-demand. Fig.~\ref{fig:ExecutionScenarios} illustrates these execution scenarios, and the explanation is given in the following subsections.

\subsubsection{Streaming Execution Scenario}
\label{sec:ActiveExecutionScenarios}
\gls{SIAT} is supposed to deploy a pool of contextual \gls{RIVA} services that are made available to the user for the subscription.  Once a video stream data source is subscribed to service in the pool of contextual \gls{RIVA} services, then the life-cycle of Streaming Execution Scenario encompasses through different stages while using distinct \gls{SIAT} components. For the ease of understandability, these components are deployed on six types of computing clusters in the cloud, which are labeled explicitly as \texttt{‘P’}, \texttt{‘V’}, \texttt{‘K’}, \texttt{‘S’}, \texttt{‘N’}, \texttt{‘I’}, as shown in Fig.~\ref{fig:ExecutionScenarios}. 

The cluster \texttt{'P'} hosts \gls{VSAS} and provides interfaces to external \gls{VSDS}. On configuration, the video streams are captured and transformed into a proper message, which is then grouped into micro-batch, compressed, and loaded to the respective queue in the cluster \texttt{'K'}. 

The cluster \texttt{'K'} deploys the distributed messaging system, where the acquired video streams, \gls{IR}, and anomalies produced by \gls{LVSM} are buffered. In this context, the cluster \texttt{‘K’} is composed of a set of three types of queues for each service, i.e., \texttt{RIVA\_ID}, \texttt{RIVA\_IR\_ID}, and \texttt{RIVA\_A\_ID}, as described in \ref{sec:RVSAS}. 

The mini-batches of video streams residing in the distributed broker's \texttt{RIVA\_ID} queue need to be persisted to the \gls{UPDDS} and \gls{ISDDS} data stores. For this purpose, the cluster 'V' deploys three types of \gls{SIAT} modules, i.e., \gls{VSCS}, \texttt{Video Processor} and \texttt{Persistence}.  The first module allows the cluster \texttt{'V'} to read the video stream mini-batches from \texttt{RIVA\_ID} topics in the Cluster \texttt{'K'}. The cluster \texttt{'V'} then processes the consumed video data, encodes it, and extracts the metadata from the video. Finally, the video stream persistence module saves the video data and the respective metadata to the \gls{UPDDS} and \gls{ISDDS}, respectively.

The cluster \texttt{'S'} is responsible for processing the video stream in near real-time while using the \gls{IVA} services. Different stream processing engines, e.g., Apache Spark Stream, can be used for \gls{RIVA}. The cluster \texttt{'S'} deploys four types of \gls{SIAT} modules. The first module is \gls{VSCS} and is used to consume the video streams from the \texttt{RIVA\_ID} queue in the cluster \texttt{'K'}. The second type of module is \gls{RIVA} services. The \gls{RIVA} service is the actual video stream analytics service that analyzes the videos. The video \gls{RIVA} service is loaded according to the \gls{RIVA} services subscription contract made by the \gls{SIAT} user. The \gls{RIVA} services can be pipelined, and the \gls{IR} might need other applications in the multi-subscription scenario. Thus the \gls{IR} producer/subscriber is used to send and receive the \gls{IR} according to the application logic to and from the \gls{IR} queue in Cluster \texttt{'K'}. 

The fourth type of module is \gls{LVSM} producer. The contextual \gls{IVAS} instance deployed in the cluster \texttt{'S'} should have some domain-specific goal and can produce anomalies if analyzed any. The \gls{SIAT} support real-time anomalies delivery system. The \gls{IVAS} sent the anomalies continuously to the \gls{LVSM} producer and the \gls{LVSM} producer to the respective anomalies queue \texttt{RIVA\_A\_ID} in the cluster \texttt{'K'}.

The cluster group \texttt{'I'} read the \gls{IR} from the \texttt{RIVA\_A\_ID} queue in cluster \texttt{'K'} continuously and sent it to the \gls{ISDDS}’s \gls{IR} data store for persistence. If the subscribed service is using the ontology then the \gls{IR} are also mapped to the VidOnto triple residing in the \texttt{Knowledge Curation Server} \texttt{‘T’}.

The final type of cluster in the Streaming Execution Scenario is cluster \texttt{'N'} and is known as \texttt{Anomalies Notification Cluster}. This cluster aims to read anomalies from the \texttt{RIVA\_A\_ID} queue in cluster \texttt{'K'} and send the same to the \gls{ISDDS} for persistence and also delivered in real-time to the video stream source owner in the form of alerts.

\subsubsection{Batch Execution Scenario}
\label{sec:PassiveExecutionScenarios}

The \gls{SIAT} architecture is also equipped with \gls{BIVA}. Unlike the \gls{RIVA}, the \gls{BIVA} is analyzed as an offline manner. The execution time of offline analytics is proportional to the video dataset size and the subscribed \gls{BIVA} service computation complexity. The Batch execution life-cycle undergoes through three types of clusters, i.e., \texttt{'R'}, \texttt{'M'}, \texttt{'B'}. 

The cluster \texttt{'R'} allows the \gls{SIAT} user to upload the batch video dataset to the \gls{SIAT} cloud and configure three types of \gls{SIAT} modules. The first type of service is Batch Video Acquisition Service, which is used to acquire batch video datasets. Once uploaded to the node buffer, the batch dataset is processed by the activated Video Processor to extract the metadata from the batch videos. After processing the batch video dataset and the respective metadata are persisted to the \gls{UPDDS} and \gls{ISDDS}, respectively. Similarly, the cluster \texttt{'M'} works the same way as that of cluster \texttt{'R'}, but this one is responsible for model management.

In the batch video analytics, the supporting layers deploy various contextual multi-domain offline \gls{BIVA} services. This cluster loads the instance of the \gls{RIVA} services as per user contract and processes the videos in an offline manner. Once subscribed, this cluster loads the batch video data set and model from the \gls{UPDDS}. Similarly, the \gls{IR} and anomalies are maintained in the \gls{ISDDS}. The acquired video streams residing in the \gls{UPDDS} is also illegible for offline analytics.

Finally, the Web Server \texttt{'W'} deploys the Web User Interface (as described in \ref{sec:WSL}), i.e., allow the users to interact with \gls{SIAT}.
\section{IVA; Constituents, and Predominant Trends}
\label{sec:IVAConstituentsPredominantTrends}
\begin{figure*}[h!] 
	\centering
	\includegraphics[width=15 cm]{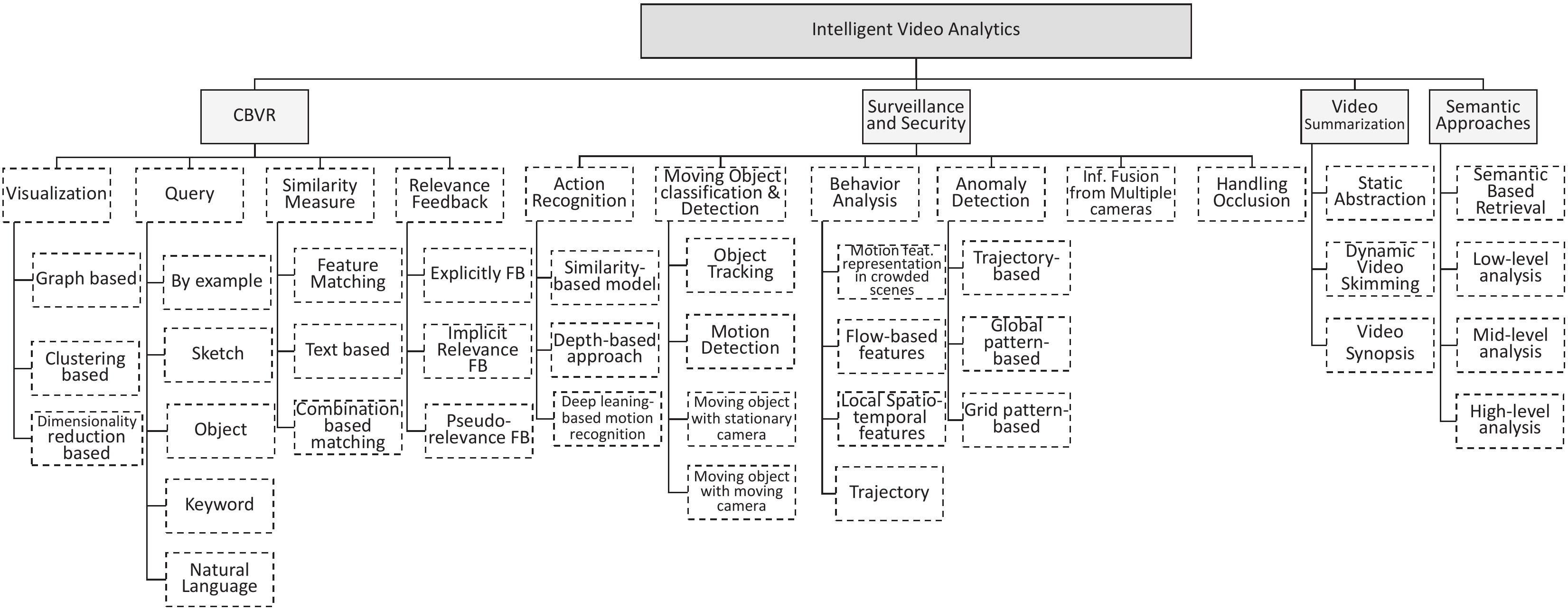}
	\caption{IVA taxonomy based on literature.}
	\label{fig:IVATaxonomy}
\end{figure*}

This section review the existing \gls{IVA} literature and can be classified into four classes under the umbrella of \gls{IVA}, i.e., \gls{CBVR}, \gls{IVA} Surveillance and Security, Video Summarization, and Semantics Approaches, as shown in Fig.~\ref{fig:IVATaxonomy}. We also show that how \gls{SIAT} can be used under these application areas.

\subsection{Content-Based Video Retrieval}
\label{sec:CBVR}
\gls{CBVR} has applications from video browsing to intelligent management of video surveillance and analysis. To uphold advancement in \gls{CBVR}, since 2001, the National Institute of Standards and Technology has been sponsoring the annual Text Retrieval Conference Video Retrieval Evaluation \cite{smeaton2010video, smeaton2009high}. The \gls{CBVR} is an active research area, and several surveys are available, i.e., \cite{hu2011survey, bhaumik2016hybrid,  smeaton2009high, hu2013internet, spolaor2020systematic }. However, here we discuss some of the scalable \gls{CBVR} system being proposed in the literature. 

In literature, some researchers tried to exploit distributed computing technology for the development of large-scale \gls{CBVR} systems. Shang et al. \cite{ shang2010real} utilize the time-oriented temporal structure of videos and the relative gray-level intensity distribution of the frames as a feature base. Their method is expensive in terms of parallel processing due to the video semantics, and then all the frames of a video must be processed within the same execution environment. Hence this approach is challenging to parallelize accurately and efficiently even for the state-of-the-art big-data frameworks. Wang et al. \cite{wang2015gpu} proposed a novel MapReduce framework called Multimedia and Intelligent Computing Cluster for near-duplicate video retrieval for large-scare multimedia data processing by joining the computing power of CPU’s and GPU’s to speed up the video data processing. They extract the keyframes using uniform sampling, store the keyframes to \gls{HDFS}, perform local feature extraction using the Hessian-Affine detector \cite{mikolajczyk2004scale} to detect interest points. K-means clustering over the feature vectors is utilized to generate visual words following the BoF \cite{sivic2003video} model, thus generating BoF-based feature vectors. Ding et al. \cite{ding2017survsurf} used big data processing technologies to design a human retrieval system on extensive surveillance video data called SurvSurf. Motion-based segments called M-clop were detected, which were utilized to remove redundant videos. Hadoop MapReduce framework was used to process M-clips for human detection and motion feature extraction. Vision algorithms were accelerated by processing only sub-areas with significant motion vectors rather than entire frames. Further, a distributed data store called V-BigTable on top of HBase was designed to structuralize M-clips’ semantic information and enables large-scale M-clips retrieval. They stated that SurvSurf outperforms the baseline solutions in terms of computational time and with acceptable human retrieval accuracy.

Authors in \cite{zhu2015marlin} proposed Marlin for video big data similarity search. They used parallel computing to extract features from the acquired video micro-batches, which are then persisted in a distributed feature indexer. The proposed indexer was able to index incremental updates and real-time queries. They designed a fine-grained resource allocation with a resource-aware data abstraction layer over streaming videos to upsurge the system throughput. They reported Marlin achieved 25X and 23X speedup against the sequential feature extraction algorithm and similarity search, respectively. The challenge of extracting distinctive features is addressed by Lv et al. \cite{lv2016efficient} for the efficiency of extraction closely related videos from the large scale data based on local and global features utilizing Spark. To balance precision and efficiency, they introduced a multi-feature based distributed system, including local and global features. They combined local feature \gls{SIFT}, Local Maximal Occurrence, and global feature Color Name. Lastly, they developed the system in a distributed environment based on Apache Spark. Further, M. N. Khan et al. \cite{numan2020FALKON} proposed FALKON for large-scale \gls{CBVR} that utilized distributed deep learning on top of Apache Spark for accuracy, efficiency, fault tolerance, and scalability. Motivated by the fact that Apache Spark, by default, does not provide native video data structure, they developed a wrapper on the top of Spark’s RDD called VidRDD. Utilizing VidRDD, first, they performed structural analysis on the videos, and then index the extracted deep spatial and temporal features in their designed distributed indexer. Finally, they evaluate their proposed system and show performance improvement in terms of scalability and accuracy. Likewise, Lin FC. et al. \cite{lin2020cloud } put forward a cloud-based face video retrieval system while utilizing deep learning. First, pre-processing operations like termination of blurry images, and face alignment are performed. Then the refined dataset is constructed and used to pre-train the \gls{CNN} models, i.e., ArcFace, FaceNet, and VGGFace for face recognition. The results of these three models are compared, and the efficient one was chosen for the retrieval system development. The input query in their proposed system is a person’s name. If the system detects a new person, it performs enrolling that person. Finally, timestamped results are returned against a query.  A prototype of the proposed face retrieval system was implemented and reported its recognition accuracy and computational time.

\subsubsection{CBVR under L-CVAS Architecture}
\begin{figure}[h!]
	\centering
	\includegraphics[width=7.3 cm]{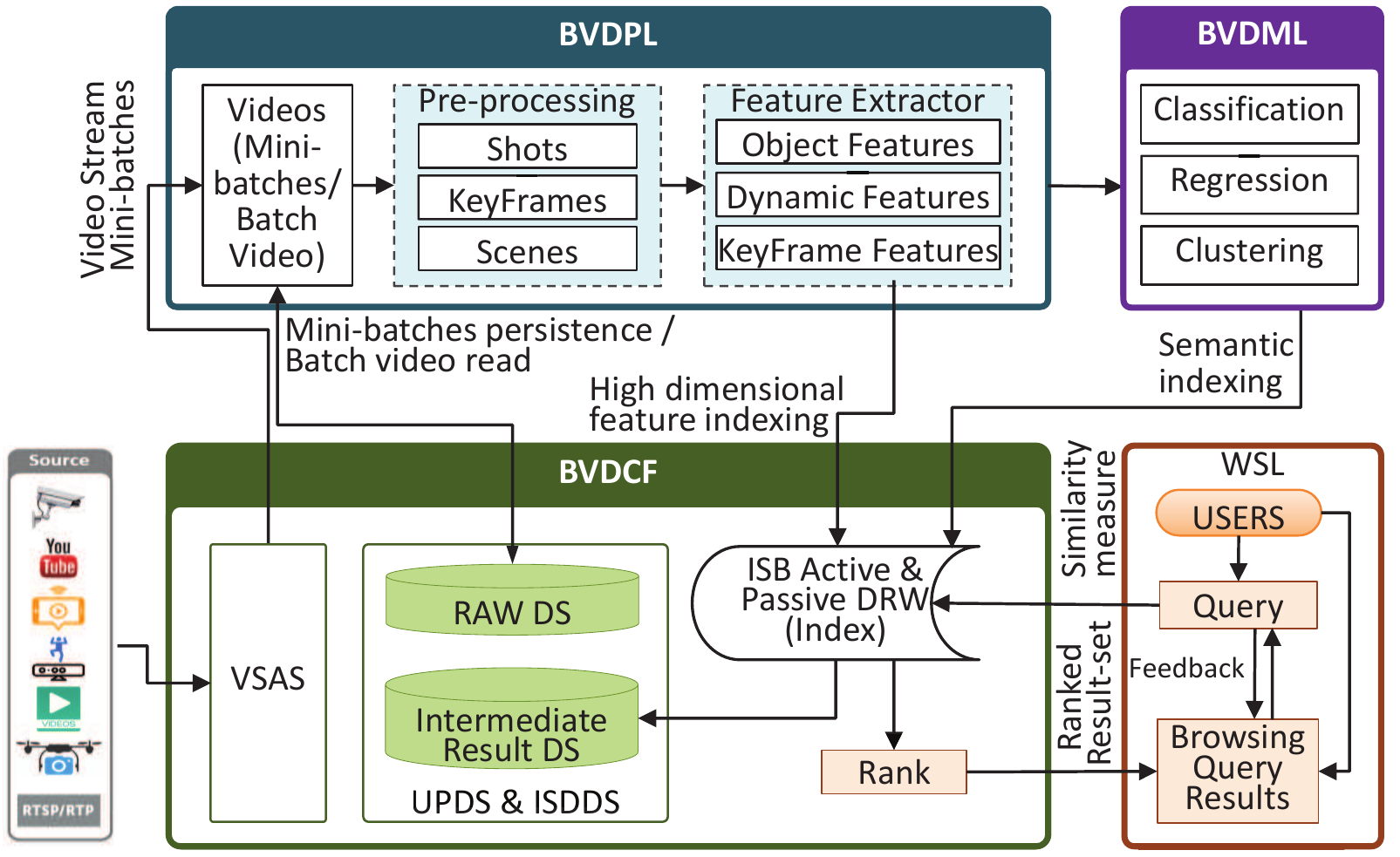}
	\caption{Components of CBVR under the Lambda CVAS architecture.}
	\label{fig:CBVR}
\end{figure}

\gls{SIAT} provides an elegant and flexible six steps solution for the implementation and customization of scalable video indexing and retrieval, as shown in Fig.~\ref{fig:CBVR}. First, the \gls{VSAS} component acquires the video streams from \gls{VSDS} in the form of mini-batches, and in the case of batch analytics, video data is loaded from the \texttt{RAW DS} and feeds to the \gls{VBDPL}. 

In the second step, \gls{VBDPL} perform pre-processing operations and feature extraction. The former one encompasses structural exploration of a video, i.e., video scenes, shots detection, frames, and keyframes extraction, while the latter one extracts the low-level features. These low-level features can be keyframes' static features (texture-based, color-based, and shape-based), object features, and/or motion features (trajectory-based, statistics-based, and objects’ spatial relationships-based). The extracted features are then handed over to the \gls{VBDML} for classification and annotation.

In the third step, the semantic and high dimensional video feature vectors’ indexes make the representative index for persisted video sequences in \gls{IR}-DS. The \gls{IR}-DS is synchronized with \texttt{RAW-DS}.

In the fourth step, through the \gls{WSL}, the users are allowed to query the desired videos. In literature for video retrieval, various types of queries have been utilized for video retrieval. The queries can be categorized as \texttt{Query by Example}, \texttt{Sketch}, \texttt{Objects}, \texttt{Keyword}, and \texttt{Natural Language}. In the\texttt{Query by Example}, similar videos or images to the given sample video or image is extracted using feature similarity. In \texttt{Query by Sketch}, the features are extracted from users drawn sketches (sketch represent the required videos) and compared against the features generated from the stored videos \cite{hu2007semantic}. In \texttt{Query by Object}, the user give an object image then the \gls{CBVR} system search for all occurrences of the object in the video database \cite{sivic2006video}. Likewise, the last two approaches use \texttt{Keywords} and \texttt{Natural Language} \cite{aytar2008utilizing} as query. Combination-based Query combines different types of queries for \gls{CBVR}, was also adopted by some researchers such as  \cite{kennedy2005automatic,yan2004learning}.

In the fifth step, the \gls{CBVR} system search for similar videos against the user query. For similarity measure approaches can be classified as \texttt{feature}, \texttt{text}, \texttt{Ontology} and \texttt{combination} based matching. The similarity measure depends on the type of query. \textbf{Feature matching} based similarity measure is the average distance between the features of the corresponding frames \cite{lienhart2000system}. Query uses uses low-level features for extracting relevant videos. The benefit of this approach is that the video similarity can be found easily from the features but is not appropriate for semantic similarity, which is conversant to users. In \texttt{text matching} based similarity measure, the name of each concept is matched with given query terms so that to find the videos that satisfy the query. The best example of this approach is that of Snoek et al. \cite{snoek2007adding}. This approach is simple, but the query text must include the relevant concepts to get pleasing search results. In \texttt{Ontology-Based Matching}, the similarity between between semantic concepts is measured utilizing ontology. Query descriptions are enriched from knowledge sources, such as the ontology of concepts. Adding extra concepts can improve the retrieval results \cite{neo2006video} but can also decline search results unexpectedly. This approached is further explained in  section~\ref{sec:SemanticBasedVideoAnalysis}. \texttt{Combination-based matching} “leverages semantic concepts from a training collection by learning the combination strategies \cite{amir2003ibm}, and query-class-dependent combination models \cite{yan2004learning}” \cite{yan2007review}. Up to some extent, using this approach, the concept weights can be automated, and hidden semantic concepts can be handled but are difficult to learn query combination models.

Finally, the ranked result-set is presented to the user for browsing. To increase efficiency and to optimize the results of the \gls{CBVR} system, many researchers use the obtained relevance feedback from the user. This feedback can be categorized as \texttt{explicit}, \texttt{implicit}, and \texttt{pseudo feedback}. In the first case, users are asked to select relevant videos from the previous results actively \cite{chen2008integrated}. The \texttt{explicit feedback} approach obtains better results than the other two approaches, but direct user involvement is required. In the second case, the retrieval results are refined by exploiting the user’s interaction with the system, i.e., clicking pattern \cite{hopfgartner2007simulated}. In the third case, there is no involvement of the users. The user’s feedback is produced through positive (closely related to the query sample) and negative sample (different from the query sample) from the previous retrieval. These samples are directed to the system for the next search. Yan et al. \cite{yan2003negative}, and Hauptmann et al. \cite{hauptmann2008video} approaches are based on \texttt{pseudo-relevance} feedback approach. This approach substantially reduces the user interaction, but the semantic gap between low and high-level features obtained from different videos does not always agree with the similarities between the user-defined videos.

\subsection{Real-time Intelligent Video Analysis and Surveillance}
\label{sec:RealTimeIntelligentVideoAnalysisandSurveillance}
In the context of growing security concerns, the surveillance meant to criminality and intrusion detection. Video surveillance is not limited to these, but it encapsulates all aspects of monitoring to capture the dynamics of diverse application areas, e.g., transportation, healthcare, retail, and service industries. A generic use case for \gls{RIVA} has been shown in section~\ref{sec:IVAServiceUseCase}, Fig.~\ref{fig:RIVAUseCasePipeline}. Likewise, domain-specific \gls{RIVA} services for security and surveillance can be created under \gls{SIAT} architecture. In this section, we discuss recent literature on how the researchers advancing the activity and behavior analytics in the video streams with the aim of intrusion and crime detection, scene monitoring, and resource tracking. 

\subsubsection{Video Segmentation for Action Recognition}
\label{sec:VideoSegmentationforActionRecognition}
A combination of numerous actions, objects, and scenes forms complex events \cite{sun2014discover}. Video analytics against complex events is a nontrivial task. Complex event detection demands the association of multiple semantic concepts because it is almost impossible to capture the complex event through a single event class label \cite{habibian2014recommendations}. Video segmentation is required to mine informative segments regarding the event happening in the video. For effective event detection in video segmentation, it is vital to take into account the temporal relations between key segments in a particular event. The event videos hold intra-class variation, and several training videos are required to consider all possible instances of event classes. Song and Wu \cite{song2017extracting} suggested a methodology to extract key segments automatically for event detection in videos by learning from a loosely tagged pool of web videos and images. For the positions of key segments and content depiction in the video, they used an adaptive latent structural \gls{SVM} model and semantic concepts, correspondingly. They also developed two types of models, i.e., Temporal Relation Model and Segment-Event Interaction Model, for the temporal relations between video segments, and for evaluating the correlation between key segments and events respectively. They adapted labeled videos and images from the Web into their model and employed \texttt{'N'} adjacent point sample consensus \cite{wang2014video} for noise elimination. A knowledge-base was produced by Zhang et al. \cite{zhang2015enhancing} to decrease the semantic gap between complex events. To effectively model event-centric semantic concepts, they used a large-scale of web images for learning noise-resistant classifiers.

Action recognition in videos encompasses both segmentation and classification. Action recognition can be tackled by sliding windows and aggregation in sequential as well as isolated manner, or by performing both tasks parallel. \cite{song2017recognizing, habibian2014recommendations} are some of the good examples of video segmentation based event detection. The related literature on video segmentation based action recognition can be classified as \texttt{action} segmentation, a \texttt{depth-based approach}, and \texttt{deep learning-based} motion recognition. 

The most popular action segmentation model is dynamic time warping scheme \cite{escalante2017principal}. \texttt{Action segmentation} from videos using an appearance-based method considers a comparison between the start and end frames of contiguous actions. Quantity of movement \cite{jiang2015multi} and KNN along with HOG \cite{wu2012one} are commonly used to identify the start and stop frames of actions. Similarly, for action recognition, \texttt{depth-based approaches} have been developed which consist of binary range-sample feature \cite{lu2014range}, capturing local motion and geometry information \cite{yang2014super}, histogram oriented 4D normal \cite{oreifej2013hon4d}, and combination of depth motion map and HOG \cite{yang2012recognizing}. \cite{wang2015convnets, wang2015actionA, wu2016deep} have applied deep learning approaches to depth-based action recognition methods. Besides, for motion recognition, \texttt{deep learning} has been utilized in many ways. One is a suboptimal way, in which video represented as still images is fed to the channel of a \gls{CNN} \cite{simonyan2014two, yue2015beyond}. Sometimes video as compact images are given as input to the already trained \gls{CNN} to achieve good performance \cite{wang2018depth}. Video in the form of a sequence of images is input to the \gls{RNN} for sequential parsing of video frames for long as well as short-term patterns \cite{veeriah2015differential, du2015hierarchical}. In order to introduce a temporal dimension, video is regarded as a volume and substitutes the 2D filters of \gls{CNN} with 3D filters \cite{tran2015learning, ji20123d}.

In many application areas, e.g., resource tracking, action recognition, human behavior recognition, and traffic control, object tracking and motion detection in videos are vital. The process of detecting moving objects in videos is known as object tracking. For object tracking, initially, the foreground information is extracted in videos, and then the background modeling of the scene is captured using a background subtraction algorithm, i.e., IAGMM \cite{guler2016real, pham2010gpu}. To increase the accuracy, subsequently, shadow elimination algorithms are applied to the foreground frame \cite{cucchiara2001detecting}. A connected component algorithm is used to determine the bounding box of an object. To ensure frame-to-frame matching of the detected object, a method such as adaptive mean shift \cite{beyan2012adaptive} can be used for comparison. Factors like size and distance are used to object matching between frames. Finally, the occlusion scheme is used to detect and resolve occlusion. On the other hand, motion detection can be detected via foreground images extracted by the Gaussian Mixture Model background and connected component algorithm for noise removal. The area of detection is refined using a connected component algorithm and produces the bounding box information of the moving objects \cite{guler2016real}. The output of the detection is a binary mask representing the moving object for each frame in a particular sequence. Object detection for moving objects is challenging, especially in the case of shadows and cloud movement \cite{risse2017visual}. The related literature for moving object detection and classification can be categorized as a \texttt{stationary camera with a moving object} and \texttt{moving objects with moving camera}.

\paragraph{Moving object detection with stationary camera} 
In fixed camera video, the background image pixels in each frame remain the same, and thus simple background subtraction techniques are required. The object detection approaches using a fixed camera can be grouped as feature-based, motion-based, classifier-based, and template-based models \cite{shantaiya2013survey}. Categorization of object tracking in videos into point tracking, kernel tracking, and silhouette tracking as well as feature-based, region-based, and contour-based was performed by \cite{shantaiya2013survey, deori2014survey} respectively. Unlike a fixed camera, moving object detection with a moving camera is relatively challenging because of camera motion and background modeling for generating foreground and background pixel fails \cite{leal2017tracking}.

\begin{itemize}
	\item \texttt{Trajectory} classification involves computing long trajectories for feature point and discriminating trajectories that belong to different objects from those backgrounds using the clustering method. Some recommended algorithms include compensating long term motion based on flow optic technique \cite{yin2015background}, bag-of word classifier, and pre-trained \gls{CNN} method for detecting moving object trajectories \cite{zhang2017tracking}. 

	\item In \texttt{background modeling} based methods, for each sequence, the frame-by-frame background is created utilizing the motion compensation method. Some popular algorithms are Mixture of Gaussian \cite{hayman2003statistical}, complex homography \cite{lenz2011sparse}, gaussian-based method \cite{wren1997pfinder}, adaptiveMoG \cite{zivkovic2006efficient}, multi-layer homography transform \cite{jin2008background}, thresholding \cite{wu2015moving}, and \gls{CNN}-based method \cite{braham2016deep} for background modeling.
	
	\item  In \texttt{extension} of background subtraction method, low rank and sparse matrix decomposition method for static camera \cite{bouwmans2017decomposition} are extended to moving camera. If there exists coherency between a set of image frames, then a low-rank representation of the matrix created by these frames contains the coherency, and the sparse matrix representation contains the outliers, which represents the moving object in these frames. Low rank and sparse decomposition involve segmenting moving objects from the fixed background by applying principal component pursuit. It is a valuable technique in background modeling. Mathematical formula and optimization of this method can be found in \cite{yazdi2018new}.
\end{itemize}

\subsubsection{Behavior Analysis}
Usually, a camera is mounted nearby the digital displays to analyze and understand human behavior by investigating user interfacing with digital display \cite{elhart2017audience}. Commercial tools have been developed to analyze audience behavior using video analytics while considering parameters like age, gender, distance from the display, and sight and spent time. The obtained data can then be used to improve advertising campaigns in combination with sales data \cite{farinella2014face}. 

Recently, crowd analytic, i.e., human detection and tracking, have attracted the attention of the researches. The exploration of both group and individual behavior to govern abnormality scope the crowd analysis. Congestion analysis, motion detection, tracking, and behavior analysis are the main attributes of crowd analytics. While performing crowd analysis, factors like terrain features, geometrical information, and crowd flow can be considered. 

For analysis of the crowded scene, motion features are vital, and can be categorized as \texttt{flow-based features}, \texttt{local spatiotemporal features}, and \texttt{trajectory features} \cite{li2014crowded}. These features have applications in crowd behavior recognition, abnormality detection in-crowd, and motion pattern segmentation.

\paragraph{Flow-based Features} 
The flow-based features are pixel-level features, and in literature, different schemes have been proposed \cite{wang2014high,mehran2010streakline}, which be further classified as \texttt{optical flow}, \texttt{partial flow}, and \texttt{streak flow}. \texttt{Optical flow} technique encompasses computing pixel-wise motion between successive frames and can handle multi-camera objection motion. It has been applied to detection crowd motion and crowd segmentation \cite{su2013large}. This approach, optical flow, is unable to capture spatiotemporal attributes of the flow and long-range dependencies. \texttt{Particle flow} contains moving a grid of particles with the optical flow and providing trajectories that maps a particle’s initial position to its future or current position. It has an application in crowd segmentation and detection of abnormal behavior \cite{li2014crowded}.  Optical flow is unable to handle spatial changes. Mehran et al. \cite{mehran2010streakline} proposed \texttt{streak flow} to overcome the shortcomings of particle flow and to analyzing crowd video while computing motion field. This approach, streak flow, captures motion information similar to particle flow; changes in the flow is faster and performs well in dynamic motion flow.

\paragraph{Local Spatiotemporal Features} 
Flow-based features fail on the very crowded scene, and resultantly local spatiotemporal features techniques are developed, which are 2D patches or 3D cubes representation of the scene. Spatiotemporal features can be categorized as \texttt{spatiotemporal gradients}, and \texttt{motion histogram}. To capture steady-state motion behavior, Kratz and Nishino \cite{kratz2011tracking} used a spatiotemporal motion pattern model and confirmed the detection of abnormal activities. On the other hand, \texttt{motion histogram} considers motion information within the local region. It is not appropriate for crowd analysis because it takes a substantial amount of time, and is subject to error. However, some improvements have been shown in the literature to motion histogram, e.g., \cite{jodoin2013meta, wang2014high}.

\paragraph{Trajectory Features} 
Trajectory features signify tracks in videos. The distance between object-based motion features can be extracted from the trajectories of objects and can be utilized to analyze crowd activities. The failure to obtain a full trajectory in dense crowd leads to the concepts of tracklet. The tracklets are extracted from the dense region and enforce the spatiotemporal correlation between them to detect patterns of behavior. Tracklet is a fragment of a trajectory obtained within a short period, and the occurrence of occlusion leads to closure. Tracklets have been used for human action recognition \cite{lewandowski2013tracklet, olatunji2019video} and for the representation of motion in crowded scenes \cite{zhou2012understanding, chongjing2013analyzing}.

\subsubsection{Anomaly Detection}
\label{sec:AnomalyDetection}
Anomaly detection an application area of crowd behavior analysis and is domain-dependent. Anomaly in a video occurs when the analyzed pattern drifts from the normal in a training video. The related literature of anomaly detection can be categorized into three, i.e., \texttt{trajectory-based}, \texttt{global pattern-based}, and \texttt{grid pattern-based} method of anomaly detection.

\paragraph{Trajectory-based method of anomaly detection} 
In trajectory-based anomaly detection, objects are formed from the segment scenes, and then the object is followed in the video. A trajectory is caused by the tracked object, which describes the behavior of the object \cite{morris2008survey}. For the evaluation of abnormality in trajectory-based methods have been used i.e., single-class \gls{SVM} \cite{piciarelli2008trajectory}, zone-based analysis \cite{cocsar2016toward}, semantic tracking \cite{song2013fully}, String kernels clustering \cite{brun2014dynamic}, Spatiotemporal path search \cite{tran2013video}, and deep learning-based approach \cite{Revathi2017} have been used.

\paragraph{Global pattern-based method of anomaly detection} 
In a global pattern-based technique, the video sequence is analyzed in whole, i.e. low, or medium-level features are extracted from video using Spatiotemporal gradients or optical flow methods \cite{6129539}. The technique is suitable for crowd analysis because it does not individually track each object in the video but is challenging while locating the position where the anomaly occurred. Approached used in the global pattern-based method are Gaussian Mixture Model \cite{7490361}, energy model \cite{XIONG2012121}, SFM \cite{6894168}, stationary-map \cite{Yi:2014:LRS:2679600.2680014}, Gaussian regression \cite{Cheng2015VideoAD}, \gls{PCA} model \cite{Lee13anomalydetection}, global motion-map \cite{KRAUSZ2012307}, motion influence map \cite{7024902}, and salient motion map \cite{6217836}. 

\paragraph{Grid pattern-based method of anomaly detection}
In a grid pattern-based method, splits frames into blocks and individually analyze pattern on a block-level basis \cite{VishwakarmaA13}. If ignoring inter-object connections that lead to processing efficiency. Spatiotemporal anomaly maps, local features probabilistic framework, joint sparsity model, mixtures of dynamic textures with Gaussian Mixture Model, low-rank and sparse decomposition, cell-based texture analysis, sparse coding and deep networks are used in evaluating grid pattern-based methods \cite{8323245}.


\subsection{Video Summarization}
\label{VidSummarization}
Video big data are facing the challenge of sparsity and redundancy, i.e., hours of videos with less meaningful information, which creates many issues for viewing, mining, browsing, and storing videos. It has motivated researchers to find ways to shorten hours of videos and led to the field of video summarization. Video summarization is a process of generating a shorter video of the original one without spoiling the capability to comprehend the meaning of the whole video \cite{sen2019video}. The video summarization can be classified as Static Video Abstracts, Dynamic Video Skimming, and Video Synopsis.

\paragraph{Static Video Abstract} These approaches include a video table of contents, a storyboard, and a pictorial video summary. For example, Xie and Wu \cite{xie2008automatic} propose an algorithm to generate a video summary for broadcasting news videos automatically. An affinity propagation-based clustering algorithm is used to group the extracted keyframes into clusters, aiming to keep the relevant keyframes that distinguish one scene from the others and remove redundant keyframes. J. Wu et al. \cite{wu2017novel} were motivated by the notion from high-density peaks search clustering algorithm. They proposed a clustering algorithm by incorporating significant properties of video to gather similar frames into clusters. Finally, all clusters’ centers were presented as a static video summary. Bhaumik et al. \cite{bhaumik2014towards} proposed a summarization technique where they detect keyframes from each shot that eliminates redundancy at the intra-shot and inter-shot levels. For frames redundancy elimination, SURF and GIST feature descriptors were extracted for computing the similarity between the frames. The quality of the summaries obtained by using SURF and GIST descriptors are also compared in terms of precision and recall.

Similarly, Zhang et al.  \cite{zhang2016summary} propose a subset selection technique that leverages supervision in the form of human-created summaries to perform automatic keyframe-based video summarization. They were motivated by the intuition that similar videos share similar summary structures. The fundamental notion is to nonparametrically transfer summary structures from annotated videos to unseen test videos. Concretely, for each fresh video, they first compute the frame-level similarity between annotated and test videos. Then the summary structures are encoded in the annotated videos  with kernel matrices made of binarized pairwise similarity among their frames. Those structures are then combined into a kernel matrix that encodes the summary structure for the test video. Finally, the summary is decoded by feeding the kernel matrix to a probabilistic model called the determinantal point process to extract a globally optimal subset of frames. M. Gygli et al. \cite{gygli2015video} used a supervised approach to learn the importance of the global characteristics in summary by extracting deep features of video frames. J. Mohan et al. \cite{mohan2019static} proposed a technique that utilizes sparse autoencoders. Motion vectors have been used for the elimination of redundant frames, and then high-level features are extracted from frames using sparse autoencoders. These high-level feature vectors are then clustered using the K-means algorithm. The frames closest to the centroid of each cluster are selected as keyframes of the input video. Ji, Zhong, et al. \cite{ji2020cross} employed tag information, i.e., titles and descriptions, as the side information for the generation of summarization. A sparse auto-encoder was used as the primary model to generate the final summary, where the input and output were multiple videos and keyframes set, respectively. They fused the visual and tag information to guide visual features, which constrained the sparse auto-encoder to select the candidate keyframes. 

\paragraph{Dynamic Video Skimming} A summary video is formed from the video segments of the original video to remove redundancy or to summarize based on object action or events. As an example of object-based skimming, Peker et al. \cite{peker2006broadcast} proposed a video skimming algorithm while utilizing face detection on broadcast video programs. In the algorithm, the attention was given to faces, as they establish the focus of most consumer video programs. Ngo et al. \cite{ngo2005video} represent a video as a complete undirected graph and exploit the normalized cut algorithm to form optimal graph clusters. One-shot was taken from each cluster of visually alike shots to remove duplicate shots. Xiao et al. \cite{xiao2008automatic} mine frequent patterns from a video. A video shot importance evaluation model is utilized to choose useful video shots to create a video summary. For personal videos, Gao et al. \cite{gao2008video} developed a video summarization technique, which encompasses a two-level redundancy detection procedure. First, they terminated redundant video content with the hierarchical agglomerative cluster method at the shot level. Then parts of shots were selected, based on the ranking of the scenes and keyframes, to generate an initial video summary. Finally, to terminate the redundant information, a repetitive frame segment detection step was utilized. They verified the proposed technique through a prototype while using TV datasets (movies and cartoons videos) and reported the performance in terms of compression ratio (81\%) and recall (87.4\%). In \cite{xu2016fast}, for user-generated video summarization, both the representativeness and the quality of the selected segments from an original video were considered. They stated that user-generated videos contain semantic and emotional content, and its preservation is vital. They have designed a scheme to pick representative segments that include consistent semantics and emotions for the whole video. To ensure the quality of the summary, they computed quality measures, i.e., motion and lighting conditions, and integrate them with the semantic and emotional clues for segment selection.

Ji, Zhong, et al. \cite{ji2019video} addresses the issue of supervised video summarization by formulating it as a sequence-to-sequence learning problem, where the input and output is a sequence of original video frames and a keyshot sequence, respectively. The notion is to learn a deep summarization network with attention mechanism to mimic the way of selecting the human keyshots. The proposed framework was called attentive encoder-decoder networks for video summarization. They utilized the BiLSTM encoder for encoding the contextual information among the input video frames. For the decoder, two attention-based LSTM networks, are explored by using additive and multiplicative objective functions, respectively. The results demonstrate the superiority of the proposed framework against the state-of-the-art approaches, with remarkable improvements. J. Wu et al. \cite{wu2020dynamic} were motivated by the fact that multi-video summarization is significant for video browsing and proposed a technique where multi-video summarization was formulated as a graph problem.  They also introduced a dynamic graph convolutional network to measure the importance and relevance of each video shot locally as well as globally. They adopted two approaches to address the inherent class imbalance issue of video summarization. Additionally, a diversity regularization to encourage the model to generate a diverse summary was introduced. The results demonstrate the effectiveness of our proposed model in generating a representative summary for multiple videos with encouraging diversity. Z. Sheng-hua et al. \cite{zhong2019video} proposed a deep learning-based dynamic video summarization model. First, they addressed the issue of the imbalanced class distribution in video summarization. The over-sampling algorithm is used to balance the class distribution on training data. They proposed two-stream deep architecture with cost-sensitive learning to handle the class imbalance problem in feature learning. RGB images are utilized to represent the appearance of video frames in the spatial stream. Likely, multi-frame motion vectors with deep learning framework are introduced to represent and extract temporal information of the input video. Moreover, they stated that the proposed method highlights the video content with the active level of arousal in effective computing tasks and can automatically preserve the connection between consecutive frames.

\begin{figure}[h!]
	\centering
	\includegraphics[width=7 cm]{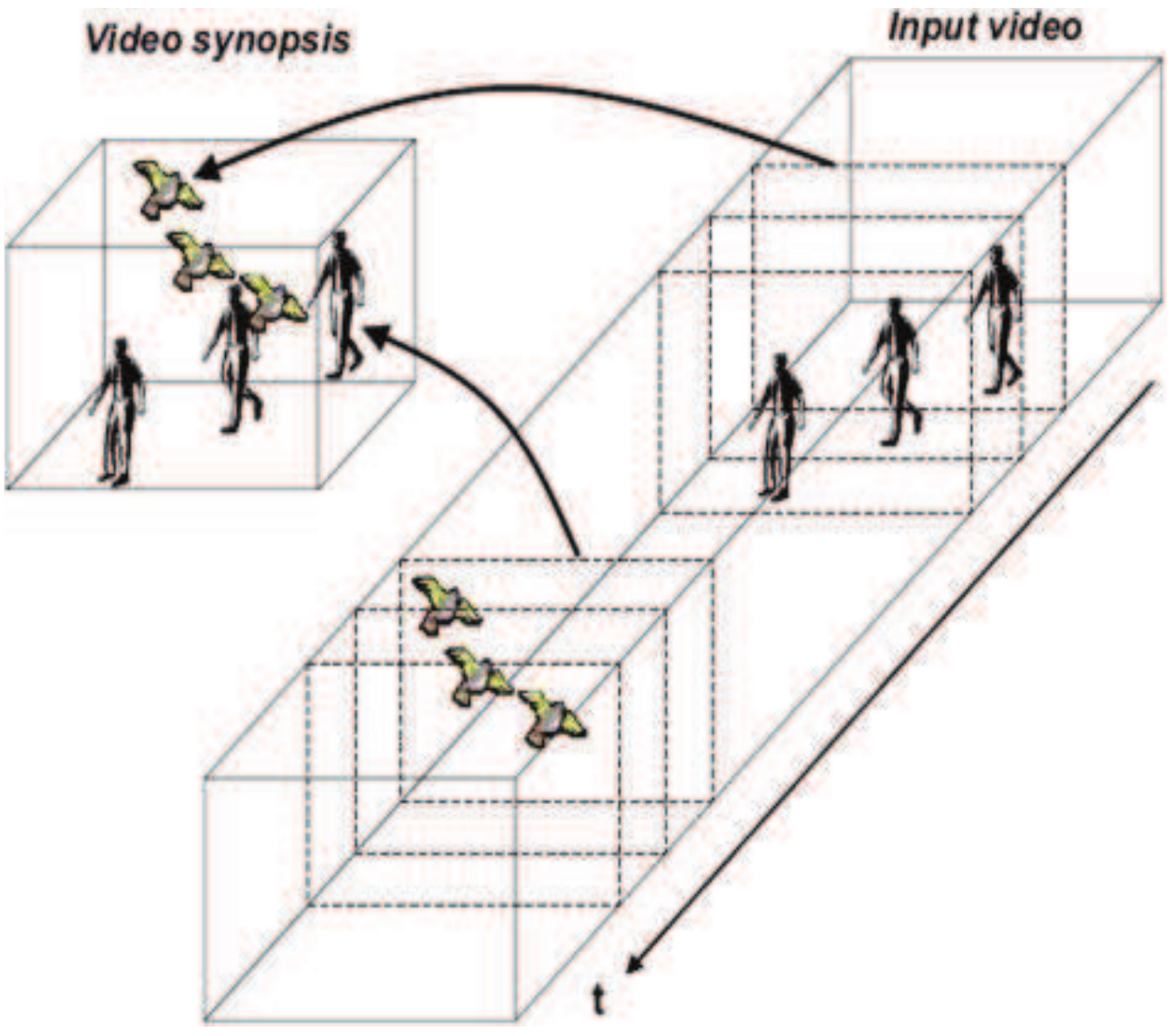}
	\caption{Simultaneous activities display in a video \cite{rav2006making}.}
	\label{fig:VideoSynopsis}
\end{figure}

\paragraph{Video Synopsis} In this approach, activities from the stated time interval are collected and moved in time to form a smaller video synopsis showing maximum activity, as shown in Figure \ref{fig:VideoSynopsis}. The notion of video synopsis was pioneered by \cite{rav2006making} in 2006 and proposed a two-phase approach, i.e., online and offline. The former phase includes the queuing of the generated activities. The later phase started after selecting a time interval of video synopsis with tube readjustment, background formation, and object stitching. A global energy function was defined and encompassing activity, temporal consistency, and collision cost. Then the simulated annealing method was applied for energy minimization. The video synopsis domain was further researched in single and multi-view scenarios. Some recent examples of a single-view are \cite{he2016fast,he2017graph,batuhan2017improved}. He et al. \cite{he2016fast,he2017graph} brought advancement in activity collision analysis by describing collision statuses between activities such as collision-free, colliding in the same direction and opposite directions. They also offered a graph-based optimization technique by considering these collision states to improve the activity density and put activity collisions at the center of their optimization strategy. Baskurt and Samet \cite{batuhan2017improved} concentrated on rising robustness of object detection by suggesting an adaptive background generation method. In another study, Baskurt and Samet \cite{bacskurt2018long} planned the object tracking method specified for video synopsis requirements. Their approach focused on long term tracking to represent each target with just one activity in video synopsis. Single view scalable approaches for video synopsis were projected by Lin et al. \cite{lin2017optimized} while utilizing distributed computing technology. Their proposed video synopsis approach encompasses steps like object detection, tracking, classification, and optimization, which were performed in a distributed environment. Ahmed, S. A. \cite{ahmed2019query} proposed a query-based method to generate a synopsis of long videos. Objects were tracked and utilized deep learning for objects classification (e.g., car, bike, etc.). Through unsupervised clustering, they identified regions in the surveillance scene. The source and the destination represented spatiotemporal object trajectories. Finally, user queries were allowed to generate video synopsis by smoothly blending the appropriate tubes over the background frame through energy minimization.

Further, the examples of multi-view video analytics are that of \cite{zhu2014key, mahapatra2016mvs}. Zhu et al. \cite{zhu2014key} proposed a framework to generate a unified synopsis of multi-view videos. The synopsis is visualized by mapping multiple views to a common ground plane. Multiple camera activities were allied via trajectory matching in overlapping camera views. The process of producing a synopsis requires a balance among minimizing the synopsis length, maximizing the information coverage, and reducing the collision among objects’ track that are presented concurrently. Likewise, Mahapatra et al. \cite{mahapatra2016mvs} proposed a multi-camera approach for an overlapping camera network and modeled the synopsis generation as a scheduling problem. They utilized three distinct methods, i.e., table-driven approach, contradictory binary graph coloring approach, and simulated annealing. Action recognition modules were integrated to recognize significant actions, i.e., walking, running, bending, jumping, hand-shaking, and one or both hands waving. The inclusion of such essential actions can help in the synopsis length reduction while preserving the value. The synopsis length was further reduced while utilizing a fuzzy inference system that computes the visibility score for each object tracking. They stated that through the contradictory binary graph coloring approach, they achieved a maximum reduction in synopsis length. Zhang, Z. et al. \cite{zhang2019multi}, tried to address the issue of video synopsis by joint object-shifting and camera view-switching to show multiple synopsis results more compactly and understandably. The input videos were synchronized and grouped the same object in different videos together. Then they shifted the grouped objects with respect to the time axis to gain multiple synopsis videos. They constructed a simultaneous object-shifting and view-switching optimization framework to achieve encouraging synopsis results. To address unified optimization, they further presented an optimization strategy composed of graph cuts and dynamic programming. 

\subsubsection{Video summarization aaS under L-CVAS}
\label{GenericVideoSummarization}

\begin{figure}[h!]
	\centering
	\includegraphics[width=7.3 cm]{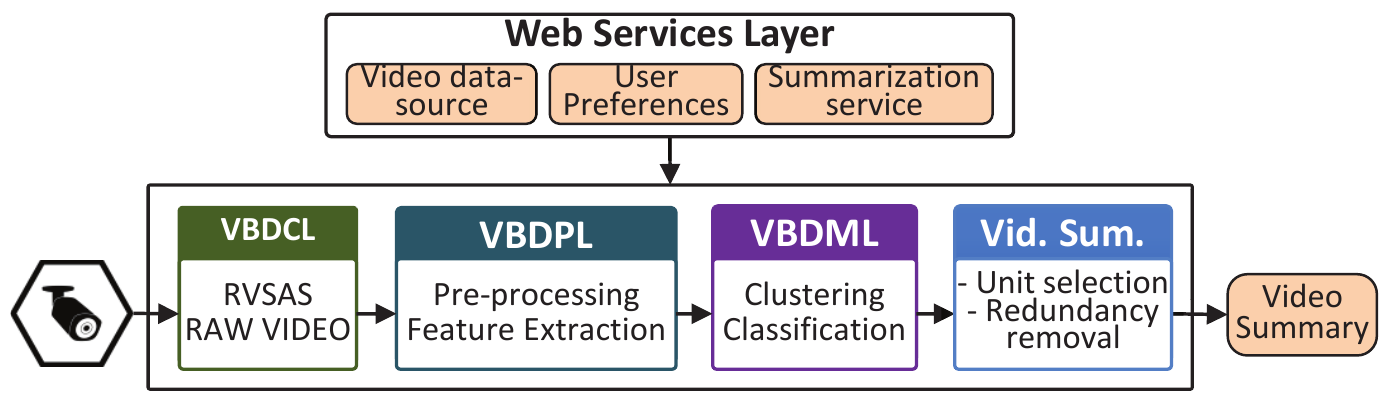}
	\caption{Generic flow of video summarization aaS under Lambda CVAS architecture.}
	\label{fig:VideoSummarizationFlow}
\end{figure}

Fig.~\ref{fig:VideoSummarizationFlow} shows the flow of the video summarization flow under the proposed \gls{SIAT} architecture. Through the \gls{WSL}, the user first subscribes to the video data-source to the video summarization service. Then the user preferences allow the users to set the parameters required for video summary service and personalization. The summary parameters encompass granularity level, type of summary to be performed (e.g., overview, highlights, synopsis, etc.) and any other as per the video summary service scenario. The personalization can be in terms of specific features of the video, like people, objects, events, etc. Through the \gls{VBDCL}, videos are acquired from the video data-source and sent to the \gls{VBDPL} for pre-processing. In pre-processing, video units are extracted (segmentation, shots, frame extraction, etc.) as per the requirements of the video summarization service. Then multiple low and high-level features, such as motion, color, aesthetics, semantics, etc., are extracted from the video units. The extracted features from the basic video units are input to the \gls{VBDPL} for object/activity identification and clustering. Once done then the next step is video summarization. The video summarization phase deploys the actual logic, i.e., video unit selection, and redundancy removal. The video unit selection and redundancy removal decide which video units should be included in the video summary based on unit significance, summary length, and other user’s parameters. This block also removes similar video units within the video summary to achieve the best possible video summary, covering the required details in the original video. Finally, the summary results are delivered to the respective user through the \gls{WSL}.

\subsection{Semantic-based Video Analysis}
\label{sec:SemanticBasedVideoAnalysis}
To bridge the semantic gap and to allow the machine to understand the visual data, Semantic Web technologies can be incorporated \cite{ballan2010video}. Semantic Web unlocked a new avenue for knowledge-based computer vision while enabling data exchange between video analysis systems in an open and extensible manner. The scientific community has exploited the Semantic Web concept for intelligent video analytics to bridge the so-called semantic gap between the low-level features and high-level human-understandable concepts. State-of-the-art scholarly work can roughly be classified as semantic-based low-level, mid-level, high-level analysis, and semantic-based video search and retrieval.

\subsubsection{Semantic-based Low-level analysis} 
\label{sec:Semantic-based-Low-level-analysis}
Semantic-based low-level analysis refers to the formalization of the extracted objects of interest from the videos. It then performs reasoning on formalization such as detection and tracking across domain-specific videos \cite{ferryman2014performance,foggia2013real}. Dasiopoulou et al. \cite{dasiopoulou2005knowledge} proposed a multimedia ontology for domain-specific video analysis. In semantic concepts, they consider object attributed, low-level features, spatial object relations, and the processing approaches while defining F-logic rules for reasoning that govern the application of analysis methods. García et al. \cite{garcia2008video} proposed a knowledge-based framework for video object segmentation, where relationships among analysis phases are utilized. The main contribution is to provide a detailed description of the scene at low, mid, and high semantic levels through an ontology. The notion is to offer the semantic rich description of a scene via an ontology that includes occurrences in the scene from high to low semantic level, controls iterative decisions on every stage. The low-level analysis modules (background subtraction, short-term change detection, point or region tracking, etc.) are provided with a structure to collaborate and achieve consistently, and contextual results. The results of the vision algorithms are mapped to the ontology, representing  low-level scene occurrences. The following stages build Point Hypothesis Maps and Region Hypothesis Maps. These were the most probable occurrences of each point and region. The points and regions are coded according to the ScenePoint and SceneRegion hierarchies of the analysis ontology. The quality of the results was evaluated through a feedback path. Gomez et al. \cite{gomez2011ontology} proposed a computer vision framework for surveillance and consists of two layers, i.e., the tracking layer and a context layer. The proposed framework relies on an ontology-based representation of the scene in combination with contextual information and sensor data. The notion is the application of logical reasoning initiating from the acquired data from a classical tracker intending to construct an ontological model of the objects and the activities happening in the area of observation. Reasoning procedures were utilized to detect and predict tracking errors, sending feedback to the tracker to adjust the low-level image processing algorithms. Vision operation, like movement detection, blob-track association, and track and trajectory generation, were performed in the tracking layer. The context layer was supposed to produce a high-level interpretation of the scene. RACER reasoner \cite{haarslev2001racer} was utilized for scene interpretation since it allows abductive reasoning. Abductive rules are defined in the proposed framework to interpret what is happening in the scene from the primary tracking data.

\subsubsection{Semantic-based Mid and High-level Analysis} 
\label{sec:SM-MidHighLevelAnalysis}
Atomic events such as loitering, fall, direction changes, group formations, and separations \cite{bird2005detection}, and  “complex” events such as aggressions, fights, thefts, and other general suspicious events \cite{laxton2007leveraging, wang2011action} falls in mid and high-level video analytics, respectively. 

The utilization of semantic technology for video event representation in surveillance videos was initiated by Video Event Representation Language \cite{francois2005verl}. The main idea is to model simple events in a hierarchical framework intended for detecting complex events. They stated that the sequence of simple events (car door opening, leaving a car,  car door closing, walking and opening a building door, and entering a building) forms complex events (a person arrived through a car and entered a building). They used Allen’s interval algebra to handle temporal relationships between subevents. They clarified their proposed application through the detection of an example event in a surveillance video, i.e., accessing a secure zone by entering behind an authorized individual. For complex events recognition in surveillance videos, Snidaro et al. \cite{snidaro2007representing} proposed an ontology and is composed of three high-level concepts, i.e., background, entities, and events. The event class, high-level in subclasses, describes simple events, spatial events, and transitive events, allowing to show how complex events can be described through simple events sequencing. The event concept is composed of sub-concepts that represent simple events, spatial events, and transitive events. The focus was on complex events, which can be achieved by sequencing simple events. SanMiguel et al. \cite{sanmiguel2009ontology} propose an ontology for representing the prior knowledge related to video event analysis. Such knowledge is described in terms of scene related entities (Object, Event, and Context), and system-related entities. The key contribution of the work is the integration of different types of knowledge in an ontology for detecting the objects and events in a video scene. In the same direction, Greco et al. \cite{greco2016abnormal} proposed a hybrid approach for simple abnormal (person falling) and complex abnormal events (person aggression) recognition using semantic web technologies. They modeled the extracted general tracking information to the proposed tracking ontology for advanced reasoning. The data from the videos were obtained using the tracking component (frames, bounding box), knowledge about the scene (static and dynamic objects, occluding objects), Situations and Events (people leaving the scene, falling ground, fighting).  For event detection, SPIN rules and functions are used while SPARQL queries are employed for analytics tasks. The system has proven to successfully recognize mid-level events (ex. people falling to the ground) and high-level events (ex. person being attacked) on the PETS2016 dataset.

Researches also utilized semantic technologies aiming to address the issue of human activity recognition in daily living. In this context, Chen et al. \cite{chen2009ontology} introduced a method for activity recognition while using ontological model, representation, and reasoning. They analyzed the nature and characteristics of daily life activities and modeling related concepts through ontologies. The authors describe the algorithms of activity recognition making full use of the reasoning power of semantic modeling and representation. They claimed that the proposed ontological models for daily life activities could easily be customized, deployed, and scaled up. Likewise, \cite{cavaliere2016towards}, an approach exploiting the synergy between the semantic technologies and tracking methods have been presented for object labeling. The work aims to augment and comprehend situation awareness, as well as critical alerting conditions. The unmanned aerial vehicles with an embedded camera were utilized to recognize moving and stationary objects along with relations between them. Contextual information was used for abnormal event detection. A prototype was designed and used a drone to capture videos on the University of Salerno. They stated that the proposed system could recognize an abnormal event by means of SWRL rules associated with mid-level activities, such as ball kicking by a human and a car passing through the same road.

\subsubsection{Semantic-based Video Retrieval} 
\label{sec:SemanticVideoRetrieval}
Some researchers exploited semantic technologies for video search and retrieval purposes. In this regard, Yao et al. \cite{ yao2010i2t} proposed the image to text framework to extract events from images (or video frames) and then provide semantic and text annotations. The And-or-Graph incorporates vocabularies of visual elements like objects and scenes along with stochastic image grammar that identifies semantic relations among the visual elements. In this way, low-level image features are linked with high-level concepts, and the parsed image can be transformed into semantic metadata to form the textual description. Video contents are expressed in both OWL and text format, and then the users are allowed to search images and video clips through keyword searching and semantic-based querying. Xue et al. \cite{ xue2012ontology} proposed an ontology-based content archive and retrieval framework for surveillance videos. A surveillance ontology was proposed that represents semantic information of video clips as a resource ontology. Such an ontology models the basic feature description in the low level, the video object description in the mid-level, and event description in the high-level. The proposed system was tested for object and event retrieval, such as walking and car parking.

Furthermore, Xu et al. in \cite{xu2013video} and \cite{xu2015semantic} propose a method to annotate video traffic events while considering their spatial and temporal relations. They introduced a hierarchical semantic data model called structural video description, which consists of three layers, i.e., pattern recognition layer (ontological representation from the video the extracted video concepts), video resources layer (links video resources with their semantic relations), and demands layer (retrieval interface). They defined various concepts in the ontology, such as persons, vehicles, and traffic signs that can be used to annotate and represent video traffic events. Besides, the spatial and temporal relationships between objects in an event are defined. As a case study, an application to annotate and search traffic events is considered. Sah et al. \cite{sah2017semantic} proposed a multimedia standard-based semantic metadata model and annotate globally inter-operable data about abnormal crowd behaviors from surveillance videos. Similar efforts are made by Sobhani et al. \cite{sobhani2015ontology} and proposed an advanced intelligent forensic retrieval system by taking advantage of an ontological knowledge representation while considering the UK riots in 2011 as a use case. Similarly, A. Alam et al. \cite{alam2020intellibvr} proposed a layered architecture for large-scale distributed intelligent video retrieval while exploiting deep-learning and semantic approaches called IntelliBVR. The base layer is responsible for large-scale video data curation. The second and third layers are supposed to process and annotate videos, respectively while using deep learning on the top of a distributed in-memory computing engine. Finally, the knowledge curation layer, where the extracted low-level and high-level features are mapped to the proposed ontology, can be searched and retrieved using semantic rich queries. Finally, they projected the effectiveness of IntelliBVR through experimental evaluation.
\section{State-of-the-art CVAS}
\label{sec:StateOfTheArtCVAS}
In this section, we review the start-of-the-art \gls{CVAS}. This discussion also supports our claim on the relationship between video big-data analytics and cloud computing. The discussion is further divided into two subsections, i.e., Scholarly \gls{CVAS} and Industrial \gls{CVAS}.    

\subsection{Scholarly CVAS}
In this subsection, we explore academic research trends (summarized in table~\ref{tab:StateOftheArtScholarlySystems}) that how scientific community investigate and proposed cloud-based \gls{IVA} solutions while utilizing big data technologies. In this direction, Ajiboye, S.O. et al. \cite{ajiboye2015hierarchical} stated that the network video recorder is already equipped with intelligent video processing capabilities but complained about its limitations, i.e., isolation, and scalability. To resolve such issues, they proposed a general high-level theoretical architecture called \gls{FVSA}. The design goals of the \gls{FVSA} were cost reduction, unify data mining, public safety, and scalable \gls{IVA}. The \gls{FVSA} architecture consists of four-layer, i.e., Application layer (responsible for system administration and user management), Services Layer (for storage and analytics), Network Layer, and Physical Layer (physical devices like camera, etc.). They guaranteed the compatibility of \gls{FVSA} with the hierarchical structure of computer networks and emerging technologies. Likewise, Lin, C.-F. et al. \cite{lin2012framework} implemented a prototype of a cloud-based video recorder system under gls{IaaS} while using big data technologies like \gls{HDFS} and Map Reduce. They showed the scalable of video recording, backup, and monitoring features only without implementing any video analytics services. Similarly, Liu, X. et al. \cite{liu2015distributed} also came out with a cloud platform for large scale video analytics and management. They stated that the existing work failed to design a versatile video management platform in a user-friendly way and to effectively use Hadoop to tune the performance of video processing. They successfully develop a cloud platform and the same big data technologies, i.e., Hadoop and MapReduce. They also managed to develop three video processing services, i.e., video summary, video encoding and decoding, and background subtraction.

Tan, H. et al. \cite{tan2014approach} used Hadoop and MapReduce for fast distributed video processing and analytics. They developed two video analytics services, i.e., face recognition and motion detection, by using JavaCV. Furthermore, Ryu, C., et al. \cite{ryu2013extensible} proposed a cloud video analytics framework using \gls{HDFS} and MapReduce along with OpenCV \cite{bradski2000opencv} and FFmpeg for video analytics. They implemented face recognition and tracking algorithm and reported the scalability of the system and the accuracy of the algorithm. Ali M. et al. \mbox{\cite{ali2020res}} proposed an edge enhanced stream analytics system for video big data called RealEdgeStream. They tried to investigate video stream analytics issues by offering filtration and identification phases to increase the value and to perform analytics on the streams, respectively. The stages are mapped onto available in-transit and cloud resources using a placement algorithm to satisfy the Quality of Service constraints recognized by a user. They demonstrate that for a 10K element data streams, with a frame rate of 15-100 per second, the job completion took 49\% less time and saves 99\% bandwidth compared to a centralized cloud-only based approach.

White et al. \cite{white2010web} researched MapReduce for \gls{IVA} services, which comprises classifier training, clustering, sliding windows, bag-of-features, image registration, and background subtraction. However, experiments were performed for the k-means clustering and Gaussian background subtraction only. Tan and Chen \cite{tan2014approach} presented face detection, motion detection, and tracking using MapReduce-based clusters on Apache Hadoop. They utilized JavaCV since Hadoop is developed and designed for Java. Pereira, R. el al. \cite{pereira2010architecture} proposed a cloud-based distributed architecture for video compression based on the Split-Merge technique while using the MapReduce framework. They stated that they optimized the Split-Merge technique against two synchronization problems. The first optimization problem was split and merge video fragments without loss in synchronization, whereas the second optimization problem is the synchronization between audio and video can be greatly affected, since the frame size of each one may not be equal. Similarly, Liu et al. \cite{liu2014distributed} used Hadoop and MapReduce for video sharing and transcoding purposes.


\begin{table*}[]
	\centering
	\caption{Scholarly state-of-the-art-work CVAS.}
	\label{tab:StateOftheArtScholarlySystems}
	\resizebox{\textwidth}{!}{%
		\begin{tabular}{|l|l|l|l|l|l|l|l|l|l|l|}
			\hline
			\multicolumn{1}{|c|}{\multirow{2}{*}{\textbf{Ref}}} &
			\multicolumn{3}{c|}{\textbf{IVA}} &
			\multicolumn{2}{c|}{\textbf{VDM}} &
			\multicolumn{3}{c|}{\textbf{Core Technology}} &
			\multicolumn{1}{c|}{\multirow{2}{*}{\textbf{
						\begin{tabular}[c]{@{}c@{}}
							Performance\\ Matrix
						\end{tabular}}}} &
			\multicolumn{1}{c|}{\multirow{2}{*}{\textbf{
						\begin{tabular}[c]{@{}c@{}}
							(a)Objectives \\
							(b)Target Domain (c) Issues
						\end{tabular}}}} \\ \cline{2-9}
			\multicolumn{1}{|c|}{} &
			\multicolumn{1}{c|}{\textbf{R}} &
			\multicolumn{1}{c|}{\textbf{B}} &
			\multicolumn{1}{c|}{\textbf{Service}} &
			\multicolumn{1}{c|}{\textbf{St}} &
			\multicolumn{1}{c|}{\textbf{S}} &
			\multicolumn{1}{c|}{\textbf{Storage}} &
			\multicolumn{1}{c|}{\textbf{DP}} &
			\multicolumn{1}{c|}{\textbf{Lib.}} &
			\multicolumn{1}{c|}{} &
			\multicolumn{1}{c|}{} 
			\\ \hline
			
				\cite{hossain2014framework} &
				\xmark &
				\cmark &
				\begin{tabular}[c]{@{}l@{}}-Face detection\\ -Motion detection\end{tabular} &
				\xmark&
				\xmark&
				\begin{tabular}[c]{@{}l@{}}MS SQL\\ Server\end{tabular} &
				\xmark&
				OpenCV &
				\begin{tabular}[c]{@{}l@{}}Workload for\\ - Face detection\\ - Storage avg.  \\ task time\end{tabular} &
				\begin{tabular}[c]{@{}l@{}}(a) CVAS Architecture, \\ Implements a prototype\\ (b) NA  \\ (c) Not scalability and SOA\end{tabular} 
			
			\\ \hline
				
				\cite{valentin2017cloud}&
				\xmark &
				\cmark&
				\begin{tabular}[c]{@{}l@{}}- Bg. Subtract\\ -Object \\ classification\\ -Object tracking\end{tabular} &
				\cmark&
				\xmark &
				- &
				- &
				- &
				\xmark &
				\begin{tabular}[c]{@{}l@{}}(a) CVAS Architecture, \\ Implements a prototype.\\ (b) Surveillance \\ (c) Not scalable, consumer \\ only, no evaluation.\end{tabular} 
		
			\\ \hline
		
				\cite{pereira2010architecture} &
				\cmark&
				\xmark &
				\begin{tabular}[c]{@{}l@{}}Video \\ compression\end{tabular} &
				\xmark &
				\cmark&
				HDFS &
				MR &
				- &
				Scalability &
				\begin{tabular}[c]{@{}l@{}}(a) Merge-Split Based \\ Video encoding. \\ (b) Storage and transmission. \\ (c) Video compression only.\end{tabular} 
				
			\\ \hline
			
				\cite{ajiboye2015hierarchical} &
				- &
				\cmark&
				- &
				\cmark&
				\cmark&
				- &
				- &
				- &
				- &
				\begin{tabular}[c]{@{}l@{}}(a) Proposed video surveillance  \\ architecture for video analytics \\ (b) Surveillance \\ (c) Based on B2C models, and\\  not SOA.\end{tabular} 
			
			\\ \hline
			
				\cite{lin2012framework} &
				\cmark&
				x &
				- &
				\cmark&
				\cmark&
				\begin{tabular}[c]{@{}l@{}}HDFS, \\ Hbase\end{tabular} &
				- &
				- &
				\begin{tabular}[c]{@{}l@{}}-Bandwidth usage\\ -Scalability\end{tabular} &
				\begin{tabular}[c]{@{}l@{}}(a) Cloud-based architecture for \\ video recordings. \\ (b) Surveillance\\ (c) IVA not supported\end{tabular} 
			
			\\ \hline
			
				\cite{liu2015distributed} &
				\cmark&
				\xmark &
				\begin{tabular}[c]{@{}l@{}}-Video summary\\ - Encoding\\ - Decoding\\ - Bg. Subtract\end{tabular} &
				\cmark&
				\cmark&
				\begin{tabular}[c]{@{}l@{}}HDFS,\\ MySQL\end{tabular} &
				MR &
				\begin{tabular}[c]{@{}l@{}}OpenCV,\\ CUDA\end{tabular} &
				\begin{tabular}[c]{@{}l@{}}-Throughput\\ -Average I/O rate\\ -Execution Time\end{tabular} &
				\begin{tabular}[c]{@{}l@{}}(a) Distributed video management \\ platform. \\ (b) Distributed video management. \\ (c) Not SOA, video stream not \\ scalable.\end{tabular} 
			\\ \hline
		
				\cite{tan2014approach} &
				\cmark&
				\cmark&
				\begin{tabular}[c]{@{}l@{}}-Face Recognition,\\ -Motion Detection\\ -Tracking\end{tabular} &
				\xmark &
				\cmark&
				HDFS &
				MR &
				\begin{tabular}[c]{@{}l@{}}OpenCV, \\ FFMPEG \\ JavaCV\end{tabular} &
				-Scalability &
				\begin{tabular}[c]{@{}l@{}}(a) Fast and distributed video \\ processing. \\ (b) Video surveillance. \\ (c) Distributed video processing \\ only.\end{tabular} 
			\\ \hline
			
			\begin{tabular}[c]{@{}l@{}}
				\cite{zhang2015deep} \\
				\cite{zhang2016deep}
			\end{tabular}	
			&
				\cmark&
				\cmark&
				\begin{tabular}[c]{@{}l@{}}Object recognition \\ and tracking \\ vehicle \\ classification\end{tabular} &
				\cmark&
				\cmark&
				\begin{tabular}[c]{@{}l@{}}HDFS,\\ HBase,\\ SploutSQL\end{tabular} &
				\begin{tabular}[c]{@{}l@{}}MR\\ Storm\end{tabular} &
				OpenCV &
				\begin{tabular}[c]{@{}l@{}}{[}8{]} -Accuracy \\ Recognition \\ {[}9{]} -Fault \\ Tolerance,\\ -Scalability\end{tabular} &
				\begin{tabular}[c]{@{}l@{}}(a) Proposed a cloud platform for\\  online and offline video analytics. \\ (b) Security, surveillance. and \\ traffic. \\ (c) Not SOA,\end{tabular} 
			\\ \hline
			
				\cite{zhang2015video} &
				\cmark&
				\cmark&
				\begin{tabular}[c]{@{}l@{}}Face recognition\\ \& counting\end{tabular} &
				\cmark&
				\cmark&
				\begin{tabular}[c]{@{}l@{}}HDFS,\\ Splout-SQL\end{tabular} &
				\begin{tabular}[c]{@{}l@{}}MR\\ Storm\end{tabular} &
				OpenCV &
				\begin{tabular}[c]{@{}l@{}}-Scalability \\ -Fault tolerance\end{tabular} &
				\begin{tabular}[c]{@{}l@{}}(a) General cloud architecture for \\ Video Analytics.\\ (b) Security and transportation \\ surveillance \\ (c) Not SOA\end{tabular} 
			
			\\ \hline
			
			\cite{uddin2019siat} &
				\cmark&
				\cmark&
				\begin{tabular}[c]{@{}l@{}}-Face detection\\ -Action recognition\end{tabular} &
				\cmark&
				\cmark&
				\begin{tabular}[c]{@{}l@{}}HDFS,\\ Splout-SQL\end{tabular} &
				\begin{tabular}[c]{@{}l@{}}MR\\ Storm\end{tabular} &
				OpenCV &
				\begin{tabular}[c]{@{}l@{}}-Scalability \\ -Fault tolerance\end{tabular} &
				\begin{tabular}[c]{@{}l@{}}(a) CVAS Platform\\ (b) Security and surveillance\end{tabular} 
			\\ \hline
			
			\multicolumn{11}{l}{\textit{
					\begin{tabular}[c]{@{}l@{}}
						\\
						B=Batch IVA, R=Real-time IVA,  
						St = Video Stream Management, 
						S = Video Storage Management,  
						DP = Distributed Processing,  MR = MapReduce
					\end{tabular}}}
		\end{tabular}%
	}
\end{table*}


Zhang, W. et al. \cite{zhang2015video} proposed a cloud-based architecture for large scale intelligent video analytics called BiF. BiF combines the merits of \gls{RIVA} and \gls{BIVA} while exploiting distributed technologies like storm and MapReduce, respectively. BiF architecture considered non-functional architectural properties and constraints, i.e., usability, scalability, reliability, fault tolerance, data immutability, re-computation, storing large objects, batch processing capabilities, streaming data access, simplicity and consistency. The BiF architecture consists of four main layers, i.e., data collection layer, batch layer, real-time layer, and serving layer. The data collection layer collects the streaming video frames from the input video sources (camera). The data collection layer forwards the video frames to the batch layer and streaming layer for batch processing and real-time analytics, respectively. The service layer is to query both batch views and real-time views and integrate them to answer queries from a client.  To evaluate the performance of the BiF architecture, they developed a video analytics algorithm, which was able to detect and count faces for a specific interval of time from the input source. During the evaluation, they showed that BiF is efficient in terms of scalability and fault tolerance. Zhang et al. \cite{zhang2016efficient} introduced Apache Kafka and Spark Streaming framework for efficient real-time video data processing. They also proposed a fine-grained online video stream task management scheme to boost resource utilization and experimented with license plate extraction and human density analysis.

Azher et al. \cite{uddin2019siat} proposed \gls{CVAS} for \gls{RIVA} and \gls{BIVA} while using Spark Stream and Spark, respectively. They implemented IVA services such as human action recognition and face recognition services, respectively. In another work, Azher et al. \cite{uddin2017human} proposed a novel feature descriptor to recognize human action on Spark while utilized the Spark MLlib \cite{meng2016mllib} to recognize the action from the feature vector generated by ALMD \cite{uddin2017human}. Wang et al. \cite{wang2015large} also performed human action recognition on Spark. The aim was to speed up some key processes, including trajectory-based feature extraction, Gaussian Mixture Model generation, and Fisher Vector encoding. Distributed video processing called streaming video engine is also introduced in \cite{huang2017sve} for distributed \gls{IVA} framework at Facebook scale against three major challenges, i.e., low latency, application-oriented flexibility, and robustness to faults and overload.  

Zhang et al. \cite{zhang2015deep,zhang2016deep} stated that the historical video data could be used with the updated video stream to know the current status of an activity, e.g., status of traffic on the road, and to predict future. To make it possible, they proposed a video cloud-based service-oriented layered architecture called Depth Awareness Framework and consists of four layers, i.e., data retrieval layer, offline video analytics layer, online video processing layer, and domain service layer. The data service layer is supposed to hander large-scale video data and Webcam Stream. The offline layer is used to perform the operation on the batch videos, whereas online processing occurs in a real-time video processing layer. On the top of the proposed cloud platform, they implemented deep convolution neural network for obtaining in-depth raw context data inside the big video, and a deep belief network-based method to predict workload status of different cloud nodes, as part of knowledge on a system running status. They prepared a dataset consisting of seven traffic videos, each of size 2GB. During the evaluation, they stated the improvement in object prediction accuracy, fault tolerance, and scalability. Zhang et al. \cite{zhang2018deep} performed pedestrian recognition on real-time video data using deep learning. Here, the \gls{CNN} network is improved to fine-CNN, which consists of a nine-layer neural network. Moreover, the Apache Storm framework, along with a GPU-based scheduling procedure, is presented.

\subsection{Industrial CVAS}
Various leading industrial organizations have successfully deployed \gls{CVAS}. Some of the most popular are briefly described in the following subsections.


\begin{table*}[!htb]
	\centering
	\caption{Industrial CVAS.}
	\label{tab:IndCVAS}
	\resizebox{\textwidth}{!}{%
		\begin{tabular}{|l|l|l|l|l|l|}
			\hline
			\textbf{CVAS}
			& \textbf{Target Domain}                                                                                                                                             & \textbf{\gls{RIVA} Services}                                                                                                                                                                                                               & \textbf{\gls{BIVA} Services}                                                                                                                                                     & \textbf{BM}                                      
			& \textbf{API} 
			\\ \hline
			
			\begin{tabular}[c]{@{}l@{}}Citilog \\ 
				(1997)	\\	\cite{CitiLog2018}
			\end{tabular}                    
			& 
			\begin{tabular}[c]{@{}l@{}}
				Security \& Surveillance \\ 
				(Transportation, Highways, \\
				Tunnels, Bridges, and smart \\
				cities).
			\end{tabular}                        
			& 
			\begin{tabular}[c]{@{}l@{}}
				Incident detection, traffic \\
				data collection, interaction \\
				control, license plate \\
				recognition
			\end{tabular}                                                                                              
			& NA                                                                                                                                                                              & 
			\begin{tabular}[c]{@{}l@{}}
				B2B, \\ B2C
			\end{tabular} 
			& \cmark             
			\\ \hline

			\begin{tabular}[c]{@{}l@{}}
				Check-\\Video\\ (1998)
				\\	\cite{checkvideo2019}	
			\end{tabular}              & 
			
			\begin{tabular}[c]{@{}l@{}}
				Security \&  Surveillance \\
				(education, guard, airports, \\
				retail, construction, services, \\ 
				residential, outdoor assets)
			\end{tabular} & 
			
			\begin{tabular}[c]{@{}l@{}}
				Basic Analytics (motion detection, \\
				zone violation, camera tamper) \\
				Object classification Business \\
				Analytics (object counting, tracks, \\
				transaction)
			\end{tabular} & NA                                                                                                                                                                              & B2C                                                 
			& \xmark             
			\\ \hline

			\begin{tabular}[c]{@{}l@{}}
				Intelli-\\ Vision\\ 
				(2002)
				\\	\cite{IntelliVision2019}
			\end{tabular}
			
			& \begin{tabular}[c]{@{}l@{}}	
				Security \& Surveillance \\
				(home \& IoT, retail, traffic \\ \& 
				transportation)
				
			\end{tabular}                                      & \begin{tabular}[c]{@{}l@{}}
				Smart home, Smart security,\\ 
				Smart retail, Smart city, \\ and ADAS
			\end{tabular}                                                                                                                            & \begin{tabular}[c]{@{}l@{}}
				Video Search, video summary \\ 
				and, recognition
			\end{tabular}                                                                                      & 
			\begin{tabular}[c]{@{}l@{}}B2B, \\ B2C\end{tabular} & 
			\cmark             
			\\ \hline

			\begin{tabular}[c]{@{}l@{}}
				Google \\ 
				Vision \\ 
				(2017)
				\\\cite{GoogleAI2019}	
			\end{tabular} & 
			
			\begin{tabular}[c]{@{}l@{}}	
				Video contents \& analytics \end{tabular} 
			
			& NA                                                                                                                                                                                                                                        & 
			\begin{tabular}[c]{@{}l@{}}
				Label Detection (detect objects), \\ 
				Content Detection (Shot change \\
				and adult contents), detection (detect \\ 
				scene changes)
			\end{tabular} & \begin{tabular}[c]{@{}l@{}}B2C,\\ B2B\end{tabular}  & \cmark             \\ \hline
			
			\begin{tabular}[c]{@{}l@{}}
				IBM \\ CVAS\\ (2017)
				\\ \cite{BlueChasm2019AI} 	
			\end{tabular}        & 
			\begin{tabular}[c]{@{}l@{}}	
				Video contents \& analytics
			\end{tabular}                                                                                                                                           & NA                                                                                                                                                                                                                                        & 
			\begin{tabular}[c]{@{}l@{}}
				Deep learning based IVA \\
				(scenes, objects,  faces)
			\end{tabular}                                                                                
			&
			\begin{tabular}[c]{@{}l@{}}
				B2B \\
				C2B
			\end{tabular}                                                 
			& \cmark             \\ \hline
			
			\begin{tabular}[c]{@{}l@{}}
				Azure \\ 
				\gls{CVAS}
				\\\cite{AzureVideoAnalytics2017}
				
			\end{tabular}
			
			& 
			\begin{tabular}[c]{@{}l@{}} 
				Video contents analytics \& \\ surveillance \end{tabular}                                                                                                                                        	
			
			& 
			\begin{tabular}[c]{@{}l@{}} 
				Motion detection, Face detection, \\ 
				Emotion detection (sadness, fear, \\ anger)
			\end{tabular}                                                                                                                                                 
			& 
			\begin{tabular}[c]{@{}l@{}}
				HyperLapse, summarization, character \\
				recognition, content moderation, \\
				face redactor
			\end{tabular}
			
			& 
			\begin{tabular}[c]{@{}l@{}}
				B2B \\	C2B
			\end{tabular} 
			
			& \cmark             \\ \hline
		\end{tabular}%
	}
\end{table*}


\paragraph{Google Vision} On March 8, 2017, at the Google Cloud Next conference in San Francisco, Google announced the release of the \gls{IVA} REST API \cite{GoogleAI2019}. The API lets the developer recognize objects in videos automatically and can detect and tag scene changes. Furthermore, it enables the users to search and discover the unstructured video contents by providing information about entities (20,000 labels). Its main features are label detection, explicit content detection, shot change detection, and regionalization \cite{GoogleAI2019Features}. It exploits deep-learning models and is built on the top of the TensorFlow framework. The Google \gls{IVA} APIs is targeting the unstructured video content analytics rather than surveillance and security. The application domains of the API can be large media organizations that want to build their media catalogs or find easy ways to manage crowd-sourced content. It can also be helpful for product recommendations, medical-image analysis, fraud detection, and many more. 

\paragraph{IBM CVAS} In April 2017 at the National Association of Broadcasters Show, IBM announced \gls{CVAS} services \cite{IBM2019AI}. BlueChasm, \cite{BlueChasm2019AI} development team, came up with a prototype app, known as “VideoRecon,” that combines \gls{IVA} via IBM Watson and IBM cloud stack. The IBM \gls{CVAS} service can extract metadata like keywords, concepts, visual imagery, tone, and emotional context from video data. The IBM \gls{CVAS} allows the users to upload video footage to the IBM Cleversafe object storage \cite{IBMCloudStorageObject2017} and subscribe to a service. When an object or event of interest is detected, the VideoRecon service creates a tag along with a timestamp of the point in the video when either the object was recognized or the event occurred. The tags are then stored in the IBM Cloudant fully managed NoSQL JSON document store \cite{IBMCloudant2017} for future use.

\paragraph{Azure CVAS} Microsoft Azure, a cloud computing service launched in 2010, started media services that enable developers to build scalable media management and delivery applications \cite{AzureVideoAnalytics2017}. Media Services is based on REST APIs that enable the users to securely manage video or audio content for both on-demand and live streaming delivery to clients. Recently, they provide \gls{CVAS} APIs to the customers for (R/B)\gls{IVA} (as shown in Table \ref{tab:IndCVAS}).

\paragraph {Citilog CVAS} Citilog \cite{CitiLog2018}, also known as CT-Cloud, provides intelligent video analytics and surveillance solutions in the domain of transportation. Citilog provides services like automatic incident detection, traffic data collection (vehicle counting, classification, average speed, occupancy and levels of service), interaction control, video management, and license plate recognition. The Citilog is an open platform, providing APIs, widgets for quick development of services. According to the Citilog, they process approximately 32000 hours of video data and detects about five incidents per minute from 900 sites worldwide. 

\paragraph{Check-Video CVAS} Check-Video \cite{checkvideo2019}, founded in 1998, is one of the initiators in the domain of \gls{IVA}. The company offers cameras, recorders, and a cloud video management solution for security dealers, integrators, and end-users. Initially, they used to provide \gls{OVAS}, but with the advancement of cloud technology, CheckVideo launched \gls{CVAS}. They provide domain-specific intelligent video analytics \gls{RIVA} solution. The main features Check-Video are \gls{RIVA}, a video search engine, cloud video storage, and an alert system. The provided services can be categorized as basic analytics, object classification, and business analytics (see table~\ref{tab:IndCVAS}). According to the company’s, they have successfully analyzed 108,458,000 hours of video and detected 61,233,000 events per month. 

\paragraph{Intelli-Vision CVAS} Intelli-Vision \cite{IntelliVision2019}, founded in 2002, is a leading and notable company in the field of \gls{AI} and deep learning-based video analytics and video cloud platform.  They are exploiting state-of-the-art technology in the area of \gls{AI} for security and monitoring purposes while targeting multiple business domains, including home, retail, transportation, and advanced driver assistance systems for cars. Intelli-Vision's analytics adds the “Brains Behind the Eyes” for cameras by analyzing the video content, extracting meta-data, sending out real-time alerts, and providing intelligence on the video. Currently, they are providing a wide range of video analytics services in the domains mentioned above, ranging from object left to night vision and enhancements (see table~\ref{tab:IndCVAS}). In Feb 2018, in a press release, the Intelli-Vision stated that they have successfully deployed four million cameras worldwide, which have been subscribed to various \gls{IVA} services.

\section{IVA Applications}
\label{sec:IVAApplications}
\gls{IVA} at scale drives many application domains ranging from security and surveillance to self-driving and healthcare. Many application areas of video big data analytics are shown, which project the significant role of big data and cloud computing in \gls{IVA}.

\paragraph{Traffic and Transportation} \gls{IVA} has been extensively used in traffic control and transportation, e.g., lane traffic counts, incident detection, illegal u-turn, and many more. One of the main reason for deaths and injuries are traffic-related misfortunes \cite{ananthanarayanan2017real}.  Proactive analytics is required to predict abnormal events so that to minimize or avoid such accidents. In this direction, VisonZero \cite{loewenherz2017video} has been developed and deployed successfully. In transportation, another application is vehicle tracking where chasing of a license plate, overspeeding, and collision cause analysis can be obtained by analyzing video data. Kestrel \cite{qiu2018kestrel} is a vehicle tracking system and uses information from various non-overlapping cameras to detect vehicle path. Gao et al. \cite{gao2009traffic}used an automatic particle filtering algorithm to track the vehicle and monitor its illegal lane changes. Chen et al. cite{chen2009machine} used hidden Markov models to determine the traffic density state probabilistically. Incident detection framework based on generative adversarial networks were proposed in \cite{lin2020automated}.

\paragraph{Intelligent Vehicle and Self-driving cars} Currently, the term self-driving cars mean that the vehicle exploits computer vision for safe and intelligent driving while assisting the driver. In an intelligent vehicle, different sensors and high definition cameras (cameras for vehicle cabin, forward roadway, and the instrument cluster) are integrated with the vehicle, which generates multi-model data, and the same is sent to the cloud for real-time analytics \cite{fridman2017autonomous, xu2017internet}. In this context, video analytics is vital with optimum algorithm accuracy. Researchers have developed several algorithms, including pedestrian detection, traffic light detection, and other driver assistance system. For example, Wang et al. \cite{wang2009improving} proposed a method for pedestrian detection in urban traffic conditions using a multilayer laser sensor mounted onboard a vehicle. An algorithm was proposed by Tsai et al. \cite{tsai2010detection} to detect three condition changes: missing, tilted, and blocked signs, using GPS data, and video log images. An innovative \gls{CNN}-based visual processing model is proposed in \cite{ayachi2020traffic} to automatically detect traffic signs and dramatically reduces the sign inventory workload. Driver decision making was improved in taking the right turn in left-hand traffic at a signalized intersection utilizing simulation \cite{callan2009neural}. They used an in-car video assistant system to present the driver's occluded view when the driver's view is occluded by truck. An effort for driver body tracking and activity analysis, posture recognition, and action predication, have been studied in \cite{yan2016driving,gite2019early} respectively.

\paragraph{HealthCare} Recently, video big data analytics is reshaping the healthcare industry and yet another vital application area that demands special focus while deploying video analytics. Surveillance video streams can help understand the tracked person's behavior, such as monitoring elderly citizens or blind people against fall detection or detecting any possible threat. Fleck and Strasser reported a prototype 24/7 system installed in a home for assisted living for several months and shows quite promising performance \cite{fleck2008smart}. Zhou et al. studied how video analytics can be used in eldercare to assist the independent living of elders and improve the efficiency of eldercare practice \cite{zhou2008activity}. Some more studies were done \cite{aertssen2011fall, brulin2012posture, sacco2012detection, yu2012posture} to analyze activities, recognize posture, and to detect falls or other substantial events. A smart gym can exploit the video cam stream to determine frequently used equipment, the duration of exercise, and time spent on a piece of particular equipment, which is useful for real-time assessment. 

\paragraph{Smart City Security (IoT) and Surveillance} In many organizations ranging from large enterprises to schools, home and law enforcement agencies where security is becoming an essential concern and is turning the security centers to video analytics to keep their premises safe. Law enforcement agents can use a body-worn camera to identify criminals in real-time while transmitting the video stream to the video analytic cloud. \gls{IVAaaS} can provide customized services that can adjust quickly to changing needs and demands. People detection, and tracking \cite{choi2012general}, motion detection, intrusion detection \cite{pragada2015intrusion}, line crossing \cite{zhao2016crossing}, object lift, loitering\cite{xu2015loitering}, and license plate recognition \cite{shivakumara2018cnn}  are the example of video analytics services for security. For security in subway stations, Krausz \cite{krausz2010metrosurv} developed as a surveillance system to detect dangerous events. Shih et al. \cite{shih2006illegal} tried to extract the color features of an employee's uniform to recognize the entry legality in a restricted area. In the context of security, abandoned object detection is indispensable and can lead to a terrorist attack. In the future, video analytics applications are developing fast, and they are changing the way the security industry works. 

\paragraph{Augmented reality and Personal Digital Assistance} Among the many, one aspect of the augmented reality is visual, where devices like special glasses, helmets, or goggles are utilized for the projection of additional information or interactive experience of the surrounding real-world environment. The visual aspect of augmented reality may encompass complex \gls{IVA} and demand powerful hardware. Likewise, vision-based digital assistants is a rising technology (e.g., personal robot Jibo) that could deeply alter our regular activities while offering personalized and interactive experiences. Such devices could be offloaded to the CVAS for low latency complex \gls{IVA} uninterruptedly.

\paragraph{Retail, Management, and Business Intelligence Analysis} Large-scale products, services, and staff management while adjusting to consumer demands can be challenging without timely and up-to-date information. Smart Retail solution takes advantage of smart cameras combined with \gls{IVA} to gather data on store operations and customer trends. Dwell analysis, face recognition, queue management, customer count, customer matrics, consumer traffic map, are some of the example services in this context \cite{singh2018applications}. Gaze analysis provides a means to learn customers' interest in merchandise by following their attention \cite{xu2010people, haritaoglu2002attentive} on a store display. Actions like reaching or grabbing products were analyzed by \cite{hu2009action, sicre2010human} to understand customers' interest. Emotion analysis can identify customers' views regarding product and interaction with the company's representative \cite{ananthanarayanan2017real}. \gls{IVA} can also be used for business intelligence analysis to answer queries like "number of people visited per unite time?" or "customer interest in items?" while utilizing the same security infrastructure. Such information is beneficial for retailers in improving customer experience and marketing strategies.

\section{Research Issues, Opportunities, and Future Directions}
\label{sec:ResearchIssues}
Intelligent video big data analytics in the cloud opens new research avenues, challenges, and opportunities. This section provides in-depth detail about such research challenges, which has been summarized in Table~\ref{tab:ResearchIssues}).

\begin{table*}[!htb]
	\centering
	\caption{Open research issues in video big data analytics in the cloud.}
	\label{tab:ResearchIssues}
	\resizebox{\textwidth}{!}{%
		\begin{tabular}{|l|l|l|l|}
			\hline
			\textbf{Component} &
			\textbf{Aspect} &
			\textbf{Layer} &
			\textbf{Open Research Issues} \\ \hline
			Video Big Data &
			Volume &
			VBDCL &
			Orchestration and Optimization of \gls{IVA} Pipeline. \\ \hline
			&
			
			&
			VBDCL, VBDPL, VBDML &
			Big dimensionality reduction, and indexing. \\ \hline
			
			&
			
			&
			
			&
			Cleaning and compressing video big data.
			\\ \hline
			&
			Velocity &
			VBDCL, VBDPL, VBDML & 
			Real-time video streams and online learning. \\ \hline
			&
			Variety &
			&
			\begin{tabular}[c]{@{}l@{}}Big dimensionality reduction.\\ 
				Data modality for single \gls{IVA} goal.\end{tabular} \\ \hline
			&
			Veracity &
			&
			\begin{tabular}[c]{@{}l@{}}Vide big data veracity assessment. 
				\\ Learning with unreliable data.\end{tabular} \\ \hline
			&
			Value &
			&
			\begin{tabular}[c]{@{}l@{}} Understandable \gls{IVA} for decision support. \end{tabular} \\ \hline
			&
			
			&
			&
			Semantic concepts extraction in distributed environment. \\ \hline
			User &
			\begin{tabular}[c]{@{}l@{}}
				Developer / \\ Researcher 
			\end{tabular}	
			&
			WSL &
			\begin{tabular}[c]{@{}l@{}}Declarative \gls{IVA}.\\ 
				\gls{IVA} and distributed computing technologies abstraction.\\ 
				Comprehensive evaluation measures for \gls{IVA} in the cloud environment. \\ 
				Visualizing video big data.\end{tabular} \\ \hline
			&
			&
			VBDCL, WSL &
			\gls{IVA} algorithm, model, and services statistics maintenance and ranking. \\ \hline
			&
			&
			&
			Effective price scheme for \gls{IVA} algorithm deployment, and subscription. \\ \hline
			&
			&
			WSL, VBDCL &
			Model management and algorithm selection. \\ \hline
			&
			Consumer &
			WSL &
			\begin{tabular}[c]{@{}l@{}}\gls{IVA} services utilization.\\ 
				Improving consumer experience based on feedback.\end{tabular} \\ \hline
			&
			&
			\begin{tabular}[c]{@{}l@{}}	
				VBDCL, VBDPL, \\ VBDML, VKCL 
			\end{tabular}
			
			&
			Effective price scheme for multiple \gls{IVA} service subscription. \\ \hline
			&
			Security and Privacy &
			- &
			Privacy preserving distributed IVA, security, and trust. \\ \hline
			Cloud System &
			Analytics engine &
			VBDPL, VBDML &
			\begin{tabular}[c]{@{}l@{}}\gls{IVA} on video big data (general big data middleware for IVA).\\ 
				Parameter Server optimization.\end{tabular} \\ \hline
			&
			Infrastructure &
			&
			\gls{IVA} on video big data (general big data middleware for \gls{IVA}). \\ \hline
		\end{tabular}%
	}
\end{table*}

\subsection{IVA on Video Big data}
Big data analytics engines are the general-purpose engine and are not mainly designed for big video analytics. Consequently, video big data analytics is challenging over such engines and demand optimization. Almost all the engines are inherently lacking the support of elementary video data structures and processing operations. Further, such engines are also not optimized, especially for iterative \gls{IVA} and dependency among processes.

Optimizing cluster resource allocations among multiple workloads of iterative algorithms often involves an approximation of their runtime, i.e., predicting the number of iterations and the processing time of each iteration \cite{popescu2013predict}. By default, Hadoop lacks iterative job support but can be handled through speculative execution. However, Spark supports not only MapReduce and fault tolerance but also cache data in memory between iterations. \gls{IVA} on video big data creates an immense space for the research community to further crack in this direction. The research community is already trying to develop basic video processing and \gls{IVA} support over big data, but it is still the beginning. How to optimize such engines for iterative IVA? It also allows us to research whether the exiting distributed computing engines fulfill the demands of the \gls{IVA} on video big data or need a specialized one.

Furthermore, the focus of the existing research on \gls{IVA} are velocity, volume, velocity, but the veracity and value have been overlooked. One promising direction in addressing video big data veracity is to research methods and techniques capable of accessing the credibility of video data sources so that untrustworthy video data can be filtered. Another way is to come up with novel \gls{ML} models that can make inferences with defective video data. Likewise, users’ assistance is required to comprehend \gls{IVA} results and the reason behind the decision to realize the value of video big data in decision support. Thus, understandable \gls{IVA} can be a significant future research area.

\subsection{IVA and Human-machine coordination}
\gls{IVA} on video big data grants a remarkable opportunity for learning with human-machine coordination for numerous reasons:
\begin{itemize}
	\item \gls{IVA} on video big data in cloud demands researchers and practitioners mastering both \gls{IVA} and distributed computing technologies. Bridging both the worlds for most analysts is challenging. Especially in an educational environment, where the researcher focuses more on the understanding, configuration, and tons of parameters rather than innovation and research contribution.  Thus there is a growing need to design such \gls{CVAS} that provide high-level abstractions to hide the underlying complexity.  
	\item IVS service to become commercially worthwhile and to achieve pervasive recognition, consumer lacking technical \gls{IVA} knowledge. The consumers should be able to configure, subscribe, and maintain \gls{IVA} services with comfort. 
	\item In traditional \gls{IVA}, consumers are usually passive. Further, research is required to build more interactive \gls{IVA} services that assist consumers in gaining insight into video big data. An efficient interactive \gls{IVA} service depends on the design of innovative interfacing practices based on an understanding of consumer abilities, behaviors, and requirements \cite{amershi2014power}. The interactive \gls{IVA} services will learn from the consumer and decrease the need for administration by a specialist. It will also enable consumers to design custom \gls{IVA} services to meet the domain-specific requirement. 
\end{itemize}

\subsection{Orchestration and Optimization of IVA Pipeline}
\label{OrchestrationAndOptimizationofIVAPipeline}
\gls{SIAT} is a service-oriented architecture, i.e., (R/B)\gls{IVAaaS}. The real-time and batch workflow are deeply dependent on the messaging middleware (Table~\ref{tab:MessageBrokers}) and distributed processing engines (Table~\ref{tab:ComparisionOfDistributedComputingEngine}). In \gls{SIAT}, the dynamic (R/B)\gls{IVA} service creation and multi-subscription environment demand the optimization and orchestration of the \gls{IVA} service pipeline while guarantees opportunities for further research. 

In literature, two types of scheduling techniques have been presented for real-time scheduling, i.e., static and dynamic \cite{kwok1999static,zhu2014analysis}. Static approaches are advantageous if the number of services and subscription sources is known priorly, but this is not the case with \gls{SIAT}. The suitability of dynamic methods is reasonable but is expensive in terms of resource utilization. Likewise, the main issue in the \gls{BIVA} service workflow on video big data is the data partitioning, scheduling, executing, and then integrate numerous predictions. The \gls{BIVA} service workflow can be affected by the data flow feature of the underlying big data engine (as shown in Table~\ref{tab:ComparisionOfDistributedComputingEngine}). As Hadoop map-reduce lacking the loops or chain of stages, and Spark support \gls{DAG} style of chaining, whereas Flink supports a \gls{CCDG}.

In the map-reduce infrastructure, a slowdown predictor can be utilized to improve the agility and timeliness of scheduling decisions \cite{bortnikov2012predicting}. Spark and Flink can accumulate a sequence of algorithms into a single pipeline but need research to examine its behavior in dynamic service creation and subscription environment. Further, concepts from the field of query and queuing optimization can be utilized while considering messaging middleware and distributed processing engines to orchestrate and optimization of \gls{IVA} service Pipeline.

\subsection{IVA and Big Dimensionality}
\label{IVAandBig Dimensionality}
The \gls{VSDS} multi-modality can produce diverse types of data streams. Similarly, the \gls{IVA} algorithm, developer, and \gls{IR} has a triangular relationship. An array of algorithms can be deployed, generating varied sorts of multi-dimensional features from the acquired data streams. The high-dimensionality factor poses many intrinsic challenges for data stream acquisition, transmission, learner, pattern recognition problems, indexing, and retrieval. In literature, it has been referred to as a “Big Dimensionality” challenge \cite{zhai2014emerging}.  

\gls{VSDS} variety leads to key challenges in acquiring and effectively processing the heterogeneous data. Most existing \gls{IVA} approaches can consider a specific input, but in many cases, for a single \gls{IVA} goal, different kinds, and formats can be considered. 

With growing features dimensionality, current algorithms quickly become computationally inflexible and, therefore, inapplicable in many real-time applications \cite{gao2017learning}. Dimension reduction approaches are still going to be a hot research topic because of data diversity, increasing volume, and complexity. Effect-learning algorithms for first-order optimization, online learning, and paralleling computing will be more preferred.

Similarly, designing a generic, efficient, and scalable multi-level distributed data model for indexing and retrieving multi-dimensional features is becoming tougher than ever because of the exponential growth and speed of video data. Considering varied situations, requirements, and parameters such as complex data type indexing (such as objects, which contains multiple types of data), multi-dimensional features (require different feature matching scheme for each type), cross scheme matching (e.g., spatiotemporal, spatial-object, object-temporal, etc.), giant search space, incremental updates, on the fly indexing, and concurrent query processing demands further investigation. It gives the research community the opportunities to optimize existing hashing schemes and indexing structures such as R-trees \cite{Guttman:1984:RDI:602259.602266}, M-tree \cite{Ciaccia:1997:MEA:645923.671005}, X-tree \cite{lin1994tv}, locality-sensitive hashing \cite{datar2004locality} etc. on the top of big data engines. Log-structured merge-tree \cite{o1996log} based distributed data stores (see Table~\ref{tab:DDS}) can be leveraged to improve multi-dimensional query performances. Additionally, \gls{ML} classification models can be used to capture the semantics by inspecting the association between features and the context among them to better index multi-dimensional data. Such methods make them more precise and effective than the non-\gls{ML} methods \cite{kraska2018case}. Thus, \gls{ML} classification models (including neural networks) can be used to capture the semantics by inspecting the association between features and the context among them to better index multi-dimensional data, that make them more precise and effective than the traditional indexing approaches \cite{kraska2018case}.

\subsection{Online learning on video big data}
The value of \gls{RIVA} is dependent on the velocity of the video streams, i.e., newness and relatedness to ongoing happenings. Though existing big data de-facto standards are lacking to deal with the changing streams \cite{de2013samoa}. The \gls{RIVA} services must address continuous and changing video streams. In this context, online learning can be utilized, representing a group of learning algorithms for constructing a predictive model incrementally from a sequence of data, e.g., Fourier Online Gradient Descent and Nystrom Online Gradient Descent algorithms \cite{lu2016large}. In this context, Nallaperuma et al. \cite{nallaperuma2019online} proposed \gls{ITS} platform utilizing unsupervised online learning and deep learning approaches. It gives further research opportunities by involving data fusion from heterogeneous data sources \cite{nallaperuma2019online}.

\subsection{Model management }
\gls{SIAT} architecture is designed to deploy an array of \gls{IVA} algorithms, i.e., both by administrator and developers. An algorithm might hold a list of parameters. The model selection process encompasses feature engineering (feature selection~\ref{sec:VBDPL}), \gls{IVA} algorithm selection, and hyperparameter tuning. Feature engineering is a laborious activity and is influenced by many key factors, e.g., domain-specific regulations, time, accuracy, video data, and \gls{IVA} properties, which resultantly slow and hinder exploration. \gls{IVA} algorithm selection is the process of choosing a model that fixes the hypothesis space of prediction function explored for a given application \cite{friedman2001elements}. This process of \gls{IVA} algorithm selection is reliant on technical and non-technical aspects, which enforce the \gls{IVA} developer to try manifold techniques at the cost of time and cloud resources. Hyperparameter is vital as they govern the trade-offs between accuracy and performance. \gls{IVA} analysts usually do ad-hoc manual tuning by iteratively choosing a set of values or using heuristics such as grid search \cite{friedman2001elements}. From \gls{IVA} analysts’ perspective, model selection is an expensive job in terms of time and resources that bringing down the video analytics lifecycle \cite{kandel2012enterprise}. Model selection is an iterative and investigative process that generally creates an endless space, and it is challenging for \gls{IVA} analysts to know a priori which combination will produce acceptable accuracy/insights. In this direction, theoretical design trade-offs are presented by Arun et al. \cite{kumar2016model}, but further research is required that how to shape a unified framework that acts as a foundation for a novel class of \gls{IVA} analytics while building the procedure of model selection easier and quicker. 

\subsection{Parameter Servers and Distributed Learning}
Developing a model (such as Stochastic Gradient Descent) for video big data analytics in a distributed environment carries an intrinsic issue of sharing and updating high-dimension parameters that can easily run into orders billions to trillions. The Parameter Server notion has been introduced to address this issue, aiming to store the parameters of an \gls{ML} model such as the weights of a neural network and serve them to clients. Parameter Server proposes a new framework for building distributed \gls{ML} algorithms, and encompass diverse design goals, e.g., efficient communication, flexible consistency, elasticity when adding resources, resource utilization, and ease of use. In literature, recently, various studies tried to optimize Parameter Server. PS2 \cite{zhang2019ps2} builds the parameter server on top of Spark. SketchML \cite{jiang2018sketchml} compresses the gradient values by a sketch-based method. FlexPS \cite{huang2018flexps} introduces a multi-stage abstraction to support flexible parallel control. The Parameter Server can be optimized further against the stated design goals and need further investigation.

\subsection{Evaluation issues and opportunities}
\label{EvaluationIssuesAndOpportunities}
Haralick \cite{haralick1992performance} initiated the discussion of \gls{IVA} performance evaluation followed by dedicated workshops \cite{forstner1997dagm}, and journals \cite{phillips1999empirical}, \cite{christensen1997special}. As a result, performance evaluation tools (ViPER5), and datasets (ETISEO \cite{nghiem2007etiseo}, TrecVID \cite{hauptmann2004successful}, i-LIDS\cite{branch2006imagery}) were introduces. It is a fact that traditional \gls{IVA} has an established set of prediction accuracy based metrics for performance evaluation that is ranging from accuracy, error rate, and precision to optimization and estimation error. Some more evaluation parameters are adopted from big data analytics when \gls{IVA} is tried on distributed computing, e.g., scalability, fault tolerance, memory usage, throughput, etc.  \cite{singh2015survey} (as shown in Table~\ref{tab:StateOftheArtScholarlySystems}. The amalgamation of two types of matrices might not be enough. 

\gls{IVA} services provided by a system like \gls{SIAT} has to accomplish predictably through an intractable number of scenarios and environmental circumstances, meeting requirements that vary according to the situation, domain, and user. For \gls{SIAT}’s consumer, the \gls{IVA} services work as a black box where the significant metrics relate to overall system performance, such as false alarms, accuracy, and detection rate. However, from the developer and researcher perspective, \gls{SIAT} consists of numerous computer vision algorithms, with complex interfaces among them. A proper performance evaluation matrix is required for \gls{IVA} service developers to comprehend these relations and to revolutionize and address novel \gls{IVA} services. Another critical issue is how to guarantee accurate and predictable \gls{IVA} service performance when porting technology between distributed algorithm development environments and deployment code environments with hardware-specific optimizations. These shortcomings result in algorithmic alterations that can influence the performance of \gls{IVA} services. 

\gls{IVA} performance evaluation is goal-oriented, and the factors should be determined carefully. Many key factors that influence the performance of video big data analytics utilizing distributed computing engines in the cloud are listed below (not limited to).

\begin{itemize}
	\item \gls{VSDS}: holds diverse types of parameters, i.e., video type (color, grey-scale, infrared, omnidirectional, depth map, etc.), property (frame-rate, field-depth), and quality (resolution, pixel depth) of the generated video as generated by the camera.
	\item \gls{VSDS}and messaging middleware parameters: encompass \gls{VSDS} connection, frames reducing and transformation, messaging queue (broker server), compression artifacts, mini-batch size, and possibly the involvement of internal and external network.
	\item \gls{VSDS}environmental parameters: some features of the configuration remain constant in a given use case, but differ between configuration, possibly influencing performance. These parameters comprise camera location (mounting height, angle, indoor or outdoor), mounting type (still or in motion), camera view (roads, water, foliage), and weather (sun, cloud, rain, snow, fog, wind).
	\item Distributed processing environment: the video processing hardware (FPGA, CPU, GPU, etc.), and network communication channels potentially impose additional limitations in terms of locality, speed, and memory.
	\item Big data analytics engines: The nature and characteristics of big data engines affect the performance of the \gls{IVA}, such as data flow, windowing, computation model, etc. Further, big video analytics in the cloud demand complex trade-offs between different evaluation criteria. In order to comprehend, assume the intricate trade-offs between accuracy and response time. Iterative tasks have an inverse relation with fault tolerance concerning scalability (e.g., MapReduce is high fault-tolerant but lacking iteration). Similarly, non-iterative \gls{IVA} algorithms scale better than iterative at the cost of performance degradation.
	\item \gls{IVA} service parameters: Application parameters are domain-specific parameters, e.g., vehicles, carts, humans, etc., tolerable miss detection and false alarm rates and their desired trade-off, \gls{IVA} type, and max acceptable latency.
	\item Computation and communication trade-off: \gls{IVA} Algorithms and services in the distributed environment should be developed and designed wisely intending to minimize computation time, which is associated with data locality and loading.
\end{itemize}

Diverse types of factors affect video big data analytics performance in the cloud, constructing a comprehensive evaluation of all use cases are almost near impossible. It further provides opportunities for the researchers to design a framework that provides a unified and generic framework that can be adapted by any \gls{CVAS}. Investigating these issues would significantly contribute to the academic and industrial communities interested in building \gls{IVA} algorithms, services, and \gls{CVAS}.

\subsection{IVA algorithm, model, and services statistics maintenance, ranking, and recommendation}
\gls{SIAT} architecture is designed under the \gls{C2C} business model. In \gls{SIAT}, a user can develop and deploy an \gls{IVA} algorithm, model, or service (here collectively we call it \gls{IVA} service) that can be either extended, utilized, or subscribed by other users.  The community members run such architecture, and rapidly, the number of \gls{IVA} services can be reached to tons of domain-dependent or independent \gls{IVA} services. This scenario develops a complex situation for the users, i.e., which \gls{IVA} service (when sharing the parallel functionalities) in a specific situation, especially during service discover. Against each \gls{IVA} service, there is a list of \gls{QoS} parameters. Some of these \gls{QoS} parameters (not limited to) are user trust, satisfaction, domain relevance, security, usability, availability, reliability, documentation, latency, response time, resource utilization, accuracy, and precision.

Such types of \gls{IVA} services against the \gls{QoS} parameters lead to the 0-1 knapsack issue. In this direction, one possible solution is utilizing multi-criteria decision-making approaches. It gives further opportunities to the research community to investigate how to rank and recommend \gls{IVA} algorithms, models, and services. Similarly, it can lead to a high-dimensional sparse matric \cite{wu2019deep}. In this direction, research is required on how to utilize such parameters for \gls{IVA} services recommendation.

\subsection{IVAaaS and Cost Model}
Recently, cloud-based analytics platforms are the key means for enterprises to provide services on the pay-as-you-go cost model.  Existing cost metrics usually are determined to utilize hardware usage comprising processing (CPU, GPU), disk space, and memory usage. These prices are often static or dynamic \cite{mazrekaj2016pricing}. The example of the former one is Amazon’s EC2, which offers tiered levels of service. In later cases, the cost model is used to determine the price of the service using analytics. This takes into account factors such as peak hours and opponent cost model etc. The hardware cost is usually minimal compared to the cost of software such as \gls{SIAT} where the cost of \gls{IVA} analytics valued more. 

\gls{SIAT} is supposed to provide \gls{IVAAaaS} and \gls{IVAaaS} in the cloud while adopting the \gls{C2C} business model. Unfortunately, current \gls{SaaS} cost models might not be applicable because of the involvement of diverse types of parameters that drastically affect the cost model. Such parameters are, business model (\gls{B2B}, \gls{B2C}, and \gls{C2C}), unite of video, user type (developer, researcher, and consumers), services (\gls{IVAAaaS} and \gls{IVAaaS}), service subscription (algorithm, \gls{IVA} service, single, multiple, dependent or independent), cloud resource utilization, user satisfaction, \gls{QoS}, location, service subscription duration, and cost model fairness. The addition of further parameters is subject to discussion, but the listed are the basic that govern \gls{SIAT} cost matrix. Additionally, the cost model demands further research and investigations to develop an effective price scheme for \gls{IVA} services while considering the stated parameters.

\subsection{Video big data management}
Despite video big data pose high value, but its management, indexing, retrieval, and mining are challenging because of its volume, velocity, and unstructuredness. \gls{IVA} has been investigated over the years (section~\ref{sec:IVAConstituentsPredominantTrends}) but still evolving and need to address diverse types of issues such as:  

\begin{itemize}
	\item In the context of video big data management, the main issue is the extraction of semantic concepts from primitive features. A general domain-independent framework is required that can extract semantic features, analyze and model the multiple semantics from the videos by using the primitive features. Further, semantic event detection is still an open research issue because of the semantic gap and the difficulty of modeling temporal and multi-modality features of video streams. The temporal information is significant in the video big data mining mainly, in pattern recognition.
	\item Motion analysis is vital, and further research is required for moving objects analysis, i.e., object tracking, handling occlusion, and moving objects with statics cameras \cite{shantaiya2013survey}, moving cameras \cite{leal2017tracking}, and multiple camera fusion.
	\item    Limited research is available on CBVR (see section~\ref{sec:CBVR}) while exploiting distributed computing. Further study is required to consider different features ranging from local to global spatiotemporal features utilizing and optimizing deep learning and distributed computing engines. 
	\item For video retrieval, semantic-based approaches have been utilized because of the semantic gap between the low-level features and high-level human-understandable concepts. Ontology adds extra concepts that can improve the retrieval results \cite{neo2006video} but can also lead to unexpected deterioration of search results. In this context, a hybrid approach can be fruitful and need to design different query planes that can fulfill diverse queries in complex situations.
	\item Insufficient research is available on graph-based video big data retrieval and analysis, opening doors for further investigation. The researchers can conduct studies to answer questions like: the formation of the video graph, tuning the similar value and its effect on the graph formation, studying properties, and meta-analytic of the formed video graph.  
	\item How the reinforcement learning and real-time feedback query expansion technique \cite{khalid2019real} can be exploited to improve the retrieval results?
	\item Recently, video query engines have been introduced to retrieve and analyze video at scale \cite{kang2019challenges,lu2016visflow,kangblazeit}. A special focus of the database community is required to design, implement, optimize, and operationalize such video query engines. 
\end{itemize}

\subsection{Privacy, Security and Trust}

Video big data, acquisition, storage, and subscriptions to shared \gls{IVA} in the cloud become mandatory, which leads to privacy concerns. For the success of such platforms, privacy, security, and trust are always central. In literature, the word ‘trust’ is commonly used as a general term for ‘security’ and ‘privacy’ \cite{ghosh2010guest}. Trust is a social phenomenon where the user has expectations from the \gls{IVA} service provider and willing to take action (subscription) on the belief based on evidence that the expected behavior occurs \cite{khusro2017social,alam2017confluence}. In the cloud environment, security and privacy are playing an active role in the trust-building. To ensure security, the \gls{CVAS} should offer different levels of privacy control. The phenomena of privacy and security are valid across \gls{VSDS}, storage security, multi-level access controls, and privacy-aware \gls{IVA} and analysis. We list some research directions that can provide opportunities for cloud security specialists. 

\begin{itemize}
	\item The video big data volume, variety, and velocity boost security threats. Recently, disputes and news were circulating regarding the misuse of user-generated content and hacking the cameras in an unauthorized manner. The \gls{CVAS} vendor, in collaboration with or under the law enforcement agencies, must come up with new rules, laws, and agreements, which can differ from country to country. Utilizing such policies, the \gls{CVAS} vendor should ensure that all the \gls{IVA} service, subscription, and storage level agreements are adequately followed. The policies can be researched whether it offers adequate protection for individuals’ data while performing video big data analytics and public monitoring.
	\item Unlike other data, videos are more valuable for the owner and can be a direct threat, e.g., live broadcast, blackmailing, etc. Likewise, the focus of traditional privacy approaches is data management that becomes absolute when it comes to data security.  Novel algorithms are required to secure user’s data both for shared \gls{IVA}, storage to make the video stream acquisition more secure.
	\item In the context of security, the blockchain (popularized by Bitcoin) \cite{crosby2016blockchain} has been studied and operationalized across academia and industry evenly. A blockchain is a modification resilient cryptography technique known as a distributed ledger, where records are linked and managed by a decentralized peer-to-peer network \cite{crosby2016blockchain}. The blockchains techniques are still in early-stage and can be researched further to form a novel automated security system for \gls{CVAS}. 
	\item \gls{ML} techniques have been matured over the years and have been successfully utilized for security, i.e., modeling attack patterns with their distinctive features. However, change in features in case of sophisticated attacks may lead to security failure. \gls{ML} could enhance the performance of security solutions to alleviate the dangers of the existing cyberattacks. The research community can further investigate how \gls{ML} techniques, especially deep learning, can be deployed to analyze logs produced by network traffic, \gls{IVA} processes, and users to recognize doubtful activities.
\end{itemize}

\section{Conclusion}
\label{sec:conclusion}

In the recent past, the number of public surveillance cameras has increased significantly, and an enormous amount of visual data is produced at an alarming rate. Such large-scale video data pose the characteristics of big data. Video big data offer opportunities to the video surveillance industry and permits them to gain insights in almost real-time. The deployment of big data technologies such as Hadoop, Spark, etc., in the cloud under \gls{aaS} paradigm to acquire, persist, process and analyze a large amount of data has been in service from last few years.  This approach has changed the context of information technology and has turned the on-demand service model's assurances into reality.

This paper provides an extensive study on intelligent video big data in the cloud. First, we define basic terminologies and establish the relation between video big data analytics and cloud computing. A comprehensive layered architecture has been proposed for intelligent video big data analytics in the cloud under the \gls{aaS} model called \gls{SIAT}. \texttt{VBDCL} is the base layer that allows the other layers to develop \gls{IVA} algorithms and services. This layer is based on the concept of \gls{IR} orchestration and takes care of data curations throughout the life cycle of an \gls{IVA} service. The \gls{VBDPL} is in-charge of pre-processing and extracting the significant features from the raw videos. The \gls{VBDML} is accountable for producing the high-level semantic result from the features generated by the \gls{VBDML}. The \gls{KCL} deploys video ontology and creates knowledge based on the extracted higher-level features obtained from \gls{VBDML}. When all these layers are pipelined in a specific context, it becomes an \gls{IVA} service to which the users can subscribe to video data sources under the \gls{IVAAaaS} paradigm.  

Furthermore, to show the significance and recent research trends of \gls{IVA} in the cloud, a broad literature review has been conducted. The research issues, opportunity, and challenges being raised by the uniqueness of the proposed \gls{SIAT}, and the triangular relation among video big data analytics, distributed computing technologies, and cloud has been reported.


\section*{Acknowledgment}
This work was supported by Institute for Information and communications Technology Promotion (IITP) grant funded by the Korea government (MSIT) (No. 2016-0-00406, SIAT CCTV Cloud Platform).


\bibliographystyle{IEEEtran}
\bibliography{references}

\begin{thebibliography}{100}
\providecommand{\url}[1]{#1}
\csname url@samestyle\endcsname
\providecommand{\newblock}{\relax}
\providecommand{\bibinfo}[2]{#2}
\providecommand{\BIBentrySTDinterwordspacing}{\spaceskip=0pt\relax}
\providecommand{\BIBentryALTinterwordstretchfactor}{4}
\providecommand{\BIBentryALTinterwordspacing}{\spaceskip=\fontdimen2\font plus
\BIBentryALTinterwordstretchfactor\fontdimen3\font minus
  \fontdimen4\font\relax}
\providecommand{\BIBforeignlanguage}[2]{{%
\expandafter\ifx\csname l@#1\endcsname\relax
\typeout{** WARNING: IEEEtran.bst: No hyphenation pattern has been}%
\typeout{** loaded for the language `#1'. Using the pattern for}%
\typeout{** the default language instead.}%
\else
\language=\csname l@#1\endcsname
\fi
#2}}
\providecommand{\BIBdecl}{\relax}
\BIBdecl

\bibitem{ApacheRocketMQ2012}
\BIBentryALTinterwordspacing
M.~INC., ``Video analytics market,'' 2012, accessed: 2019-12-20. [Online].
  Available: \url{https://github.com/apache/rocketmq}
\BIBentrySTDinterwordspacing

\bibitem{Facebook2020}
\BIBentryALTinterwordspacing
Facebook.Com, ``Facebook,'' 2020, accessed: 2019-12-20. [Online]. Available:
  \url{https://www.facebook.com}
\BIBentrySTDinterwordspacing

\bibitem{YouTube2019}
\BIBentryALTinterwordspacing
youtube.com, ``Youtube statistics,'' 2019, accessed: 2019-12-20. [Online].
  Available: \url{https://www.youtube.com/about/press/}
\BIBentrySTDinterwordspacing

\bibitem{Netflix2020}
\BIBentryALTinterwordspacing
Netflix.Com, ``Netflix,'' 2020, accessed: 2019-12-20. [Online]. Available:
  \url{https://www.netflix.com/}
\BIBentrySTDinterwordspacing

\bibitem{pouyanfar2018multimedia}
S.~Pouyanfar, Y.~Yang, S.-C. Chen, M.-L. Shyu, and S.~Iyengar, ``Multimedia big
  data analytics: A survey,'' \emph{ACM Computing Surveys (CSUR)}, vol.~51,
  no.~1, p.~10, 2018.

\bibitem{olatunji2018dynamic}
I.~E. Olatunji and C.-H. Cheng, ``Dynamic threshold for resource tracking in
  observed scenes,'' in \emph{2018 9th International Conference on Information,
  Intelligence, Systems and Applications (IISA)}.\hskip 1em plus 0.5em minus
  0.4em\relax IEEE, 2018, pp. 1--6.

\bibitem{idg2019}
\BIBentryALTinterwordspacing
I.~D. Corporation, ``The growth in connected iot devices,'' 2019, accessed:
  2019-12-07. [Online]. Available:
  \url{https://www.idc.com/getdoc.jsp?containerId=prUS45213219}
\BIBentrySTDinterwordspacing

\bibitem{huang2014surveillance}
T.~Huang, ``Surveillance video: The biggest big data,'' \emph{Computing Now},
  vol.~7, no.~2, pp. 82--91, 2014.

\bibitem{mell2011nist}
P.~Mell, T.~Grance \emph{et~al.}, ``The nist definition of cloud computing,''
  2011.

\bibitem{zikopoulos2012harness}
P.~Zikopoulos, D.~Deroos, K.~Parasuraman, T.~Deutsch, J.~Giles, and
  D.~Corrigan, \emph{Harness the power of big data The IBM big data
  platform}.\hskip 1em plus 0.5em minus 0.4em\relax McGraw Hill Professional,
  2012.

\bibitem{amazon2015amazon}
E.~Amazon, ``Amazon web services,'' \emph{Available in: http://aws. amazon.
  com/es/ec2/(November 2012)}, 2015.

\bibitem{copeland2015overview}
M.~Copeland, J.~Soh, A.~Puca, M.~Manning, and D.~Gollob, ``Overview of
  microsoft azure services,'' in \emph{Microsoft Azure}.\hskip 1em plus 0.5em
  minus 0.4em\relax Springer, 2015, pp. 27--69.

\bibitem{dijcks2012oracle}
J.-P. Dijcks, ``Oracle: Big data for the enterprise,'' \emph{Oracle white
  paper}, p.~16, 2012.

\bibitem{mashey1997big}
J.~R. Mashey, ``Big data and the next wave of infras-tress,'' in \emph{Computer
  Science Division Seminar, University of California, Berkeley}, 1997.

\bibitem{snijders2012big}
C.~Snijders, U.~Matzat, and U.-D. Reips, ``" big data": big gaps of knowledge
  in the field of internet science,'' \emph{International Journal of Internet
  Science}, vol.~7, no.~1, pp. 1--5, 2012.

\bibitem{hashem2015rise}
I.~A.~T. Hashem, I.~Yaqoob, N.~B. Anuar, S.~Mokhtar, A.~Gani, and S.~U. Khan,
  ``The rise of “big data” on cloud computing: Review and open research
  issues,'' \emph{Information Systems}, vol.~47, pp. 98--115, 2015.

\bibitem{elgendy2014big}
N.~Elgendy and A.~Elragal, ``Big data analytics: a literature review paper,''
  in \emph{Industrial Conference on Data Mining}.\hskip 1em plus 0.5em minus
  0.4em\relax Springer, 2014, pp. 214--227.

\bibitem{ozkose2015yesterday}
H.~{\"O}zk{\"o}se, E.~S. Ar{\i}, and C.~Gencer, ``Yesterday, today and tomorrow
  of big data,'' \emph{Procedia-Social and Behavioral Sciences}, vol. 195, pp.
  1042--1050, 2015.

\bibitem{fadiya2014advancing}
S.~O. Fadiya, S.~Saydam, and V.~V. Zira, ``Advancing big data for humanitarian
  needs,'' \emph{Procedia Engineering}, vol.~78, pp. 88--95, 2014.

\bibitem{marr2015big}
B.~Marr, \emph{Big Data: Using SMART big data, analytics and metrics to make
  better decisions and improve performance}.\hskip 1em plus 0.5em minus
  0.4em\relax John Wiley \& Sons, 2015.

\bibitem{o2013artificial}
D.~E. O'Leary, ``Artificial intelligence and big data,'' \emph{IEEE Intelligent
  Systems}, vol.~28, no.~2, pp. 96--99, 2013.

\bibitem{berman2013principles}
J.~J. Berman, \emph{Principles of big data: preparing, sharing, and analyzing
  complex information}.\hskip 1em plus 0.5em minus 0.4em\relax Newnes, 2013.

\bibitem{john2014big}
S.~John~Walker, ``Big data: A revolution that will transform how we live, work,
  and think,'' 2014.

\bibitem{bernard2015bigdata}
\BIBentryALTinterwordspacing
B.~Marr. (2015) Why only one of the 5 vs of big data really matters. [Online].
  Available:
  \url{http://www.ibmbigdatahub.com/blog/why-only-one-5-vs-big-data-really-matters}
\BIBentrySTDinterwordspacing

\bibitem{pfister1998search}
G.~F. Pfister, \emph{In search of clusters}.\hskip 1em plus 0.5em minus
  0.4em\relax Prentice Hall PTR Englewood Cliffs, 1998, vol.~2.

\bibitem{buyya1999high}
R.~Buyya \emph{et~al.}, ``High performance cluster computing: Architectures and
  systems (volume 1),'' \emph{Prentice Hall, Upper SaddleRiver, NJ, USA},
  vol.~1, p. 999, 1999.

\bibitem{brian2008cloud}
H.~Brian, T.~Brunschwiler, H.~Dill, H.~Christ, B.~Falsafi, M.~Fischer, S.~G.
  Grivas, C.~Giovanoli, R.~E. Gisi, R.~Gutmann \emph{et~al.}, ``Cloud
  computing,'' \emph{Communications of the ACM}, vol.~51, no.~7, pp. 9--11,
  2008.

\bibitem{aceto2013cloud}
G.~Aceto, A.~Botta, W.~De~Donato, and A.~Pescap{\`e}, ``Cloud monitoring: A
  survey,'' \emph{Computer Networks}, vol.~57, no.~9, pp. 2093--2115, 2013.

\bibitem{marz2015big}
N.~Marz and J.~Warren, \emph{Big Data: Principles and best practices of
  scalable realtime data systems}.\hskip 1em plus 0.5em minus 0.4em\relax New
  York; Manning Publications Co., 2015.

\bibitem{liu2013intelligent}
H.~Liu, S.~Chen, and N.~Kubota, ``Intelligent video systems and analytics: A
  survey,'' \emph{IEEE Transactions on Industrial Informatics}, vol.~9, no.~3,
  pp. 1222--1233, 2013.

\bibitem{olatunji2019video}
I.~E. Olatunji and C.-H. Cheng, ``Video analytics for visual surveillance and
  applications: An overview and survey,'' in \emph{Machine Learning
  Paradigms}.\hskip 1em plus 0.5em minus 0.4em\relax Springer, 2019, pp.
  475--515.

\bibitem{hu2011survey}
W.~Hu, N.~Xie, L.~Li, X.~Zeng, and S.~Maybank, ``A survey on visual
  content-based video indexing and retrieval,'' \emph{IEEE Transactions on
  Systems, Man, and Cybernetics, Part C (Applications and Reviews)}, vol.~41,
  no.~6, pp. 797--819, 2011.

\bibitem{patel2012content}
B.~Patel and B.~Meshram, ``Content based video retrieval systems,'' \emph{arXiv
  preprint arXiv:1205.1641}, 2012.

\bibitem{haseyama2013survey}
M.~Haseyama, T.~Ogawa, and N.~Yagi, ``[survey paper] a review of video
  retrieval based on image and video semantic understanding,'' \emph{ITE
  Transactions on Media Technology and Applications}, vol.~1, no.~1, pp. 2--9,
  2013.

\bibitem{tian2014hier}
B.~Tian, B.~T. Morris, M.~Tang, Y.~Liu, Y.~Yao, C.~Gou, D.~Shen, and S.~Tang,
  ``Hierarchical and networked vehicle surveillance in its: a survey,''
  \emph{IEEE transactions on intelligent transportation systems}, vol.~16,
  no.~2, pp. 557--580, 2014.

\bibitem{mabrouk2018abnr}
A.~B. Mabrouk and E.~Zagrouba, ``Abnormal behavior recognition for intelligent
  video surveillance systems: A review,'' \emph{Expert Systems with
  Applications}, vol.~91, pp. 480--491, 2018.

\bibitem{khan2014big}
N.~Khan, I.~Yaqoob, I.~A.~T. Hashem, Z.~Inayat, M.~Ali, W.~Kamaleldin, M.~Alam,
  M.~Shiraz, and A.~Gani, ``Big data: survey, technologies, opportunities, and
  challenges,'' \emph{The Scientific World Journal}, vol. 2014, 2014.

\bibitem{tsai2015big}
C.-W. Tsai, C.-F. Lai, H.-C. Chao, and A.~V. Vasilakos, ``Big data analytics: a
  survey,'' \emph{Journal of Big data}, vol.~2, no.~1, p.~21, 2015.

\bibitem{agrawal2011big}
D.~Agrawal, S.~Das, and A.~El~Abbadi, ``Big data and cloud computing: current
  state and future opportunities,'' in \emph{Proceedings of the 14th
  International Conference on Extending Database Technology}.\hskip 1em plus
  0.5em minus 0.4em\relax ACM, 2011, pp. 530--533.

\bibitem{zhou2017machine}
L.~Zhou, S.~Pan, J.~Wang, and A.~V. Vasilakos, ``Machine learning on big data:
  Opportunities and challenges,'' \emph{Neurocomputing}, vol. 237, pp.
  350--361, 2017.

\bibitem{che2013big}
D.~Che, M.~Safran, and Z.~Peng, ``From big data to big data mining: challenges,
  issues, and opportunities,'' in \emph{International conference on database
  systems for advanced applications}.\hskip 1em plus 0.5em minus 0.4em\relax
  Springer, 2013, pp. 1--15.

\bibitem{pouyan2018mul}
S.~Pouyanfar, Y.~Yang, S.-C. Chen, M.-L. Shyu, and S.~Iyengar, ``Multimedia big
  data analytics: A survey,'' \emph{ACM Computing Surveys (CSUR)}, vol.~51,
  no.~1, p.~10, 2018.

\bibitem{zhu2018bigITS}
L.~Zhu, F.~R. Yu, Y.~Wang, B.~Ning, and T.~Tang, ``Big data analytics in
  intelligent transportation systems: A survey,'' \emph{IEEE Transactions on
  Intelligent Transportation Systems}, vol.~20, no.~1, pp. 383--398, 2018.

\bibitem{zahid2019big}
H.~Zahid, T.~Mahmood, A.~Morshed, and T.~Sellis, ``Big data analytics in
  telecommunications: literature review and architecture recommendations,''
  \emph{IEEE/CAA Journal of Automatica Sinica}, vol.~7, no.~1, pp. 18--38,
  2019.

\bibitem{elliott2010intelligent}
D.~Elliott, ``Intelligent video solution: A definition,'' \emph{Security},
  vol.~47, no.~6, 2010.

\bibitem{tian2008ibm}
Y.-l. Tian, L.~Brown, A.~Hampapur, M.~Lu, A.~Senior, and C.-f. Shu, ``Ibm smart
  surveillance system (s3): event based video surveillance system with an open
  and extensible framework,'' \emph{Machine Vision and Applications}, vol.~19,
  no. 5-6, pp. 315--327, 2008.

\bibitem{sujana2004zoneminder}
M.~A. Sujana, ``Zoneminder video surveillance by linux platform,'' Ph.D.
  dissertation, Universiti Teknologi MARA, 2004.

\bibitem{alam2020tornado}
A.~Alam and Y.-K. Lee, ``Tornado: intermediate results orchestration based
  service-oriented data curation framework for intelligent video big data
  analytics in the cloud,'' \emph{Sensors}, vol.~20, no.~12, p. 3581, 2020.

\bibitem{kreps2011kafka}
J.~Kreps, N.~Narkhede, J.~Rao \emph{et~al.}, ``Kafka: A distributed messaging
  system for log processing,'' in \emph{Proceedings of the NetDB}, 2011, pp.
  1--7.

\bibitem{videla2012rabbitmq}
A.~Videla and J.~J. Williams, \emph{RabbitMQ in action: distributed messaging
  for everyone}.\hskip 1em plus 0.5em minus 0.4em\relax Manning, 2012.

\bibitem{snyder2011activemq}
B.~Snyder, D.~Bosnanac, and R.~Davies, \emph{ActiveMQ in action}.\hskip 1em
  plus 0.5em minus 0.4em\relax Manning Greenwich Conn., 2011, vol.~47.

\bibitem{zhang2015video}
W.~Zhang, L.~Xu, P.~Duan, W.~Gong, Q.~Lu, and S.~Yang, ``A video cloud platform
  combing online and offline cloud computing technologies,'' \emph{Personal and
  Ubiquitous Computing}, vol.~19, no.~7, pp. 1099--1110, 2015.

\bibitem{10.1007/3-540-48169-9_1}
G.~Banavar, T.~Chandra, R.~Strom, and D.~Sturman, ``A case for message oriented
  middleware,'' in \emph{Distributed Computing}, P.~Jayanti, Ed.\hskip 1em plus
  0.5em minus 0.4em\relax Berlin, Heidelberg: Springer Berlin Heidelberg, 1999,
  pp. 1--17.

\bibitem{ejsmont2015web}
A.~Ejsmont, ``Web scalability for startup engineers,'' 2015.

\bibitem{2015:WSS:2935460}
------, \emph{Web Scalability for Startup Engineers}, 1st~ed.\hskip 1em plus
  0.5em minus 0.4em\relax McGraw-Hill Education Group, 2015.

\bibitem{gamma1995design}
E.~Gamma, \emph{Design patterns: elements of reusable object-oriented
  software}.\hskip 1em plus 0.5em minus 0.4em\relax Pearson Education India,
  1995.

\bibitem{Weil:2006:CSH:1298455.1298485}
\BIBentryALTinterwordspacing
S.~A. Weil, S.~A. Brandt, E.~L. Miller, D.~D.~E. Long, and C.~Maltzahn, ``Ceph:
  A scalable, high-performance distributed file system,'' in \emph{Proceedings
  of the 7th Symposium on Operating Systems Design and Implementation}, ser.
  OSDI '06.\hskip 1em plus 0.5em minus 0.4em\relax Berkeley, CA, USA: USENIX
  Association, 2006, pp. 307--320. [Online]. Available:
  \url{http://dl.acm.org/citation.cfm?id=1298455.1298485}
\BIBentrySTDinterwordspacing

\bibitem{boyer2012glusterfs}
E.~B. Boyer, M.~C. Broomfield, and T.~A. Perrotti, ``Glusterfs one storage
  server to rule them all,'' Los Alamos National Lab.(LANL), Los Alamos, NM
  (United States), Tech. Rep., 2012.

\bibitem{5496972}
K.~Shvachko, H.~Kuang, S.~Radia, and R.~Chansler, ``The hadoop distributed file
  system,'' in \emph{2010 IEEE 26th Symposium on Mass Storage Systems and
  Technologies (MSST)}, May 2010, pp. 1--10.

\bibitem{braam2002lustre}
P.~J. Braam and R.~Zahir, ``Lustre: A scalable, high performance file system.
  cluster file systems,'' \emph{Inc. technic white paper}, 2002.

\bibitem{GASTON20091768}
\BIBentryALTinterwordspacing
D.~Gaston, C.~Newman, G.~Hansen, and D.~Lebrun-Grandié, ``Moose: A parallel
  computational framework for coupled systems of nonlinear equations,''
  \emph{Nuclear Engineering and Design}, vol. 239, no.~10, pp. 1768 -- 1778,
  2009. [Online]. Available:
  \url{http://www.sciencedirect.com/science/article/pii/S0029549309002635}
\BIBentrySTDinterwordspacing

\bibitem{Ovsiannikov:2013:QFS:2536222.2536234}
\BIBentryALTinterwordspacing
M.~Ovsiannikov, S.~Rus, D.~Reeves, P.~Sutter, S.~Rao, and J.~Kelly, ``The
  quantcast file system,'' \emph{Proc. VLDB Endow.}, vol.~6, no.~11, pp.
  1092--1101, Aug. 2013. [Online]. Available:
  \url{http://dx.doi.org/10.14778/2536222.2536234}
\BIBentrySTDinterwordspacing

\bibitem{GANESHCHANDRA201513}
\BIBentryALTinterwordspacing
D.~G. Chandra, ``Base analysis of nosql database,'' \emph{Future Generation
  Computer Systems}, vol.~52, pp. 13 -- 21, 2015, special Section: Cloud
  Computing: Security, Privacy and Practice. [Online]. Available:
  \url{http://www.sciencedirect.com/science/article/pii/S0167739X15001788}
\BIBentrySTDinterwordspacing

\bibitem{moniruzzaman2013nosql}
A.~Moniruzzaman and S.~A. Hossain, ``Nosql database: New era of databases for
  big data analytics-classification, characteristics and comparison,''
  \emph{arXiv preprint arXiv:1307.0191}, 2013.

\bibitem{Cattell:2011:SSN:1978915.1978919}
\BIBentryALTinterwordspacing
R.~Cattell, ``Scalable sql and nosql data stores,'' \emph{SIGMOD Rec.},
  vol.~39, no.~4, pp. 12--27, May 2011. [Online]. Available:
  \url{http://doi.acm.org/10.1145/1978915.1978919}
\BIBentrySTDinterwordspacing

\bibitem{Brewer:2012:PCS:2360751.2360957}
\BIBentryALTinterwordspacing
E.~Brewer, ``Pushing the cap: Strategies for consistency and availability,''
  \emph{Computer}, vol.~45, no.~2, pp. 23--29, Feb. 2012. [Online]. Available:
  \url{https://doi.org/10.1109/MC.2012.37}
\BIBentrySTDinterwordspacing

\bibitem{6133253}
------, ``Cap twelve years later: How the "rules" have changed,''
  \emph{Computer}, vol.~45, no.~2, pp. 23--29, Feb 2012.

\bibitem{Carlson:2013:RA:2505464}
J.~L. Carlson, \emph{Redis in Action}.\hskip 1em plus 0.5em minus 0.4em\relax
  Greenwich, CT, USA: Manning Publications Co., 2013.

\bibitem{Banker:2011:MA:2207997}
K.~Banker, \emph{MongoDB in Action}.\hskip 1em plus 0.5em minus 0.4em\relax
  Greenwich, CT, USA: Manning Publications Co., 2011.

\bibitem{vora2011hadoop}
M.~N. Vora, ``Hadoop-hbase for large-scale data,'' in \emph{Computer science
  and network technology (ICCSNT), 2011 international conference on},
  vol.~1.\hskip 1em plus 0.5em minus 0.4em\relax IEEE, 2011, pp. 601--605.

\bibitem{Chang:2008:BDS:1365815.1365816}
\BIBentryALTinterwordspacing
F.~Chang, J.~Dean, S.~Ghemawat, W.~C. Hsieh, D.~A. Wallach, M.~Burrows,
  T.~Chandra, A.~Fikes, and R.~E. Gruber, ``Bigtable: A distributed storage
  system for structured data,'' \emph{ACM Trans. Comput. Syst.}, vol.~26,
  no.~2, pp. 4:1--4:26, Jun. 2008. [Online]. Available:
  \url{http://doi.acm.org/10.1145/1365815.1365816}
\BIBentrySTDinterwordspacing

\bibitem{stonebraker2013voltdb}
M.~Stonebraker and A.~Weisberg, ``The voltdb main memory dbms.'' \emph{IEEE
  Data Eng. Bull.}, vol.~36, no.~2, pp. 21--27, 2013.

\bibitem{Ghemawat:2003:GFS:945445.945450}
\BIBentryALTinterwordspacing
S.~Ghemawat, H.~Gobioff, and S.-T. Leung, ``The google file system,'' in
  \emph{Proceedings of the Nineteenth ACM Symposium on Operating Systems
  Principles}, ser. SOSP '03.\hskip 1em plus 0.5em minus 0.4em\relax New York,
  NY, USA: ACM, 2003, pp. 29--43. [Online]. Available:
  \url{http://doi.acm.org/10.1145/945445.945450}
\BIBentrySTDinterwordspacing

\bibitem{xie2008event}
L.~Xie, H.~Sundaram, and M.~Campbell, ``Event mining in multimedia streams,''
  \emph{Proceedings of the IEEE}, vol.~96, no.~4, pp. 623--647, 2008.

\bibitem{shyu2008video}
M.-L. Shyu, Z.~Xie, M.~Chen, and S.-C. Chen, ``Video semantic event/concept
  detection using a subspace-based multimedia data mining framework,''
  \emph{IEEE Transactions on Multimedia}, vol.~10, no.~2, pp. 252--259, 2008.

\bibitem{regazzoni2010video}
C.~S. Regazzoni, A.~Cavallaro, Y.~Wu, J.~Konrad, and A.~Hampapur, ``Video
  analytics for surveillance: Theory and practice [from the guest editors],''
  \emph{IEEE Signal Processing Magazine}, vol.~27, no.~5, pp. 16--17, 2010.

\bibitem{buitinck2013api}
L.~Buitinck, G.~Louppe, M.~Blondel, F.~Pedregosa, A.~Mueller, O.~Grisel,
  V.~Niculae, P.~Prettenhofer, A.~Gramfort, J.~Grobler \emph{et~al.}, ``Api
  design for machine learning software: experiences from the scikit-learn
  project,'' \emph{arXiv preprint arXiv:1309.0238}, 2013.

\bibitem{schultz1996extraction}
R.~R. Schultz and R.~L. Stevenson, ``Extraction of high-resolution frames from
  video sequences,'' \emph{IEEE transactions on image processing}, vol.~5,
  no.~6, pp. 996--1011, 1996.

\bibitem{saravanan2010color}
C.~Saravanan, ``Color image to grayscale image conversion,'' in \emph{2010
  Second International Conference on Computer Engineering and Applications},
  vol.~2.\hskip 1em plus 0.5em minus 0.4em\relax IEEE, 2010, pp. 196--199.

\bibitem{boreczky1996comparison}
J.~S. Boreczky and L.~A. Rowe, ``Comparison of video shot boundary detection
  techniques,'' \emph{Journal of Electronic Imaging}, vol.~5, no.~2, pp.
  122--129, 1996.

\bibitem{hampapur1994digital}
A.~Hampapur, T.~Weymouth, and R.~Jain, ``Digital video segmentation,'' in
  \emph{Proceedings of the second ACM international conference on
  Multimedia}.\hskip 1em plus 0.5em minus 0.4em\relax ACM, 1994, pp. 357--364.

\bibitem{vetro2003video}
A.~Vetro, C.~Christopoulos, and H.~Sun, ``Video transcoding architectures and
  techniques: an overview,'' \emph{IEEE Signal processing magazine}, vol.~20,
  no.~2, pp. 18--29, 2003.

\bibitem{zhuang1998adaptive}
Y.~Zhuang, Y.~Rui, T.~S. Huang, and S.~Mehrotra, ``Adaptive key frame
  extraction using unsupervised clustering,'' in \emph{Proceedings 1998
  International Conference on Image Processing. ICIP98 (Cat. No. 98CB36269)},
  vol.~1.\hskip 1em plus 0.5em minus 0.4em\relax IEEE, 1998, pp. 866--870.

\bibitem{bengio2013representation}
Y.~Bengio, A.~Courville, and P.~Vincent, ``Representation learning: A review
  and new perspectives,'' \emph{IEEE transactions on pattern analysis and
  machine intelligence}, vol.~35, no.~8, pp. 1798--1828, 2013.

\bibitem{baumann2014computation}
F.~Baumann, A.~Ehlers, B.~Rosenhahn, and J.~Liao, ``Computation strategies for
  volume local binary patterns applied to action recognition,'' in \emph{2014
  11th IEEE International Conference on Advanced Video and Signal Based
  Surveillance (AVSS)}.\hskip 1em plus 0.5em minus 0.4em\relax IEEE, 2014, pp.
  68--73.

\bibitem{baumann2016recognizing}
------, ``Recognizing human actions using novel space-time volume binary
  patterns,'' \emph{Neurocomputing}, vol. 173, pp. 54--63, 2016.

\bibitem{amir2003ibm}
A.~Amir, M.~Berg, S.-F. Chang, W.~Hsu, G.~Iyengar, C.-Y. Lin, M.~Naphade,
  A.~Natsev, C.~Neti, H.~Nock \emph{et~al.}, ``Ibm research trecvid-2003 video
  retrieval system,'' \emph{NIST TRECVID-2003}, vol.~7, no.~8, p.~36, 2003.

\bibitem{adcock2004fxpal}
J.~Adcock, A.~Girgensohn, M.~Cooper, T.~Liu, L.~Wilcox, and E.~Rieffel, ``Fxpal
  experiments for trecvid 2004,'' \emph{Proceedings of the TREC Video Retrieval
  Evaluation (TRECVID)}, pp. 70--81, 2004.

\bibitem{yan2007review}
R.~Yan and A.~G. Hauptmann, ``A review of text and image retrieval approaches
  for broadcast news video,'' \emph{Information Retrieval}, vol.~10, no. 4-5,
  pp. 445--484, 2007.

\bibitem{sivic2005person}
J.~Sivic, M.~Everingham, and A.~Zisserman, ``Person spotting: video shot
  retrieval for face sets,'' in \emph{International conference on image and
  video retrieval}.\hskip 1em plus 0.5em minus 0.4em\relax Springer, 2005, pp.
  226--236.

\bibitem{visser2002object}
R.~Visser, N.~Sebe, and E.~Bakker, ``Object recognition for video retrieval,''
  in \emph{International Conference on Image and Video Retrieval}.\hskip 1em
  plus 0.5em minus 0.4em\relax Springer, 2002, pp. 262--270.

\bibitem{mattivi2009human}
R.~Mattivi and L.~Shao, ``Human action recognition using lbp-top as sparse
  spatio-temporal feature descriptor,'' in \emph{International Conference on
  Computer Analysis of Images and Patterns}.\hskip 1em plus 0.5em minus
  0.4em\relax Springer, 2009, pp. 740--747.

\bibitem{zhao2007dynamic}
G.~Zhao and M.~Pietikainen, ``Dynamic texture recognition using local binary
  patterns with an application to facial expressions,'' \emph{IEEE transactions
  on pattern analysis and machine intelligence}, vol.~29, no.~6, pp. 915--928,
  2007.

\bibitem{nanni2011local}
L.~Nanni, S.~Brahnam, and A.~Lumini, ``Local ternary patterns from three
  orthogonal planes for human action classification,'' \emph{Expert Systems
  with Applications}, vol.~38, no.~5, pp. 5125--5128, 2011.

\bibitem{yi2017realistic}
Y.~Yi, Z.~Zheng, and M.~Lin, ``Realistic action recognition with salient
  foreground trajectories,'' \emph{Expert Systems with Applications}, vol.~75,
  pp. 44--55, 2017.

\bibitem{uddin2019feature}
M.~A. Uddin and Y.-K. Lee, ``Feature fusion of deep spatial features and
  handcrafted spatiotemporal features for human action recognition,''
  \emph{Sensors}, vol.~19, no.~7, p. 1599, 2019.

\bibitem{simonyan2014two}
K.~Simonyan and A.~Zisserman, ``Two-stream convolutional networks for action
  recognition in videos,'' in \emph{Advances in neural information processing
  systems}, 2014, pp. 568--576.

\bibitem{karpathy2014large}
A.~Karpathy, G.~Toderici, S.~Shetty, T.~Leung, R.~Sukthankar, and L.~Fei-Fei,
  ``Large-scale video classification with convolutional neural networks,'' in
  \emph{Proceedings of the IEEE conference on Computer Vision and Pattern
  Recognition}, 2014, pp. 1725--1732.

\bibitem{wang2015action}
L.~Wang, Y.~Qiao, and X.~Tang, ``Action recognition with trajectory-pooled
  deep-convolutional descriptors,'' in \emph{Proceedings of the IEEE conference
  on computer vision and pattern recognition}, 2015, pp. 4305--4314.

\bibitem{lu2019action}
X.~Lu, H.~Yao, S.~Zhao, X.~Sun, and S.~Zhang, ``Action recognition with
  multi-scale trajectory-pooled 3d convolutional descriptors,''
  \emph{Multimedia Tools and Applications}, vol.~78, no.~1, pp. 507--523, 2019.

\bibitem{yao2019learning}
G.~Yao, T.~Lei, J.~Zhong, and P.~Jiang, ``Learning multi-temporal-scale deep
  information for action recognition,'' \emph{Applied Intelligence}, vol.~49,
  no.~6, pp. 2017--2029, 2019.

\bibitem{wang2018action}
L.~Wang, J.~Zang, Q.~Zhang, Z.~Niu, G.~Hua, and N.~Zheng, ``Action recognition
  by an attention-aware temporal weighted convolutional neural network,''
  \emph{Sensors}, vol.~18, no.~7, p. 1979, 2018.

\bibitem{girdhar2017actionvlad}
R.~Girdhar, D.~Ramanan, A.~Gupta, J.~Sivic, and B.~Russell, ``Actionvlad:
  Learning spatio-temporal aggregation for action classification,'' in
  \emph{Proceedings of the IEEE Conference on Computer Vision and Pattern
  Recognition}, 2017, pp. 971--980.

\bibitem{zhao2017pooling}
S.~Zhao, Y.~Liu, Y.~Han, R.~Hong, Q.~Hu, and Q.~Tian, ``Pooling the
  convolutional layers in deep convnets for video action recognition,''
  \emph{IEEE Transactions on Circuits and Systems for Video Technology},
  vol.~28, no.~8, pp. 1839--1849, 2017.

\bibitem{hoi2012online}
S.~C. Hoi, J.~Wang, P.~Zhao, and R.~Jin, ``Online feature selection for mining
  big data,'' in \emph{Proceedings of the 1st international workshop on big
  data, streams and heterogeneous source mining: Algorithms, systems,
  programming models and applications}.\hskip 1em plus 0.5em minus 0.4em\relax
  ACM, 2012, pp. 93--100.

\bibitem{tan2014approach}
H.~Tan and L.~Chen, ``An approach for fast and parallel video processing on
  apache hadoop clusters,'' in \emph{Multimedia and Expo (ICME), 2014 IEEE
  International Conference on}.\hskip 1em plus 0.5em minus 0.4em\relax IEEE,
  2014, pp. 1--6.

\bibitem{cong2016udsfs}
Y.~Cong, S.~Wang, B.~Fan, Y.~Yang, and H.~Yu, ``Udsfs: Unsupervised deep sparse
  feature selection,'' \emph{Neurocomputing}, vol. 196, pp. 150--158, 2016.

\bibitem{dekel2009online}
O.~Dekel, ``From online to batch learning with cutoff-averaging,'' in
  \emph{Advances in neural information processing systems}, 2009, pp. 377--384.

\bibitem{abdel2006csift}
A.~E. Abdel-Hakim and A.~A. Farag, ``Csift: A sift descriptor with color
  invariant characteristics,'' in \emph{2006 IEEE Computer Society Conference
  on Computer Vision and Pattern Recognition (CVPR'06)}, vol.~2.\hskip 1em plus
  0.5em minus 0.4em\relax Ieee, 2006, pp. 1978--1983.

\bibitem{fang2008improving}
Y.~Fang and Z.~Wang, ``Improving lbp features for gender classification,'' in
  \emph{2008 International Conference on Wavelet Analysis and Pattern
  Recognition}, vol.~1.\hskip 1em plus 0.5em minus 0.4em\relax IEEE, 2008, pp.
  373--377.

\bibitem{deniz2011face}
O.~D{\'e}niz, G.~Bueno, J.~Salido, and F.~De~la Torre, ``Face recognition using
  histograms of oriented gradients,'' \emph{Pattern Recognition Letters},
  vol.~32, no.~12, pp. 1598--1603, 2011.

\bibitem{borisyuk2018rosetta}
F.~Borisyuk, A.~Gordo, and V.~Sivakumar, ``Rosetta: Large scale system for text
  detection and recognition in images,'' in \emph{Proceedings of the 24th ACM
  SIGKDD International Conference on Knowledge Discovery \& Data Mining}.\hskip
  1em plus 0.5em minus 0.4em\relax ACM, 2018, pp. 71--79.

\bibitem{lecun2015deep}
Y.~LeCun, Y.~Bengio, and G.~Hinton, ``Deep learning,'' \emph{nature}, vol. 521,
  no. 7553, p. 436, 2015.

\bibitem{dean2012large}
J.~Dean, G.~Corrado, R.~Monga, K.~Chen, M.~Devin, M.~Mao, M.~Ranzato,
  A.~Senior, P.~Tucker, K.~Yang \emph{et~al.}, ``Large scale distributed deep
  networks,'' in \emph{Advances in neural information processing systems},
  2012, pp. 1223--1231.

\bibitem{hestness2017deep}
J.~Hestness, S.~Narang, N.~Ardalani, G.~Diamos, H.~Jun, H.~Kianinejad,
  M.~Patwary, M.~Ali, Y.~Yang, and Y.~Zhou, ``Deep learning scaling is
  predictable, empirically,'' \emph{arXiv preprint arXiv:1712.00409}, 2017.

\bibitem{zhang2017poseidon}
H.~Zhang, Z.~Zheng, S.~Xu, W.~Dai, Q.~Ho, X.~Liang, Z.~Hu, J.~Wei, P.~Xie, and
  E.~P. Xing, ``Poseidon: An efficient communication architecture for
  distributed deep learning on $\{$GPU$\}$ clusters,'' in \emph{2017
  $\{$USENIX$\}$ Annual Technical Conference ($\{$USENIX$\}$$\{$ATC$\}$ 17)},
  2017, pp. 181--193.

\bibitem{yue2015beyond}
J.~Yue-Hei~Ng, M.~Hausknecht, S.~Vijayanarasimhan, O.~Vinyals, R.~Monga, and
  G.~Toderici, ``Beyond short snippets: Deep networks for video
  classification,'' in \emph{Proceedings of the IEEE conference on computer
  vision and pattern recognition}, 2015, pp. 4694--4702.

\bibitem{krizhevsky2012imagenet}
A.~Krizhevsky, I.~Sutskever, and G.~E. Hinton, ``Imagenet classification with
  deep convolutional neural networks,'' in \emph{Advances in neural information
  processing systems}, 2012, pp. 1097--1105.

\bibitem{lecun1998gradient}
Y.~LeCun, L.~Bottou, Y.~Bengio, P.~Haffner \emph{et~al.}, ``Gradient-based
  learning applied to document recognition,'' \emph{Proceedings of the IEEE},
  vol.~86, no.~11, pp. 2278--2324, 1998.

\bibitem{szegedy2015going}
C.~Szegedy, W.~Liu, Y.~Jia, P.~Sermanet, S.~Reed, D.~Anguelov, D.~Erhan,
  V.~Vanhoucke, and A.~Rabinovich, ``Going deeper with convolutions,'' in
  \emph{Proceedings of the IEEE conference on computer vision and pattern
  recognition}, 2015, pp. 1--9.

\bibitem{43022GoogleNet2015}
\BIBentryALTinterwordspacing
------, ``Going deeper with convolutions,'' in \emph{Computer Vision and
  Pattern Recognition (CVPR)}, 2015. [Online]. Available:
  \url{http://arxiv.org/abs/1409.4842}
\BIBentrySTDinterwordspacing

\bibitem{simonyan2014very}
K.~Simonyan and A.~Zisserman, ``Very deep convolutional networks for
  large-scale image recognition,'' \emph{arXiv preprint arXiv:1409.1556}, 2014.

\bibitem{he2016deep}
K.~He, X.~Zhang, S.~Ren, and J.~Sun, ``Deep residual learning for image
  recognition,'' in \emph{Proceedings of the IEEE conference on computer vision
  and pattern recognition}, 2016, pp. 770--778.

\bibitem{abadi2016tensorflow}
M.~Abadi, P.~Barham, J.~Chen, Z.~Chen, A.~Davis, J.~Dean, M.~Devin,
  S.~Ghemawat, G.~Irving, M.~Isard \emph{et~al.}, ``Tensorflow: A system for
  large-scale machine learning.'' in \emph{OSDI}, vol.~16, 2016, pp. 265--283.

\bibitem{jia2014caffe}
Y.~Jia, E.~Shelhamer, J.~Donahue, S.~Karayev, J.~Long, R.~Girshick,
  S.~Guadarrama, and T.~Darrell, ``Caffe: Convolutional architecture for fast
  feature embedding,'' in \emph{Proceedings of the 22nd ACM international
  conference on Multimedia}.\hskip 1em plus 0.5em minus 0.4em\relax ACM, 2014,
  pp. 675--678.

\bibitem{ketkar2017introduction}
N.~Ketkar, ``Introduction to pytorch,'' in \emph{Deep learning with
  python}.\hskip 1em plus 0.5em minus 0.4em\relax Springer, 2017, pp. 195--208.

\bibitem{chen2015mxnet}
T.~Chen, M.~Li, Y.~Li, M.~Lin, N.~Wang, M.~Wang, T.~Xiao, B.~Xu, C.~Zhang, and
  Z.~Zhang, ``Mxnet: A flexible and efficient machine learning library for
  heterogeneous distributed systems,'' \emph{arXiv preprint arXiv:1512.01274},
  2015.

\bibitem{seide2016cntk}
F.~Seide and A.~Agarwal, ``Cntk: Microsoft's open-source deep-learning
  toolkit,'' in \emph{Proceedings of the 22nd ACM SIGKDD International
  Conference on Knowledge Discovery and Data Mining}, 2016, pp. 2135--2135.

\bibitem{mathworkdeep}
M.~Mathwork, ``Deep learning toolbox.''

\bibitem{gao2018dendritic}
S.~Gao, M.~Zhou, Y.~Wang, J.~Cheng, H.~Yachi, and J.~Wang, ``Dendritic neuron
  model with effective learning algorithms for classification, approximation,
  and prediction,'' \emph{IEEE transactions on neural networks and learning
  systems}, vol.~30, no.~2, pp. 601--614, 2018.

\bibitem{mayer2019scalable}
R.~Mayer and H.-A. Jacobsen, ``Scalable deep learning on distributed
  infrastructures: Challenges, techniques and tools,'' \emph{arXiv preprint
  arXiv:1903.11314}, 2019.

\bibitem{deshpande2018artificial}
A.~Deshpande and M.~Kumar, \emph{Artificial Intelligence for Big Data: Complete
  Guide to Automating Big Data Solutions Using Artificial Intelligence
  Techniques}.\hskip 1em plus 0.5em minus 0.4em\relax Packt Publishing Ltd,
  2018.

\bibitem{huang2019gpipe}
Y.~Huang, Y.~Cheng, A.~Bapna, O.~Firat, D.~Chen, M.~Chen, H.~Lee, J.~Ngiam,
  Q.~V. Le, Y.~Wu \emph{et~al.}, ``Gpipe: Efficient training of giant neural
  networks using pipeline parallelism,'' in \emph{Advances in Neural
  Information Processing Systems}, 2019, pp. 103--112.

\bibitem{dean2008mapreduce}
J.~Dean and S.~Ghemawat, ``Mapreduce: simplified data processing on large
  clusters,'' \emph{Communications of the ACM}, vol.~51, no.~1, pp. 107--113,
  2008.

\bibitem{zaharia2016apache}
M.~Zaharia, R.~S. Xin, P.~Wendell, T.~Das, M.~Armbrust, A.~Dave, X.~Meng,
  J.~Rosen, S.~Venkataraman, M.~J. Franklin \emph{et~al.}, ``Apache spark: a
  unified engine for big data processing,'' \emph{Communications of the ACM},
  vol.~59, no.~11, pp. 56--65, 2016.

\bibitem{carbone2015apache}
P.~Carbone, A.~Katsifodimos, S.~Ewen, V.~Markl, S.~Haridi, and K.~Tzoumas,
  ``Apache flink: Stream and batch processing in a single engine,''
  \emph{Bulletin of the IEEE Computer Society Technical Committee on Data
  Engineering}, vol.~36, no.~4, 2015.

\bibitem{toshniwal2014storm}
A.~Toshniwal, S.~Taneja, A.~Shukla, K.~Ramasamy, J.~M. Patel, S.~Kulkarni,
  J.~Jackson, K.~Gade, M.~Fu, J.~Donham \emph{et~al.}, ``Storm@ twitter,'' in
  \emph{Proceedings of the 2014 ACM SIGMOD international conference on
  Management of data}.\hskip 1em plus 0.5em minus 0.4em\relax ACM, 2014, pp.
  147--156.

\bibitem{wang2015building}
G.~Wang, J.~Koshy, S.~Subramanian, K.~Paramasivam, M.~Zadeh, N.~Narkhede,
  J.~Rao, J.~Kreps, and J.~Stein, ``Building a replicated logging system with
  apache kafka,'' \emph{Proceedings of the VLDB Endowment}, vol.~8, no.~12, pp.
  1654--1655, 2015.

\bibitem{owen2012mahout}
S.~Owen and S.~Owen, ``Mahout in action,'' 2012.

\bibitem{team2016deeplearning4j}
D.~Team \emph{et~al.}, ``Deeplearning4j: Open-source distributed deep learning
  for the jvm,'' \emph{Apache Software Foundation License}, vol.~2, 2016.

\bibitem{chollet2015keras}
F.~Chollet \emph{et~al.}, ``Keras,'' 2015.

\bibitem{dai2018bigdl}
J.~Dai, Y.~Wang, X.~Qiu, D.~Ding, Y.~Zhang, Y.~Wang, X.~Jia, C.~Zhang, Y.~Wan,
  Z.~Li \emph{et~al.}, ``Bigdl: A distributed deep learning framework for big
  data,'' \emph{arXiv preprint arXiv:1804.05839}, 2018.

\bibitem{meng2016mllib}
X.~Meng, J.~Bradley, B.~Yavuz, E.~Sparks, S.~Venkataraman, D.~Liu, J.~Freeman,
  D.~Tsai, M.~Amde, S.~Owen \emph{et~al.}, ``Mllib: Machine learning in apache
  spark,'' \emph{The Journal of Machine Learning Research}, vol.~17, no.~1, pp.
  1235--1241, 2016.

\bibitem{morales2015samoa}
G.~D.~F. Morales and A.~Bifet, ``Samoa: scalable advanced massive online
  analysis.'' \emph{Journal of Machine Learning Research}, vol.~16, no.~1, pp.
  149--153, 2015.

\bibitem{dean2010mapreduce}
J.~Dean and S.~Ghemawat, ``Mapreduce: a flexible data processing tool,''
  \emph{Communications of the ACM}, vol.~53, no.~1, pp. 72--77, 2010.

\bibitem{zaharia2010spark}
M.~Zaharia, M.~Chowdhury, M.~J. Franklin, S.~Shenker, and I.~Stoica, ``Spark:
  Cluster computing with working sets.'' \emph{HotCloud}, vol.~10, no. 10-10,
  p.~95, 2010.

\bibitem{zaharia2013discretized}
M.~Zaharia, T.~Das, H.~Li, T.~Hunter, S.~Shenker, and I.~Stoica, ``Discretized
  streams: Fault-tolerant streaming computation at scale,'' in
  \emph{Proceedings of the twenty-fourth ACM symposium on operating systems
  principles}.\hskip 1em plus 0.5em minus 0.4em\relax ACM, 2013, pp. 423--438.

\bibitem{cook2016practical}
D.~Cook, \emph{Practical machine learning with H2O: powerful, scalable
  techniques for deep learning and AI}.\hskip 1em plus 0.5em minus 0.4em\relax
  " O'Reilly Media, Inc.", 2016.

\bibitem{deng2009imagenet}
J.~Deng, W.~Dong, R.~Socher, L.-J. Li, K.~Li, and L.~Fei-Fei, ``Imagenet: A
  large-scale hierarchical image database,'' in \emph{2009 IEEE conference on
  computer vision and pattern recognition}.\hskip 1em plus 0.5em minus
  0.4em\relax Ieee, 2009, pp. 248--255.

\bibitem{thomee2015yfcc100m}
B.~Thomee, D.~A. Shamma, G.~Friedland, B.~Elizalde, K.~Ni, D.~Poland, D.~Borth,
  and L.-J. Li, ``Yfcc100m: The new data in multimedia research,'' \emph{arXiv
  preprint arXiv:1503.01817}, 2015.

\bibitem{trecvid2019}
\BIBentryALTinterwordspacing
Trecvid, ``Trec video retrieval evaluation: Trecvid,'' 2019, accessed:
  2019-12-07. [Online]. Available: \url{https://trecvid.nist.gov/}
\BIBentrySTDinterwordspacing

\bibitem{soomro2012ucf101}
K.~Soomro, A.~R. Zamir, and M.~Shah, ``Ucf101: A dataset of 101 human actions
  classes from videos in the wild,'' \emph{arXiv preprint arXiv:1212.0402},
  2012.

\bibitem{kay2017kinetics}
W.~Kay, J.~Carreira, K.~Simonyan, B.~Zhang, C.~Hillier, S.~Vijayanarasimhan,
  F.~Viola, T.~Green, T.~Back, P.~Natsev \emph{et~al.}, ``The kinetics human
  action video dataset,'' \emph{arXiv preprint arXiv:1705.06950}, 2017.

\bibitem{MediaEval2015}
\BIBentryALTinterwordspacing
CEUR, ``Multimedia benchmark workshop,'' 2015, accessed: 2019-12-07. [Online].
  Available: \url{http://ceur-ws.org/Vol-1436/}
\BIBentrySTDinterwordspacing

\bibitem{nambiar2014multi}
A.~Nambiar, M.~Taiana, D.~Figueira, J.~C. Nascimento, and A.~Bernardino, ``A
  multi-camera video dataset for research on high-definition surveillance,''
  \emph{International Journal of Machine Intelligence and Sensory Signal
  Processing}, vol.~1, no.~3, pp. 267--286, 2014.

\bibitem{caba2015activitynet}
B.~G. Fabian Caba~Heilbron, Victor~Escorcia and J.~C. Niebles, ``Activitynet: A
  large-scale video benchmark for human activity understanding,'' in
  \emph{Proceedings of the IEEE Conference on Computer Vision and Pattern
  Recognition}, 2015, pp. 961--970.

\bibitem{kuehne2011hmdb}
H.~Kuehne, H.~Jhuang, E.~Garrote, T.~Poggio, and T.~Serre, ``Hmdb: a large
  video database for human motion recognition,'' in \emph{2011 International
  Conference on Computer Vision}.\hskip 1em plus 0.5em minus 0.4em\relax IEEE,
  2011, pp. 2556--2563.

\bibitem{gu2018ava}
C.~Gu, C.~Sun, D.~A. Ross, C.~Vondrick, C.~Pantofaru, Y.~Li,
  S.~Vijayanarasimhan, G.~Toderici, S.~Ricco, R.~Sukthankar \emph{et~al.},
  ``Ava: A video dataset of spatio-temporally localized atomic visual
  actions,'' in \emph{Proceedings of the IEEE Conference on Computer Vision and
  Pattern Recognition}, 2018, pp. 6047--6056.

\bibitem{miech2019howto100m}
A.~Miech, D.~Zhukov, J.-B. Alayrac, M.~Tapaswi, I.~Laptev, and J.~Sivic,
  ``Howto100m: Learning a text-video embedding by watching hundred million
  narrated video clips,'' \emph{arXiv preprint arXiv:1906.03327}, 2019.

\bibitem{abu2016youtube}
S.~Abu-El-Haija, N.~Kothari, J.~Lee, P.~Natsev, G.~Toderici, B.~Varadarajan,
  and S.~Vijayanarasimhan, ``Youtube-8m: A large-scale video classification
  benchmark,'' \emph{arXiv preprint arXiv:1609.08675}, 2016.

\bibitem{ballan2010video}
L.~Ballan, M.~Bertini, A.~Del~Bimbo, and G.~Serra, ``Video annotation and
  retrieval using ontologies and rule learning,'' \emph{IEEE MultiMedia},
  vol.~17, no.~4, pp. 80--88, 2010.

\bibitem{smeaton2010video}
A.~F. Smeaton, P.~Over, and A.~R. Doherty, ``Video shot boundary detection:
  Seven years of trecvid activity,'' \emph{Computer Vision and Image
  Understanding}, vol. 114, no.~4, pp. 411--418, 2010.

\bibitem{smeaton2009high}
A.~F. Smeaton, P.~Over, and W.~Kraaij, ``High-level feature detection from
  video in trecvid: a 5-year retrospective of achievements,'' in
  \emph{Multimedia content analysis}.\hskip 1em plus 0.5em minus 0.4em\relax
  Springer, 2009, pp. 1--24.

\bibitem{bhaumik2016hybrid}
H.~Bhaumik, S.~Bhattacharyya, M.~D. Nath, and S.~Chakraborty, ``Hybrid soft
  computing approaches to content based video retrieval: A brief review,''
  \emph{Applied Soft Computing}, vol.~46, pp. 1008--1029, 2016.

\bibitem{hu2013internet}
S.-M. Hu, T.~Chen, K.~Xu, M.-M. Cheng, and R.~R. Martin, ``Internet visual
  media processing: a survey with graphics and vision applications,'' \emph{The
  Visual Computer}, vol.~29, no.~5, pp. 393--405, 2013.

\bibitem{spolaor2020systematic}
N.~Spola{\^o}r, H.~D. Lee, W.~S.~R. Takaki, L.~A. Ensina, C.~S.~R. Coy, and
  F.~C. Wu, ``A systematic review on content-based video retrieval,''
  \emph{Engineering Applications of Artificial Intelligence}, vol.~90, p.
  103557, 2020.

\bibitem{shang2010real}
L.~Shang, L.~Yang, F.~Wang, K.-P. Chan, and X.-S. Hua, ``Real-time large scale
  near-duplicate web video retrieval,'' in \emph{Proceedings of the 18th ACM
  international conference on Multimedia}, 2010, pp. 531--540.

\bibitem{wang2015gpu}
H.~Wang, F.~Zhu, B.~Xiao, L.~Wang, and Y.-G. Jiang, ``Gpu-based mapreduce for
  large-scale near-duplicate video retrieval,'' \emph{Multimedia Tools and
  Applications}, vol.~74, no.~23, pp. 10\,515--10\,534, 2015.

\bibitem{mikolajczyk2004scale}
K.~Mikolajczyk and C.~Schmid, ``Scale \& affine invariant interest point
  detectors,'' \emph{International journal of computer vision}, vol.~60, no.~1,
  pp. 63--86, 2004.

\bibitem{sivic2003video}
J.~Sivic and A.~Zisserman, ``Video google: A text retrieval approach to object
  matching in videos,'' in \emph{null}.\hskip 1em plus 0.5em minus 0.4em\relax
  IEEE, 2003, p. 1470.

\bibitem{ding2017survsurf}
S.~Ding, G.~Li, Y.~Li, X.~Li, Q.~Zhai, A.~C. Champion, J.~Zhu, D.~Xuan, and
  Y.~F. Zheng, ``Survsurf: human retrieval on large surveillance video data,''
  \emph{Multimedia Tools and Applications}, vol.~76, no.~5, pp. 6521--6549,
  2017.

\bibitem{zhu2015marlin}
N.~Zhu, W.~He, Y.~Hua, and Y.~Chen, ``Marlin: Taming the big streaming data in
  large scale video similarity search,'' in \emph{Big Data (Big Data), 2015
  IEEE International Conference on}.\hskip 1em plus 0.5em minus 0.4em\relax
  IEEE, 2015, pp. 1755--1764.

\bibitem{lv2016efficient}
J.~Lv, B.~Wu, S.~Yang, B.~Jia, and P.~Qiu, ``Efficient large scale
  near-duplicate video detection base on spark,'' in \emph{Big Data (Big Data),
  2016 IEEE International Conference on}.\hskip 1em plus 0.5em minus
  0.4em\relax IEEE, 2016, pp. 957--962.

\bibitem{numan2020FALKON}
M.~N. Khan, A.~Alam, and Y.-K. Lee, ``Falkon: Large-scale content-based video
  retrieval utilizing deep-features and distributed in-memory computing,'' in
  \emph{2020 IEEE International Conference on Big Data and Smart Computing
  (BigComp)}.\hskip 1em plus 0.5em minus 0.4em\relax IEEE, 2020, pp. 36--43.

\bibitem{lin2020cloud}
F.-C. Lin, H.-H. Ngo, and C.-R. Dow, ``A cloud-based face video retrieval
  system with deep learning,'' \emph{The Journal of Supercomputing}, pp. 1--21,
  2020.

\bibitem{hu2007semantic}
W.~Hu, D.~Xie, Z.~Fu, W.~Zeng, and S.~Maybank, ``Semantic-based surveillance
  video retrieval,'' \emph{IEEE Transactions on image processing}, vol.~16,
  no.~4, pp. 1168--1181, 2007.

\bibitem{sivic2006video}
J.~Sivic and A.~Zisserman, ``Video google: Efficient visual search of videos,''
  in \emph{Toward category-level object recognition}.\hskip 1em plus 0.5em
  minus 0.4em\relax Springer, 2006, pp. 127--144.

\bibitem{aytar2008utilizing}
Y.~Aytar, M.~Shah, and J.~Luo, ``Utilizing semantic word similarity measures
  for video retrieval,'' in \emph{2008 IEEE Conference on Computer Vision and
  Pattern Recognition}.\hskip 1em plus 0.5em minus 0.4em\relax IEEE, 2008, pp.
  1--8.

\bibitem{kennedy2005automatic}
L.~S. Kennedy, A.~P. Natsev, and S.-F. Chang, ``Automatic discovery of
  query-class-dependent models for multimodal search,'' in \emph{Proceedings of
  the 13th annual ACM international conference on Multimedia}.\hskip 1em plus
  0.5em minus 0.4em\relax ACM, 2005, pp. 882--891.

\bibitem{yan2004learning}
R.~Yan, J.~Yang, and A.~G. Hauptmann, ``Learning query-class dependent weights
  in automatic video retrieval,'' in \emph{Proceedings of the 12th annual ACM
  international conference on Multimedia}.\hskip 1em plus 0.5em minus
  0.4em\relax ACM, 2004, pp. 548--555.

\bibitem{lienhart2000system}
R.~Lienhart, ``A system for effortless content annotation to unfold the
  semantics in videos,'' in \emph{cbaivl}.\hskip 1em plus 0.5em minus
  0.4em\relax IEEE, 2000, p.~45.

\bibitem{snoek2007adding}
C.~G. Snoek, B.~Huurnink, L.~Hollink, M.~De~Rijke, G.~Schreiber, and
  M.~Worring, ``Adding semantics to detectors for video retrieval,'' \emph{IEEE
  Transactions on multimedia}, vol.~9, no.~5, pp. 975--986, 2007.

\bibitem{neo2006video}
S.-Y. Neo, J.~Zhao, M.-Y. Kan, and T.-S. Chua, ``Video retrieval using high
  level features: Exploiting query matching and confidence-based weighting,''
  in \emph{International Conference on Image and Video Retrieval}.\hskip 1em
  plus 0.5em minus 0.4em\relax Springer, 2006, pp. 143--152.

\bibitem{chen2008integrated}
L.-H. Chen, K.-H. Chin, and H.-Y. Liao, ``An integrated approach to video
  retrieval,'' in \emph{Proceedings of the nineteenth conference on
  Australasian database-Volume 75}.\hskip 1em plus 0.5em minus 0.4em\relax
  Australian Computer Society, Inc., 2008, pp. 49--55.

\bibitem{hopfgartner2007simulated}
F.~Hopfgartner, J.~Urban, R.~Villa, and J.~Jose, ``Simulated testing of an
  adaptive multimedia information retrieval system,'' in \emph{2007
  International Workshop on Content-Based Multimedia Indexing}.\hskip 1em plus
  0.5em minus 0.4em\relax IEEE, 2007, pp. 328--335.

\bibitem{yan2003negative}
R.~Yan, A.~G. Hauptmann, and R.~Jin, ``Negative pseudo-relevance feedback in
  content-based video retrieval,'' in \emph{Proceedings of the eleventh ACM
  international conference on Multimedia}.\hskip 1em plus 0.5em minus
  0.4em\relax ACM, 2003, pp. 343--346.

\bibitem{hauptmann2008video}
A.~G. Hauptmann, M.~G. Christel, and R.~Yan, ``Video retrieval based on
  semantic concepts,'' \emph{Proceedings of the IEEE}, vol.~96, no.~4, pp.
  602--622, 2008.

\bibitem{sun2014discover}
C.~Sun and R.~Nevatia, ``Discover: Discovering important segments for
  classification of video events and recounting,'' in \emph{Proceedings of the
  IEEE Conference on Computer Vision and Pattern Recognition}, 2014, pp.
  2569--2576.

\bibitem{habibian2014recommendations}
A.~Habibian and C.~G. Snoek, ``Recommendations for recognizing video events by
  concept vocabularies,'' \emph{Computer Vision and Image Understanding}, vol.
  124, pp. 110--122, 2014.

\bibitem{song2017extracting}
H.~Song, X.~Wu, W.~Yu, and Y.~Jia, ``Extracting key segments of videos for
  event detection by learning from web sources,'' \emph{IEEE Transactions on
  Multimedia}, vol.~20, no.~5, pp. 1088--1100, 2017.

\bibitem{wang2014video}
H.~Wang, X.~Wu, and Y.~Jia, ``Video annotation via image groups from the web,''
  \emph{IEEE Transactions on Multimedia}, vol.~16, no.~5, pp. 1282--1291, 2014.

\bibitem{zhang2015enhancing}
X.~Zhang, Y.~Yang, Y.~Zhang, H.~Luan, J.~Li, H.~Zhang, and T.-S. Chua,
  ``Enhancing video event recognition using automatically constructed
  semantic-visual knowledge base,'' \emph{IEEE Transactions on Multimedia},
  vol.~17, no.~9, pp. 1562--1575, 2015.

\bibitem{song2017recognizing}
H.~Song, X.~Wu, W.~Liang, and Y.~Jia, ``Recognizing key segments of videos for
  video annotation by learning from web image sets,'' \emph{Multimedia Tools
  and Applications}, vol.~76, no.~5, pp. 6111--6126, 2017.

\bibitem{escalante2017principal}
H.~J. Escalante, I.~Guyon, V.~Athitsos, P.~Jangyodsuk, and J.~Wan, ``Principal
  motion components for one-shot gesture recognition,'' \emph{Pattern Analysis
  and Applications}, vol.~20, no.~1, pp. 167--182, 2017.

\bibitem{jiang2015multi}
F.~Jiang, S.~Zhang, S.~Wu, Y.~Gao, and D.~Zhao, ``Multi-layered gesture
  recognition with kinect,'' \emph{The Journal of Machine Learning Research},
  vol.~16, no.~1, pp. 227--254, 2015.

\bibitem{wu2012one}
D.~Wu, F.~Zhu, and L.~Shao, ``One shot learning gesture recognition from rgbd
  images,'' in \emph{2012 IEEE Computer Society Conference on Computer Vision
  and Pattern Recognition Workshops}.\hskip 1em plus 0.5em minus 0.4em\relax
  IEEE, 2012, pp. 7--12.

\bibitem{lu2014range}
C.~Lu, J.~Jia, and C.-K. Tang, ``Range-sample depth feature for action
  recognition,'' in \emph{Proceedings of the IEEE conference on computer vision
  and pattern recognition}, 2014, pp. 772--779.

\bibitem{yang2014super}
X.~Yang and Y.~Tian, ``Super normal vector for activity recognition using depth
  sequences,'' in \emph{Proceedings of the IEEE conference on computer vision
  and pattern recognition}, 2014, pp. 804--811.

\bibitem{oreifej2013hon4d}
O.~Oreifej and Z.~Liu, ``Hon4d: Histogram of oriented 4d normals for activity
  recognition from depth sequences,'' in \emph{Proceedings of the IEEE
  conference on computer vision and pattern recognition}, 2013, pp. 716--723.

\bibitem{yang2012recognizing}
X.~Yang, C.~Zhang, and Y.~Tian, ``Recognizing actions using depth motion
  maps-based histograms of oriented gradients,'' in \emph{Proceedings of the
  20th ACM international conference on Multimedia}.\hskip 1em plus 0.5em minus
  0.4em\relax ACM, 2012, pp. 1057--1060.

\bibitem{wang2015convnets}
P.~Wang, W.~Li, Z.~Gao, C.~Tang, J.~Zhang, and P.~Ogunbona, ``Convnets-based
  action recognition from depth maps through virtual cameras and
  pseudocoloring,'' in \emph{Proceedings of the 23rd ACM international
  conference on Multimedia}.\hskip 1em plus 0.5em minus 0.4em\relax ACM, 2015,
  pp. 1119--1122.

\bibitem{wang2015actionA}
P.~Wang, W.~Li, Z.~Gao, J.~Zhang, C.~Tang, and P.~O. Ogunbona, ``Action
  recognition from depth maps using deep convolutional neural networks,''
  \emph{IEEE Transactions on Human-Machine Systems}, vol.~46, no.~4, pp.
  498--509, 2015.

\bibitem{wu2016deep}
D.~Wu, L.~Pigou, P.-J. Kindermans, N.~D.-H. Le, L.~Shao, J.~Dambre, and J.-M.
  Odobez, ``Deep dynamic neural networks for multimodal gesture segmentation
  and recognition,'' \emph{IEEE transactions on pattern analysis and machine
  intelligence}, vol.~38, no.~8, pp. 1583--1597, 2016.

\bibitem{wang2018depth}
P.~Wang, W.~Li, Z.~Gao, C.~Tang, and P.~O. Ogunbona, ``Depth pooling based
  large-scale 3-d action recognition with convolutional neural networks,''
  \emph{IEEE Transactions on Multimedia}, vol.~20, no.~5, pp. 1051--1061, 2018.

\bibitem{veeriah2015differential}
V.~Veeriah, N.~Zhuang, and G.-J. Qi, ``Differential recurrent neural networks
  for action recognition,'' in \emph{Proceedings of the IEEE international
  conference on computer vision}, 2015, pp. 4041--4049.

\bibitem{du2015hierarchical}
Y.~Du, W.~Wang, and L.~Wang, ``Hierarchical recurrent neural network for
  skeleton based action recognition,'' in \emph{Proceedings of the IEEE
  conference on computer vision and pattern recognition}, 2015, pp. 1110--1118.

\bibitem{tran2015learning}
D.~Tran, L.~Bourdev, R.~Fergus, L.~Torresani, and M.~Paluri, ``Learning
  spatiotemporal features with 3d convolutional networks,'' in
  \emph{Proceedings of the IEEE international conference on computer vision},
  2015, pp. 4489--4497.

\bibitem{ji20123d}
S.~Ji, W.~Xu, M.~Yang, and K.~Yu, ``3d convolutional neural networks for human
  action recognition,'' \emph{IEEE transactions on pattern analysis and machine
  intelligence}, vol.~35, no.~1, pp. 221--231, 2012.

\bibitem{guler2016real}
P.~Guler, D.~Emeksiz, A.~Temizel, M.~Teke, and T.~T. Temizel, ``Real-time
  multi-camera video analytics system on gpu,'' \emph{Journal of Real-Time
  Image Processing}, vol.~11, no.~3, pp. 457--472, 2016.

\bibitem{pham2010gpu}
V.~Pham, P.~Vo, V.~T. Hung \emph{et~al.}, ``Gpu implementation of extended
  gaussian mixture model for background subtraction,'' in \emph{2010 IEEE RIVF
  International Conference on Computing \& Communication Technologies,
  Research, Innovation, and Vision for the Future (RIVF)}.\hskip 1em plus 0.5em
  minus 0.4em\relax IEEE, 2010, pp. 1--4.

\bibitem{cucchiara2001detecting}
R.~Cucchiara, C.~Grana, M.~Piccardi, and A.~Prati, ``Detecting objects, shadows
  and ghosts in video streams by exploiting color and motion information,'' in
  \emph{Proceedings 11th international conference on image analysis and
  processing}.\hskip 1em plus 0.5em minus 0.4em\relax IEEE, 2001, pp. 360--365.

\bibitem{beyan2012adaptive}
C.~Beyan and A.~Temizel, ``Adaptive mean-shift for automated multi object
  tracking,'' \emph{IET computer vision}, vol.~6, no.~1, pp. 1--12, 2012.

\bibitem{risse2017visual}
B.~Risse, M.~Mangan, L.~Del~Pero, and B.~Webb, ``Visual tracking of small
  animals in cluttered natural environments using a freely moving camera,'' in
  \emph{Proceedings of the IEEE International Conference on Computer Vision},
  2017, pp. 2840--2849.

\bibitem{shantaiya2013survey}
S.~Shantaiya, K.~Verma, and K.~Mehta, ``A survey on approaches of object
  detection,'' \emph{International Journal of Computer Applications}, vol.~65,
  no.~18, 2013.

\bibitem{deori2014survey}
B.~Deori and D.~M. Thounaojam, ``A survey on moving object tracking in video,''
  \emph{International Journal on Information Theory (IJIT)}, vol.~3, no.~3, pp.
  31--46, 2014.

\bibitem{leal2017tracking}
L.~Leal-Taix{\'e}, A.~Milan, K.~Schindler, D.~Cremers, I.~Reid, and S.~Roth,
  ``Tracking the trackers: an analysis of the state of the art in multiple
  object tracking,'' \emph{arXiv preprint arXiv:1704.02781}, 2017.

\bibitem{yin2015background}
X.~Yin, B.~Wang, W.~Li, Y.~Liu, and M.~Zhang, ``Background subtraction for
  moving cameras based on trajectory-controlled segmentation and label
  inference.'' \emph{Ksii Transactions on Internet \& Information Systems},
  vol.~9, no.~10, 2015.

\bibitem{zhang2017tracking}
S.~Zhang, J.-B. Huang, J.~Lim, Y.~Gong, J.~Wang, N.~Ahuja, and M.-H. Yang,
  ``Tracking persons-of-interest via unsupervised representation adaptation,''
  \emph{arXiv preprint arXiv:1710.02139}, 2017.

\bibitem{hayman2003statistical}
E.~Hayman and J.-O. Eklundh, ``Statistical background subtraction for a mobile
  observer,'' in \emph{null}.\hskip 1em plus 0.5em minus 0.4em\relax IEEE,
  2003, p.~67.

\bibitem{lenz2011sparse}
P.~Lenz, J.~Ziegler, A.~Geiger, and M.~Roser, ``Sparse scene flow segmentation
  for moving object detection in urban environments,'' in \emph{2011 IEEE
  Intelligent Vehicles Symposium (IV)}.\hskip 1em plus 0.5em minus 0.4em\relax
  IEEE, 2011, pp. 926--932.

\bibitem{wren1997pfinder}
C.~R. Wren, A.~Azarbayejani, T.~Darrell, and A.~P. Pentland, ``Pfinder:
  Real-time tracking of the human body,'' \emph{IEEE Transactions on pattern
  analysis and machine intelligence}, vol.~19, no.~7, pp. 780--785, 1997.

\bibitem{zivkovic2006efficient}
Z.~Zivkovic and F.~Van Der~Heijden, ``Efficient adaptive density estimation per
  image pixel for the task of background subtraction,'' \emph{Pattern
  recognition letters}, vol.~27, no.~7, pp. 773--780, 2006.

\bibitem{jin2008background}
Y.~Jin, L.~Tao, H.~Di, N.~I. Rao, and G.~Xu, ``Background modeling from a
  free-moving camera by multi-layer homography algorithm,'' in \emph{2008 15th
  IEEE International Conference on Image Processing}.\hskip 1em plus 0.5em
  minus 0.4em\relax IEEE, 2008, pp. 1572--1575.

\bibitem{wu2015moving}
Y.~Wu, X.~He, and T.~Q. Nguyen, ``Moving object detection with a freely moving
  camera via background motion subtraction,'' \emph{IEEE Transactions on
  Circuits and Systems for Video Technology}, vol.~27, no.~2, pp. 236--248,
  2015.

\bibitem{braham2016deep}
M.~Braham and M.~Van~Droogenbroeck, ``Deep background subtraction with
  scene-specific convolutional neural networks,'' in \emph{2016 international
  conference on systems, signals and image processing (IWSSIP)}.\hskip 1em plus
  0.5em minus 0.4em\relax IEEE, 2016, pp. 1--4.

\bibitem{bouwmans2017decomposition}
T.~Bouwmans, A.~Sobral, S.~Javed, S.~K. Jung, and E.-H. Zahzah, ``Decomposition
  into low-rank plus additive matrices for background/foreground separation: A
  review for a comparative evaluation with a large-scale dataset,''
  \emph{Computer Science Review}, vol.~23, pp. 1--71, 2017.

\bibitem{yazdi2018new}
M.~Yazdi and T.~Bouwmans, ``New trends on moving object detection in video
  images captured by a moving camera: A survey,'' \emph{Computer Science
  Review}, vol.~28, pp. 157--177, 2018.

\bibitem{elhart2017audience}
I.~Elhart, M.~Mikusz, C.~G. Mora, M.~Langheinrich, and N.~Davies, ``Audience
  monitor: an open source tool for tracking audience mobility in front of
  pervasive displays,'' in \emph{Proceedings of the 6th ACM International
  Symposium on Pervasive Displays}.\hskip 1em plus 0.5em minus 0.4em\relax ACM,
  2017, p.~10.

\bibitem{farinella2014face}
G.~M. Farinella, G.~Farioli, S.~Battiato, S.~Leonardi, and G.~Gallo, ``Face
  re-identification for digital signage applications,'' in \emph{International
  Workshop on Video Analytics for Audience Measurement in Retail and Digital
  Signage}.\hskip 1em plus 0.5em minus 0.4em\relax Springer, 2014, pp. 40--52.

\bibitem{li2014crowded}
T.~Li, H.~Chang, M.~Wang, B.~Ni, R.~Hong, and S.~Yan, ``Crowded scene analysis:
  A survey,'' \emph{IEEE transactions on circuits and systems for video
  technology}, vol.~25, no.~3, pp. 367--386, 2014.

\bibitem{wang2014high}
X.~Wang, X.~Yang, X.~He, Q.~Teng, and M.~Gao, ``A high accuracy flow
  segmentation method in crowded scenes based on streakline,''
  \emph{Optik-International Journal for Light and Electron Optics}, vol. 125,
  no.~3, pp. 924--929, 2014.

\bibitem{mehran2010streakline}
R.~Mehran, B.~E. Moore, and M.~Shah, ``A streakline representation of flow in
  crowded scenes,'' in \emph{European conference on computer vision}.\hskip 1em
  plus 0.5em minus 0.4em\relax Springer, 2010, pp. 439--452.

\bibitem{su2013large}
H.~Su, H.~Yang, S.~Zheng, Y.~Fan, and S.~Wei, ``The large-scale crowd behavior
  perception based on spatio-temporal viscous fluid field,'' \emph{IEEE
  Transactions on Information Forensics and security}, vol.~8, no.~10, pp.
  1575--1589, 2013.

\bibitem{kratz2011tracking}
L.~Kratz and K.~Nishino, ``Tracking pedestrians using local spatio-temporal
  motion patterns in extremely crowded scenes,'' \emph{IEEE transactions on
  pattern analysis and machine intelligence}, vol.~34, no.~5, pp. 987--1002,
  2011.

\bibitem{jodoin2013meta}
P.-M. Jodoin, Y.~Benezeth, and Y.~Wang, ``Meta-tracking for video scene
  understanding,'' in \emph{2013 10th IEEE International Conference on Advanced
  Video and Signal Based Surveillance}.\hskip 1em plus 0.5em minus 0.4em\relax
  IEEE, 2013, pp. 1--6.

\bibitem{lewandowski2013tracklet}
M.~Lewandowski, D.~Simonnet, D.~Makris, S.~A. Velastin, and J.~Orwell,
  ``Tracklet reidentification in crowded scenes using bag of spatio-temporal
  histograms of oriented gradients,'' in \emph{Mexican Conference on Pattern
  Recognition}.\hskip 1em plus 0.5em minus 0.4em\relax Springer, 2013, pp.
  94--103.

\bibitem{zhou2012understanding}
B.~Zhou, X.~Wang, and X.~Tang, ``Understanding collective crowd behaviors:
  Learning a mixture model of dynamic pedestrian-agents,'' in \emph{2012 IEEE
  Conference on Computer Vision and Pattern Recognition}.\hskip 1em plus 0.5em
  minus 0.4em\relax IEEE, 2012, pp. 2871--2878.

\bibitem{chongjing2013analyzing}
W.~Chongjing, Z.~Xu, Z.~Yi, and L.~Yuncai, ``Analyzing motion patterns in
  crowded scenes via automatic tracklets clustering,'' \emph{china
  communications}, vol.~10, no.~4, pp. 144--154, 2013.

\bibitem{morris2008survey}
B.~T. Morris and M.~M. Trivedi, ``A survey of vision-based trajectory learning
  and analysis for surveillance,'' \emph{IEEE transactions on circuits and
  systems for video technology}, vol.~18, no.~8, pp. 1114--1127, 2008.

\bibitem{piciarelli2008trajectory}
C.~Piciarelli, C.~Micheloni, and G.~L. Foresti, ``Trajectory-based anomalous
  event detection,'' \emph{IEEE Transactions on Circuits and Systems for video
  Technology}, vol.~18, no.~11, pp. 1544--1554, 2008.

\bibitem{cocsar2016toward}
S.~Co{\c{s}}ar, G.~Donatiello, V.~Bogorny, C.~Garate, L.~O. Alvares, and
  F.~Br{\'e}mond, ``Toward abnormal trajectory and event detection in video
  surveillance,'' \emph{IEEE Transactions on Circuits and Systems for Video
  Technology}, vol.~27, no.~3, pp. 683--695, 2016.

\bibitem{song2013fully}
X.~Song, X.~Shao, Q.~Zhang, R.~Shibasaki, H.~Zhao, J.~Cui, and H.~Zha, ``A
  fully online and unsupervised system for large and high-density area
  surveillance: Tracking, semantic scene learning and abnormality detection,''
  \emph{ACM Transactions on Intelligent Systems and Technology (TIST)}, vol.~4,
  no.~2, p.~35, 2013.

\bibitem{brun2014dynamic}
L.~Brun, A.~Saggese, and M.~Vento, ``Dynamic scene understanding for behavior
  analysis based on string kernels,'' \emph{IEEE Transactions on Circuits and
  Systems for Video Technology}, vol.~24, no.~10, pp. 1669--1681, 2014.

\bibitem{tran2013video}
D.~Tran, J.~Yuan, and D.~Forsyth, ``Video event detection: From subvolume
  localization to spatiotemporal path search,'' \emph{IEEE transactions on
  pattern analysis and machine intelligence}, vol.~36, no.~2, pp. 404--416,
  2013.

\bibitem{Revathi2017}
\BIBentryALTinterwordspacing
A.~R. Revathi and D.~Kumar, ``An efficient system for anomaly detection using
  deep learning classifier,'' \emph{Signal, Image and Video Processing},
  vol.~11, no.~2, pp. 291--299, Feb 2017. [Online]. Available:
  \url{https://doi.org/10.1007/s11760-016-0935-0}
\BIBentrySTDinterwordspacing

\bibitem{6129539}
O.~P. {Popoola} and K.~{Wang}, ``Video-based abnormal human behavior
  recognition—a review,'' \emph{IEEE Transactions on Systems, Man, and
  Cybernetics, Part C (Applications and Reviews)}, vol.~42, no.~6, pp.
  865--878, Nov 2012.

\bibitem{7490361}
Y.~{Yuan}, Y.~{Feng}, and X.~{Lu}, ``Statistical hypothesis detector for
  abnormal event detection in crowded scenes,'' \emph{IEEE Transactions on
  Cybernetics}, vol.~47, no.~11, pp. 3597--3608, Nov 2017.

\bibitem{XIONG2012121}
\BIBentryALTinterwordspacing
G.~Xiong, J.~Cheng, X.~Wu, Y.-L. Chen, Y.~Ou, and Y.~Xu, ``An energy model
  approach to people counting for abnormal crowd behavior detection,''
  \emph{Neurocomputing}, vol.~83, pp. 121 -- 135, 2012. [Online]. Available:
  \url{http://www.sciencedirect.com/science/article/pii/S0925231211006990}
\BIBentrySTDinterwordspacing

\bibitem{6894168}
Y.~{Zhang}, L.~{Qin}, R.~{Ji}, H.~{Yao}, and Q.~{Huang}, ``Social
  attribute-aware force model: Exploiting richness of interaction for abnormal
  crowd detection,'' \emph{IEEE Transactions on Circuits and Systems for Video
  Technology}, vol.~25, no.~7, pp. 1231--1245, July 2015.

\bibitem{Yi:2014:LRS:2679600.2680014}
\BIBentryALTinterwordspacing
S.~Yi, X.~Wang, C.~Lu, and J.~Jia, ``L0 regularized stationary time estimation
  for crowd group analysis,'' in \emph{Proceedings of the 2014 IEEE Conference
  on Computer Vision and Pattern Recognition}, ser. CVPR '14.\hskip 1em plus
  0.5em minus 0.4em\relax Washington, DC, USA: IEEE Computer Society, 2014, pp.
  2219--2226. [Online]. Available: \url{https://doi.org/10.1109/CVPR.2014.284}
\BIBentrySTDinterwordspacing

\bibitem{Cheng2015VideoAD}
K.-W. Cheng, Y.-T. Chen, and W.-H. Fang, ``Video anomaly detection and
  localization using hierarchical feature representation and gaussian process
  regression,'' \emph{2015 IEEE Conference on Computer Vision and Pattern
  Recognition (CVPR)}, pp. 2909--2917, 2015.

\bibitem{Lee13anomalydetection}
Y.~jye Lee, Y.~ren Yeh, and Y.~chiang Frank~Wang, ``Anomaly detection via
  online oversampling principal component analysis,'' \emph{IEEE Trans.
  Knowledge and Data Eng}, 2013.

\bibitem{KRAUSZ2012307}
\BIBentryALTinterwordspacing
B.~Krausz and C.~Bauckhage, ``Loveparade 2010: Automatic video analysis of a
  crowd disaster,'' \emph{Computer Vision and Image Understanding}, vol. 116,
  no.~3, pp. 307 -- 319, 2012, special issue on Semantic Understanding of Human
  Behaviors in Image Sequences. [Online]. Available:
  \url{http://www.sciencedirect.com/science/article/pii/S1077314211002037}
\BIBentrySTDinterwordspacing

\bibitem{7024902}
D.~{Lee}, H.~{Suk}, S.~{Park}, and S.~{Lee}, ``Motion influence map for unusual
  human activity detection and localization in crowded scenes,'' \emph{IEEE
  Transactions on Circuits and Systems for Video Technology}, vol.~25, no.~10,
  pp. 1612--1623, Oct 2015.

\bibitem{6217836}
C.~C. {Loy}, T.~{Xiang}, and S.~{Gong}, ``Salient motion detection in crowded
  scenes,'' in \emph{2012 5th International Symposium on Communications,
  Control and Signal Processing}, May 2012, pp. 1--4.

\bibitem{VishwakarmaA13}
\BIBentryALTinterwordspacing
S.~Vishwakarma and A.~Agrawal, ``A survey on activity recognition and behavior
  understanding in video surveillance,'' \emph{The Visual Computer}, vol.~29,
  no.~10, pp. 983--1009, 2013. [Online]. Available:
  \url{https://doi.org/10.1007/s00371-012-0752-6}
\BIBentrySTDinterwordspacing

\bibitem{8323245}
K.~{Xu}, X.~{Jiang}, and T.~{Sun}, ``Anomaly detection based on stacked sparse
  coding with intraframe classification strategy,'' \emph{IEEE Transactions on
  Multimedia}, vol.~20, no.~5, pp. 1062--1074, May 2018.

\bibitem{sen2019video}
D.~Sen, B.~Raman \emph{et~al.}, ``Video skimming: taxonomy and comprehensive
  survey,'' \emph{ACM Computing Surveys (CSUR)}, vol.~52, no.~5, p. 106, 2019.

\bibitem{xie2008automatic}
X.-n. Xie and F.~Wu, ``Automatic video summarization by affinity propagation
  clustering and semantic content mining,'' in \emph{2008 International
  Symposium on Electronic Commerce and Security}.\hskip 1em plus 0.5em minus
  0.4em\relax IEEE, 2008, pp. 203--208.

\bibitem{wu2017novel}
J.~Wu, S.-h. Zhong, J.~Jiang, and Y.~Yang, ``A novel clustering method for
  static video summarization,'' \emph{Multimedia Tools and Applications},
  vol.~76, no.~7, pp. 9625--9641, 2017.

\bibitem{bhaumik2014towards}
H.~Bhaumik, S.~Bhattacharyya, S.~Dutta, and S.~Chakraborty, ``Towards
  redundancy reduction in storyboard representation for static video
  summarization,'' in \emph{2014 International Conference on Advances in
  Computing, Communications and Informatics (ICACCI)}.\hskip 1em plus 0.5em
  minus 0.4em\relax IEEE, 2014, pp. 344--350.

\bibitem{zhang2016summary}
K.~Zhang, W.-L. Chao, F.~Sha, and K.~Grauman, ``Summary transfer:
  Exemplar-based subset selection for video summarization,'' in
  \emph{Proceedings of the IEEE conference on computer vision and pattern
  recognition}, 2016, pp. 1059--1067.

\bibitem{gygli2015video}
M.~Gygli, H.~Grabner, and L.~Van~Gool, ``Video summarization by learning
  submodular mixtures of objectives,'' in \emph{Proceedings of the IEEE
  Conference on Computer Vision and Pattern Recognition}, 2015, pp. 3090--3098.

\bibitem{mohan2019static}
J.~Mohan and M.~S. Nair, ``Static video summarization using sparse
  autoencoders,'' in \emph{2019 IEEE International Conference on Electrical,
  Computer and Communication Technologies (ICECCT)}.\hskip 1em plus 0.5em minus
  0.4em\relax IEEE, 2019, pp. 1--8.

\bibitem{ji2020cross}
Z.~Ji, Y.~Zhao, Y.~Pang, and X.~Li, ``Cross-modal guidance based auto-encoder
  for multi-video summarization,'' \emph{Pattern Recognition Letters}, 2020.

\bibitem{peker2006broadcast}
K.~A. Peker, I.~Otsuka, and A.~Divakaran, ``Broadcast video program
  summarization using face tracks,'' in \emph{2006 IEEE International
  Conference on Multimedia and Expo}.\hskip 1em plus 0.5em minus 0.4em\relax
  IEEE, 2006, pp. 1053--1056.

\bibitem{ngo2005video}
C.-W. Ngo, Y.-F. Ma, and H.-J. Zhang, ``Video summarization and scene detection
  by graph modeling,'' \emph{IEEE Transactions on Circuits and Systems for
  Video Technology}, vol.~15, no.~2, pp. 296--305, 2005.

\bibitem{xiao2008automatic}
R.-g. Xiao, Y.-y. Wang, H.~Pan, and F.~Wu, ``Automatic video summarization by
  spatio-temporal analysis and non-trivial repeating pattern detection,'' in
  \emph{2008 Congress on Image and Signal Processing}, vol.~4.\hskip 1em plus
  0.5em minus 0.4em\relax IEEE, 2008, pp. 555--559.

\bibitem{gao2008video}
Y.~Gao, W.-B. Wang, and J.-H. Yong, ``A video summarization tool using
  two-level redundancy detection for personal video recorders,'' \emph{IEEE
  Transactions on Consumer Electronics}, vol.~54, no.~2, pp. 521--526, 2008.

\bibitem{xu2016fast}
B.~Xu, X.~Wang, and Y.-G. Jiang, ``Fast summarization of user-generated videos:
  Exploiting semantic, emotional, and quality clues,'' \emph{IEEE MultiMedia},
  vol.~23, no.~3, pp. 23--33, 2016.

\bibitem{ji2019video}
Z.~Ji, K.~Xiong, Y.~Pang, and X.~Li, ``Video summarization with attention-based
  encoder-decoder networks,'' \emph{IEEE Transactions on Circuits and Systems
  for Video Technology}, 2019.

\bibitem{wu2020dynamic}
J.~Wu, S.-h. Zhong, and Y.~Liu, ``Dynamic graph convolutional network for
  multi-video summarization,'' \emph{Pattern Recognition}, p. 107382, 2020.

\bibitem{zhong2019video}
S.-h. Zhong, J.~Wu, and J.~Jiang, ``Video summarization via spatio-temporal
  deep architecture,'' \emph{Neurocomputing}, vol. 332, pp. 224--235, 2019.

\bibitem{rav2006making}
A.~Rav-Acha, Y.~Pritch, and S.~Peleg, ``Making a long video short: Dynamic
  video synopsis,'' in \emph{2006 IEEE Computer Society Conference on Computer
  Vision and Pattern Recognition (CVPR'06)}, vol.~1.\hskip 1em plus 0.5em minus
  0.4em\relax IEEE, 2006, pp. 435--441.

\bibitem{he2016fast}
Y.~He, Z.~Qu, C.~Gao, and N.~Sang, ``Fast online video synopsis based on
  potential collision graph,'' \emph{IEEE Signal Processing Letters}, vol.~24,
  no.~1, pp. 22--26, 2016.

\bibitem{he2017graph}
Y.~He, C.~Gao, N.~Sang, Z.~Qu, and J.~Han, ``Graph coloring based surveillance
  video synopsis,'' \emph{Neurocomputing}, vol. 225, pp. 64--79, 2017.

\bibitem{batuhan2017improved}
K.~Batuhan~Baskurt and R.~Same, ``Improved adaptive background subtraction
  method using pixel-based segmenter,'' 2017.

\bibitem{bacskurt2018long}
K.~B. BA{\c{S}}KURT and R.~Samet, ``Long-term multiobject tracking using
  alternative correlation filters,'' \emph{Turkish Journal of Electrical
  Engineering and Computer Science}, vol.~26, no.~5, pp. 2246--2259, 2018.

\bibitem{lin2017optimized}
L.~Lin, W.~Lin, W.~Xiao, and S.~Huang, ``An optimized video synopsis algorithm
  and its distributed processing model,'' \emph{Soft computing}, vol.~21,
  no.~4, pp. 935--947, 2017.

\bibitem{ahmed2019query}
S.~A. Ahmed, D.~P. Dogra, S.~Kar, R.~Patnaik, S.-C. Lee, H.~Choi, G.~P. Nam,
  and I.-J. Kim, ``Query-based video synopsis for intelligent traffic
  monitoring applications,'' \emph{IEEE Transactions on Intelligent
  Transportation Systems}, 2019.

\bibitem{zhu2014key}
X.~Zhu, J.~Liu, J.~Wang, and H.~Lu, ``Key observation selection-based effective
  video synopsis for camera network,'' \emph{Machine vision and applications},
  vol.~25, no.~1, pp. 145--157, 2014.

\bibitem{mahapatra2016mvs}
A.~Mahapatra, P.~K. Sa, B.~Majhi, and S.~Padhy, ``Mvs: A multi-view video
  synopsis framework,'' \emph{Signal Processing: Image Communication}, vol.~42,
  pp. 31--44, 2016.

\bibitem{zhang2019multi}
Z.~Zhang, Y.~Nie, H.~Sun, Q.~Zhang, Q.~Lai, G.~Li, and M.~Xiao, ``Multi-view
  video synopsis via simultaneous object-shifting and view-switching
  optimization,'' \emph{IEEE Transactions on Image Processing}, vol.~29, pp.
  971--985, 2019.

\bibitem{ferryman2014performance}
J.~Ferryman and A.-L. Ellis, ``Performance evaluation of crowd image analysis
  using the pets2009 dataset,'' \emph{Pattern Recognition Letters}, vol.~44,
  pp. 3--15, 2014.

\bibitem{foggia2013real}
P.~Foggia, G.~Percannella, A.~Saggese, and M.~Vento, ``Real-time tracking of
  single people and groups simultaneously by contextual graph-based reasoning
  dealing complex occlusions,'' in \emph{2013 IEEE International Workshop on
  Performance Evaluation of Tracking and Surveillance (PETS)}.\hskip 1em plus
  0.5em minus 0.4em\relax IEEE, 2013, pp. 29--36.

\bibitem{dasiopoulou2005knowledge}
S.~Dasiopoulou, V.~Mezaris, I.~Kompatsiaris, V.-K. Papastathis, and M.~G.
  Strintzis, ``Knowledge-assisted semantic video object detection,'' \emph{IEEE
  Transactions on Circuits and Systems for Video Technology}, vol.~15, no.~10,
  pp. 1210--1224, 2005.

\bibitem{garcia2008video}
A.~Garc{\'\i}a and J.~Besc{\'o}s, ``Video object segmentation based on feedback
  schemes guided by a low-level scene ontology,'' in \emph{International
  Conference on Advanced Concepts for Intelligent Vision Systems}.\hskip 1em
  plus 0.5em minus 0.4em\relax Springer, 2008, pp. 322--333.

\bibitem{gomez2011ontology}
J.~G{\'o}mez-Romero, M.~A. Patricio, J.~Garc{\'\i}a, and J.~M. Molina,
  ``Ontology-based context representation and reasoning for object tracking and
  scene interpretation in video,'' \emph{Expert Systems with Applications},
  vol.~38, no.~6, pp. 7494--7510, 2011.

\bibitem{haarslev2001racer}
V.~Haarslev and R.~M{\"o}ller, ``Racer system description,'' in
  \emph{International Joint Conference on Automated Reasoning}.\hskip 1em plus
  0.5em minus 0.4em\relax Springer, 2001, pp. 701--705.

\bibitem{bird2005detection}
N.~D. Bird, O.~Masoud, N.~P. Papanikolopoulos, and A.~Isaacs, ``Detection of
  loitering individuals in public transportation areas,'' \emph{IEEE
  Transactions on intelligent transportation systems}, vol.~6, no.~2, pp.
  167--177, 2005.

\bibitem{laxton2007leveraging}
B.~Laxton, J.~Lim, and D.~Kriegman, ``Leveraging temporal, contextual and
  ordering constraints for recognizing complex activities in video,'' in
  \emph{2007 IEEE Conference on Computer Vision and Pattern Recognition}.\hskip
  1em plus 0.5em minus 0.4em\relax IEEE, 2007, pp. 1--8.

\bibitem{wang2011action}
J.~Wang, Z.~Chen, and Y.~Wu, ``Action recognition with multiscale
  spatio-temporal contexts,'' in \emph{CVPR 2011}.\hskip 1em plus 0.5em minus
  0.4em\relax IEEE, 2011, pp. 3185--3192.

\bibitem{francois2005verl}
A.~R. Francois, R.~Nevatia, J.~Hobbs, R.~C. Bolles, and J.~R. Smith, ``Verl: an
  ontology framework for representing and annotating video events,'' \emph{IEEE
  multimedia}, vol.~12, no.~4, pp. 76--86, 2005.

\bibitem{snidaro2007representing}
L.~Snidaro, M.~Belluz, and G.~L. Foresti, ``Representing and recognizing
  complex events in surveillance applications,'' in \emph{2007 IEEE Conference
  on Advanced Video and Signal Based Surveillance}.\hskip 1em plus 0.5em minus
  0.4em\relax IEEE, 2007, pp. 493--498.

\bibitem{sanmiguel2009ontology}
J.~C. SanMiguel, J.~M. Martinez, and {\'A}.~Garcia, ``An ontology for event
  detection and its application in surveillance video,'' in \emph{2009 Sixth
  IEEE International Conference on Advanced Video and Signal Based
  Surveillance}.\hskip 1em plus 0.5em minus 0.4em\relax IEEE, 2009, pp.
  220--225.

\bibitem{greco2016abnormal}
L.~Greco, P.~Ritrovato, A.~Saggese, and M.~Vento, ``Abnormal event recognition:
  A hybrid approach using semanticweb technologies,'' in \emph{2016 IEEE
  Conference on Computer Vision and Pattern Recognition Workshops
  (CVPRW)}.\hskip 1em plus 0.5em minus 0.4em\relax IEEE, 2016, pp. 1297--1304.

\bibitem{chen2009ontology}
L.~Chen and C.~Nugent, ``Ontology-based activity recognition in intelligent
  pervasive environments,'' \emph{International Journal of Web Information
  Systems}, vol.~5, no.~4, pp. 410--430, 2009.

\bibitem{cavaliere2016towards}
D.~Cavaliere, S.~Senatore, M.~Vento, and V.~Loia, ``Towards semantic
  context-aware drones for aerial scenes understanding,'' in \emph{2016 13th
  IEEE International Conference on Advanced Video and Signal Based Surveillance
  (AVSS)}.\hskip 1em plus 0.5em minus 0.4em\relax IEEE, 2016, pp. 115--121.

\bibitem{yao2010i2t}
B.~Z. Yao, X.~Yang, L.~Lin, M.~W. Lee, and S.-C. Zhu, ``I2t: Image parsing to
  text description,'' \emph{Proceedings of the IEEE}, vol.~98, no.~8, pp.
  1485--1508, 2010.

\bibitem{xue2012ontology}
M.~Xue, S.~Zheng, and C.~Zhang, ``Ontology-based surveillance video archive and
  retrieval system,'' in \emph{2012 IEEE Fifth International Conference on
  Advanced Computational Intelligence (ICACI)}.\hskip 1em plus 0.5em minus
  0.4em\relax IEEE, 2012, pp. 84--89.

\bibitem{xu2013video}
Z.~Xu, L.~Mei, Y.~Liu, and C.~Hu, ``Video structural description: a semantic
  based model for representing and organizing video surveillance big data,'' in
  \emph{2013 IEEE 16th International Conference on Computational Science and
  Engineering}.\hskip 1em plus 0.5em minus 0.4em\relax IEEE, 2013, pp.
  802--809.

\bibitem{xu2015semantic}
Z.~Xu, Y.~Liu, L.~Mei, C.~Hu, and L.~Chen, ``Semantic based representing and
  organizing surveillance big data using video structural description
  technology,'' \emph{Journal of Systems and Software}, vol. 102, pp. 217--225,
  2015.

\bibitem{sah2017semantic}
M.~Sah and C.~Direkoglu, ``Semantic annotation of surveillance videos for
  abnormal crowd behaviour search and analysis,'' in \emph{2017 14th IEEE
  International Conference on Advanced Video and Signal Based Surveillance
  (AVSS)}.\hskip 1em plus 0.5em minus 0.4em\relax IEEE, 2017, pp. 1--6.

\bibitem{sobhani2015ontology}
F.~Sobhani, N.~F. Kahar, and Q.~Zhang, ``An ontology framework for automated
  visual surveillance system,'' in \emph{2015 13th International Workshop on
  Content-Based Multimedia Indexing (CBMI)}.\hskip 1em plus 0.5em minus
  0.4em\relax IEEE, 2015, pp. 1--7.

\bibitem{alam2020intellibvr}
A.~Alam, M.~N. Khan, J.~Khan, and Y.-K. Lee, ``Intellibvr-intelligent
  large-scale video retrieval for objects and events utilizing distributed
  deep-learning and semantic approaches,'' in \emph{2020 IEEE International
  Conference on Big Data and Smart Computing (BigComp)}.\hskip 1em plus 0.5em
  minus 0.4em\relax IEEE, 2020, pp. 28--35.

\bibitem{ajiboye2015hierarchical}
S.~O. Ajiboye, P.~Birch, C.~Chatwin, and R.~Young, ``Hierarchical video
  surveillance architecture: a chassis for video big data analytics and
  exploration,'' in \emph{Video Surveillance and Transportation Imaging
  Applications 2015}, vol. 9407.\hskip 1em plus 0.5em minus 0.4em\relax
  International Society for Optics and Photonics, 2015, p. 94070K.

\bibitem{lin2012framework}
C.-F. Lin, S.-M. Yuan, M.-C. Leu, and C.-T. Tsai, ``A framework for scalable
  cloud video recorder system in surveillance environment,'' in
  \emph{Ubiquitous intelligence \& computing and 9th international conference
  on autonomic \& trusted computing (UIC/ATC), 2012 9th international
  conference on}.\hskip 1em plus 0.5em minus 0.4em\relax IEEE, 2012, pp.
  655--660.

\bibitem{liu2015distributed}
X.~Liu, D.~Zhao, L.~Xu, W.~Zhang, J.~Yin, and X.~Chen, ``A distributed video
  management cloud platform using hadoop,'' \emph{IEEE Access}, vol.~3, pp.
  2637--2643, 2015.

\bibitem{ryu2013extensible}
C.~Ryu, D.~Lee, M.~Jang, C.~Kim, and E.~Seo, ``Extensible video processing
  framework in apache hadoop,'' in \emph{Cloud Computing Technology and Science
  (CloudCom), 2013 IEEE 5th International Conference on}, vol.~2.\hskip 1em
  plus 0.5em minus 0.4em\relax IEEE, 2013, pp. 305--310.

\bibitem{bradski2000opencv}
G.~Bradski and A.~Kaehler, ``Opencv,'' \emph{Dr. Dobb’s journal of software
  tools}, vol.~3, 2000.

\bibitem{ali2020res}
M.~Ali, A.~Anjum, O.~Rana, A.~R. Zamani, D.~Balouek-Thomert, and M.~Parashar,
  ``Res: Real-time video stream analytics using edge enhanced clouds,''
  \emph{IEEE Transactions on Cloud Computing}, 2020.

\bibitem{white2010web}
B.~White, T.~Yeh, J.~Lin, and L.~Davis, ``Web-scale computer vision using
  mapreduce for multimedia data mining,'' in \emph{Proceedings of the Tenth
  International Workshop on Multimedia Data Mining}.\hskip 1em plus 0.5em minus
  0.4em\relax ACM, 2010, p.~9.

\bibitem{pereira2010architecture}
R.~Pereira, M.~Azambuja, K.~Breitman, and M.~Endler, ``An architecture for
  distributed high performance video processing in the cloud,'' in \emph{Cloud
  Computing (CLOUD), 2010 IEEE 3rd International Conference on}.\hskip 1em plus
  0.5em minus 0.4em\relax IEEE, 2010, pp. 482--489.

\bibitem{liu2014distributed}
C.~Liu, K.~Fan, Z.~Yang, and J.~Xiu, ``A distributed video share system based
  on hadoop,'' in \emph{Cloud Computing and Intelligence Systems (CCIS), 2014
  IEEE 3rd International Conference on}.\hskip 1em plus 0.5em minus 0.4em\relax
  IEEE, 2014, pp. 587--590.

\bibitem{hossain2014framework}
M.~A. Hossain, ``Framework for a cloud-based multimedia surveillance system,''
  \emph{International Journal of Distributed Sensor Networks}, vol.~10, no.~5,
  p. 135257, 2014.

\bibitem{valentin2017cloud}
L.~Valent{\'\i}n, S.~A. Serrano, R.~O. Garc{\'\i}a, A.~Andrade, M.~A.
  Palacios-Alonso, and L.~E. Sucar, ``a cloud-based architecture for smart
  video surveillance,'' \emph{The International Archives of Photogrammetry,
  Remote Sensing and Spatial Information Sciences}, vol.~42, p.~99, 2017.

\bibitem{zhang2015deep}
W.~Zhang, P.~Duan, Z.~Li, Q.~Lu, W.~Gong, and S.~Yang, ``A deep awareness
  framework for pervasive video cloud.'' \emph{IEEE Access}, vol.~3, pp.
  2227--2237, 2015.

\bibitem{zhang2016deep}
W.~Zhang, L.~Xu, Z.~Li, Q.~Lu, and Y.~Liu, ``A deep-intelligence framework for
  online video processing,'' \emph{IEEE Software}, vol.~33, no.~2, pp. 44--51,
  2016.

\bibitem{uddin2019siat}
M.~A. Uddin, A.~Alam, N.~A. Tu, M.~S. Islam, and Y.-K. Lee, ``Siat: A
  distributed video analytics framework for intelligent video surveillance,''
  \emph{Symmetry}, vol.~11, no.~7, p. 911, 2019.

\bibitem{zhang2016efficient}
H.~Zhang, J.~Yan, and Y.~Kou, ``Efficient online surveillance video processing
  based on spark framework,'' in \emph{International Conference on Big Data
  Computing and Communications}.\hskip 1em plus 0.5em minus 0.4em\relax
  Springer, 2016, pp. 309--318.

\bibitem{uddin2017human}
M.~A. Uddin, J.~B. Joolee, A.~Alam, and Y.-K. Lee, ``Human action recognition
  using adaptive local motion descriptor in spark,'' \emph{IEEE Access},
  vol.~5, pp. 21\,157--21\,167, 2017.

\bibitem{wang2015large}
H.~Wang, X.~Zheng, and B.~Xiao, ``Large-scale human action recognition with
  spark,'' in \emph{Multimedia Signal Processing (MMSP), 2015 IEEE 17th
  International Workshop on}.\hskip 1em plus 0.5em minus 0.4em\relax IEEE,
  2015, pp. 1--6.

\bibitem{huang2017sve}
Q.~Huang, P.~Ang, P.~Knowles, T.~Nykiel, I.~Tverdokhlib, A.~Yajurvedi,
  P.~Dapolito~IV, X.~Yan, M.~Bykov, C.~Liang \emph{et~al.}, ``Sve: Distributed
  video processing at facebook scale,'' in \emph{Proceedings of the 26th
  Symposium on Operating Systems Principles}.\hskip 1em plus 0.5em minus
  0.4em\relax ACM, 2017, pp. 87--103.

\bibitem{zhang2018deep}
W.~Zhang, Z.~Wang, X.~Liu, H.~Sun, J.~Zhou, Y.~Liu, and W.~Gong, ``Deep
  learning-based real-time fine-grained pedestrian recognition using stream
  processing,'' \emph{IET Intelligent Transport Systems}, vol.~12, no.~7, pp.
  602--609, 2018.

\bibitem{CitiLog2018}
Citilog, ``Citilog,'' 1997.

\bibitem{checkvideo2019}
\BIBentryALTinterwordspacing
checkvideo, ``checkvideo,'' 1998, accessed: 2018-12-07. [Online]. Available:
  \url{https://www.checkvideo.com/}
\BIBentrySTDinterwordspacing

\bibitem{IntelliVision2019}
\BIBentryALTinterwordspacing
I.~Vision, ``Intelli vision,'' 2002. [Online]. Available:
  \url{https://www.intelli-vision.com/}
\BIBentrySTDinterwordspacing

\bibitem{GoogleAI2019}
\BIBentryALTinterwordspacing
G.~Inc, ``Video ai,'' 2017. [Online]. Available:
  \url{https://cloud.google.com/video-intelligence/}
\BIBentrySTDinterwordspacing

\bibitem{BlueChasm2019AI}
\BIBentryALTinterwordspacing
B.~LLC, ``Usher in the new cognitive era,'' 2015. [Online]. Available:
  \url{https://bluechasm.com/}
\BIBentrySTDinterwordspacing

\bibitem{AzureVideoAnalytics2017}
\BIBentryALTinterwordspacing
J.~H. Julia~Kornich, Craig~Casey, ``Media analytics on the media services
  platform,'' 2010. [Online]. Available:
  \url{https://docs.microsoft.com/en-gb/azure/media-services/previous/media-services-analytics-overview}
\BIBentrySTDinterwordspacing

\bibitem{GoogleAI2019Features}
\BIBentryALTinterwordspacing
G.~Inc, ``Features,'' 2017. [Online]. Available:
  \url{https://cloud.google.com/video-intelligence/docs/features}
\BIBentrySTDinterwordspacing

\bibitem{IBM2019AI}
\BIBentryALTinterwordspacing
J.~G. Collier, ``Ibm uses watson’s power to unlock insights from video in the
  cloud,'' 2017. [Online]. Available:
  \url{https://www-03.ibm.com/press/us/en/pressrelease/52126.wss}
\BIBentrySTDinterwordspacing

\bibitem{IBMCloudStorageObject2017}
\BIBentryALTinterwordspacing
I.~Cloud, ``Ibm cloud object storage,'' 2017. [Online]. Available:
  \url{https://www.ibm.com/cloud/object-storage}
\BIBentrySTDinterwordspacing

\bibitem{IBMCloudant2017}
\BIBentryALTinterwordspacing
------, ``Cloudant,'' 2017. [Online]. Available:
  \url{https://www.ibm.com/cloud/cloudant}
\BIBentrySTDinterwordspacing

\bibitem{ananthanarayanan2017real}
G.~Ananthanarayanan, P.~Bahl, P.~Bod{\'\i}k, K.~Chintalapudi, M.~Philipose,
  L.~Ravindranath, and S.~Sinha, ``Real-time video analytics: The killer app
  for edge computing,'' \emph{computer}, vol.~50, no.~10, pp. 58--67, 2017.

\bibitem{loewenherz2017video}
F.~Loewenherz, V.~Bahl, and Y.~Wang, ``Video analytics towards vision zero,''
  \emph{Institute of Transportation Engineers. ITE Journal}, vol.~87, no.~3,
  p.~25, 2017.

\bibitem{qiu2018kestrel}
H.~Qiu, X.~Liu, S.~Rallapalli, A.~J. Bency, K.~Chan, R.~Urgaonkar,
  B.~Manjunath, and R.~Govindan, ``Kestrel: Video analytics for augmented
  multi-camera vehicle tracking,'' in \emph{2018 IEEE/ACM Third International
  Conference on Internet-of-Things Design and Implementation (IoTDI)}.\hskip
  1em plus 0.5em minus 0.4em\relax IEEE, 2018, pp. 48--59.

\bibitem{gao2009traffic}
T.~Gao, Z.-G. Liu, S.-H. Yue, J.-Q. Mei, and J.~Zhang, ``Traffic video-based
  moving vehicle detection and tracking in the complex environment,''
  \emph{Cybernetics and Systems: An International Journal}, vol.~40, no.~7, pp.
  569--588, 2009.

\bibitem{lin2020automated}
Y.~Lin, L.~Li, H.~Jing, B.~Ran, and D.~Sun, ``Automated traffic incident
  detection with a smaller dataset based on generative adversarial networks,''
  \emph{Accident Analysis \& Prevention}, vol. 144, p. 105628, 2020.

\bibitem{fridman2017autonomous}
L.~Fridman, D.~E. Brown, M.~Glazer, W.~Angell, S.~Dodd, B.~Jenik,
  J.~Terwilliger, J.~Kindelsberger, L.~Ding, S.~Seaman \emph{et~al.}, ``Mit
  autonomous vehicle technology study: Large-scale deep learning based analysis
  of driver behavior and interaction with automation,'' \emph{arXiv preprint
  arXiv:1711.06976}, 2017.

\bibitem{xu2017internet}
W.~Xu, H.~Zhou, N.~Cheng, F.~Lyu, W.~Shi, J.~Chen, and X.~Shen, ``Internet of
  vehicles in big data era,'' \emph{IEEE/CAA Journal of Automatica Sinica},
  vol.~5, no.~1, pp. 19--35, 2017.

\bibitem{wang2009improving}
J.~Wang, G.~Bebis, M.~Nicolescu, M.~Nicolescu, and R.~Miller, ``Improving
  target detection by coupling it with tracking,'' \emph{Machine Vision and
  Applications}, vol.~20, no.~4, pp. 205--223, 2009.

\bibitem{tsai2010detection}
Y.~J. Tsai, Z.~Hu, and C.~Alberti, ``Detection of roadway sign condition
  changes using multi-scale sign image matching (m-sim),''
  \emph{Photogrammetric Engineering \& Remote Sensing}, vol.~76, no.~4, pp.
  391--405, 2010.

\bibitem{ayachi2020traffic}
R.~Ayachi, M.~Afif, Y.~Said, and M.~Atri, ``Traffic signs detection for
  real-world application of an advanced driving assisting system using deep
  learning,'' \emph{Neural Processing Letters}, vol.~51, no.~1, pp. 837--851,
  2020.

\bibitem{callan2009neural}
A.~M. Callan, R.~Osu, Y.~Yamagishi, D.~E. Callan, and N.~Inoue, ``Neural
  correlates of resolving uncertainty in driver's decision making,''
  \emph{Human brain mapping}, vol.~30, no.~9, pp. 2804--2812, 2009.

\bibitem{yan2016driving}
C.~Yan, F.~Coenen, and B.~Zhang, ``Driving posture recognition by convolutional
  neural networks,'' \emph{IET Computer Vision}, vol.~10, no.~2, pp. 103--114,
  2016.

\bibitem{gite2019early}
S.~Gite and H.~Agrawal, ``Early prediction of driver's action using deep neural
  networks,'' \emph{International Journal of Information Retrieval Research
  (IJIRR)}, vol.~9, no.~2, pp. 11--27, 2019.

\bibitem{fleck2008smart}
S.~Fleck and W.~Stra{\ss}er, ``Smart camera based monitoring system and its
  application to assisted living,'' \emph{Proceedings of the IEEE}, vol.~96,
  no.~10, pp. 1698--1714, 2008.

\bibitem{zhou2008activity}
Z.~Zhou, X.~Chen, Y.-C. Chung, Z.~He, T.~X. Han, and J.~M. Keller, ``Activity
  analysis, summarization, and visualization for indoor human activity
  monitoring,'' \emph{IEEE Transactions on Circuits and Systems for Video
  Technology}, vol.~18, no.~11, pp. 1489--1498, 2008.

\bibitem{aertssen2011fall}
J.~Aertssen, M.~Rudinac, and P.~Jonker, ``Fall and action detection in elderly
  homes,'' \emph{AAATE: Maastricht, The Netherlands}, 2011.

\bibitem{brulin2012posture}
D.~Brulin, Y.~Benezeth, and E.~Courtial, ``Posture recognition based on fuzzy
  logic for home monitoring of the elderly,'' \emph{IEEE transactions on
  information technology in biomedicine}, vol.~16, no.~5, pp. 974--982, 2012.

\bibitem{sacco2012detection}
G.~Sacco, V.~Joumier, N.~Darmon, A.~Dechamps, A.~Derreumaux, J.-H. Lee,
  J.~Piano, N.~Bordone, A.~Konig, B.~Teboul \emph{et~al.}, ``Detection of
  activities of daily living impairment in alzheimer’s disease and mild
  cognitive impairment using information and communication technology,''
  \emph{Clinical interventions in aging}, vol.~7, p. 539, 2012.

\bibitem{yu2012posture}
M.~Yu, A.~Rhuma, S.~M. Naqvi, L.~Wang, and J.~Chambers, ``A posture
  recognition-based fall detection system for monitoring an elderly person in a
  smart home environment,'' \emph{IEEE transactions on information technology
  in biomedicine}, vol.~16, no.~6, pp. 1274--1286, 2012.

\bibitem{choi2012general}
W.~Choi, C.~Pantofaru, and S.~Savarese, ``A general framework for tracking
  multiple people from a moving camera,'' \emph{IEEE transactions on pattern
  analysis and machine intelligence}, vol.~35, no.~7, pp. 1577--1591, 2012.

\bibitem{pragada2015intrusion}
V.~Pragada, ``Intrusion detection and video surveillance activation and
  processing,'' Aug.~6 2015, uS Patent App. 14/172,880.

\bibitem{zhao2016crossing}
Z.~Zhao, H.~Li, R.~Zhao, and X.~Wang, ``Crossing-line crowd counting with
  two-phase deep neural networks,'' in \emph{European Conference on Computer
  Vision}.\hskip 1em plus 0.5em minus 0.4em\relax Springer, 2016, pp. 712--726.

\bibitem{xu2015loitering}
G.~Xu and W.~K. Cobb, ``Loitering detection in a video surveillance system,''
  Dec.~8 2015, uS Patent 9,208,675.

\bibitem{shivakumara2018cnn}
P.~Shivakumara, D.~Tang, M.~Asadzadehkaljahi, T.~Lu, U.~Pal, and M.~H. Anisi,
  ``Cnn-rnn based method for license plate recognition,'' \emph{CAAI
  Transactions on Intelligence Technology}, vol.~3, no.~3, pp. 169--175, 2018.

\bibitem{krausz2010metrosurv}
B.~Krausz and R.~Herpers, ``Metrosurv: detecting events in subway stations,''
  \emph{Multimedia Tools and Applications}, vol.~50, no.~1, pp. 123--147, 2010.

\bibitem{shih2006illegal}
J.-L. Shih, C.-c. Han, and K.-C. Yan, ``Illegal entrant detection at a
  restricted area in open spaces using color features,'' in \emph{Proceedings
  40th Annual 2006 International Carnahan Conference on Security
  Technology}.\hskip 1em plus 0.5em minus 0.4em\relax IEEE, 2006, pp. 68--74.

\bibitem{singh2018applications}
H.~Singh, ``Applications of intelligent video analytics in the field of retail
  management: A study,'' in \emph{Supply Chain Management Strategies and Risk
  Assessment in Retail Environments}.\hskip 1em plus 0.5em minus 0.4em\relax
  IGI Global, 2018, pp. 42--59.

\bibitem{xu2010people}
H.~Xu, P.~Lv, and L.~Meng, ``A people counting system based on head-shoulder
  detection and tracking in surveillance video,'' in \emph{2010 International
  Conference On Computer Design and Applications}, vol.~1.\hskip 1em plus 0.5em
  minus 0.4em\relax IEEE, 2010, pp. V1--394.

\bibitem{haritaoglu2002attentive}
I.~Haritaoglu and M.~Flickner, ``Attentive billboards: Towards to video based
  customer behavior understanding,'' in \emph{Sixth IEEE Workshop on
  Applications of Computer Vision, 2002.(WACV 2002). Proceedings.}\hskip 1em
  plus 0.5em minus 0.4em\relax IEEE, 2002, pp. 127--131.

\bibitem{hu2009action}
Y.~Hu, L.~Cao, F.~Lv, S.~Yan, Y.~Gong, and T.~S. Huang, ``Action detection in
  complex scenes with spatial and temporal ambiguities,'' in \emph{2009 IEEE
  12th International Conference on Computer Vision}.\hskip 1em plus 0.5em minus
  0.4em\relax IEEE, 2009, pp. 128--135.

\bibitem{sicre2010human}
R.~Sicre and H.~Nicolas, ``Human behaviour analysis and event recognition at a
  point of sale,'' in \emph{2010 Fourth Pacific-Rim Symposium on Image and
  Video Technology}.\hskip 1em plus 0.5em minus 0.4em\relax IEEE, 2010, pp.
  127--132.

\bibitem{popescu2013predict}
A.~D. Popescu, A.~Balmin, V.~Ercegovac, and A.~Ailamaki, ``Predict: towards
  predicting the runtime of large scale iterative analytics,''
  \emph{Proceedings of the VLDB Endowment}, vol.~6, no.~14, pp. 1678--1689,
  2013.

\bibitem{amershi2014power}
S.~Amershi, M.~Cakmak, W.~B. Knox, and T.~Kulesza, ``Power to the people: The
  role of humans in interactive machine learning,'' \emph{AI Magazine},
  vol.~35, no.~4, pp. 105--120, 2014.

\bibitem{kwok1999static}
Y.-K. Kwok and I.~Ahmad, ``Static scheduling algorithms for allocating directed
  task graphs to multiprocessors,'' \emph{ACM Computing Surveys (CSUR)},
  vol.~31, no.~4, pp. 406--471, 1999.

\bibitem{zhu2014analysis}
X.~Zhu, J.~Wang, J.~Wang, and X.~Qin, ``Analysis and design of fault-tolerant
  scheduling for real-time tasks on earth-observation satellites,'' in
  \emph{2014 43rd International Conference on Parallel Processing}.\hskip 1em
  plus 0.5em minus 0.4em\relax IEEE, 2014, pp. 491--500.

\bibitem{bortnikov2012predicting}
E.~Bortnikov, A.~Frank, E.~Hillel, and S.~Rao, ``Predicting execution
  bottlenecks in map-reduce clusters,'' in \emph{Presented as part of the},
  2012.

\bibitem{zhai2014emerging}
Y.~Zhai, Y.-S. Ong, and I.~W. Tsang, ``The emerging" big dimensionality",''
  2014.

\bibitem{gao2017learning}
L.~Gao, J.~Song, X.~Liu, J.~Shao, J.~Liu, and J.~Shao, ``Learning in
  high-dimensional multimedia data: the state of the art,'' \emph{Multimedia
  Systems}, vol.~23, no.~3, pp. 303--313, 2017.

\bibitem{Guttman:1984:RDI:602259.602266}
\BIBentryALTinterwordspacing
A.~Guttman, ``R-trees: A dynamic index structure for spatial searching,'' in
  \emph{Proceedings of the 1984 ACM SIGMOD International Conference on
  Management of Data}, ser. SIGMOD '84.\hskip 1em plus 0.5em minus 0.4em\relax
  New York, NY, USA: ACM, 1984, pp. 47--57. [Online]. Available:
  \url{http://doi.acm.org/10.1145/602259.602266}
\BIBentrySTDinterwordspacing

\bibitem{Ciaccia:1997:MEA:645923.671005}
\BIBentryALTinterwordspacing
P.~Ciaccia, M.~Patella, and P.~Zezula, ``M-tree: An efficient access method for
  similarity search in metric spaces,'' in \emph{Proceedings of the 23rd
  International Conference on Very Large Data Bases}, ser. VLDB '97.\hskip 1em
  plus 0.5em minus 0.4em\relax San Francisco, CA, USA: Morgan Kaufmann
  Publishers Inc., 1997, pp. 426--435. [Online]. Available:
  \url{http://dl.acm.org/citation.cfm?id=645923.671005}
\BIBentrySTDinterwordspacing

\bibitem{lin1994tv}
K.-I. Lin, H.~V. Jagadish, and C.~Faloutsos, ``The tv-tree: An index structure
  for high-dimensional data,'' \emph{The VLDB Journal}, vol.~3, no.~4, pp.
  517--542, 1994.

\bibitem{datar2004locality}
M.~Datar, N.~Immorlica, P.~Indyk, and V.~S. Mirrokni, ``Locality-sensitive
  hashing scheme based on p-stable distributions,'' in \emph{Proceedings of the
  twentieth annual symposium on Computational geometry}.\hskip 1em plus 0.5em
  minus 0.4em\relax ACM, 2004, pp. 253--262.

\bibitem{o1996log}
P.~O’Neil, E.~Cheng, D.~Gawlick, and E.~O’Neil, ``The log-structured
  merge-tree (lsm-tree),'' \emph{Acta Informatica}, vol.~33, no.~4, pp.
  351--385, 1996.

\bibitem{kraska2018case}
T.~Kraska, A.~Beutel, E.~H. Chi, J.~Dean, and N.~Polyzotis, ``The case for
  learned index structures,'' in \emph{Proceedings of the 2018 International
  Conference on Management of Data}.\hskip 1em plus 0.5em minus 0.4em\relax
  ACM, 2018, pp. 489--504.

\bibitem{de2013samoa}
G.~De~Francisci~Morales, ``Samoa: A platform for mining big data streams,'' in
  \emph{Proceedings of the 22nd International Conference on World Wide
  Web}.\hskip 1em plus 0.5em minus 0.4em\relax ACM, 2013, pp. 777--778.

\bibitem{lu2016large}
J.~Lu, S.~C. Hoi, J.~Wang, P.~Zhao, and Z.-Y. Liu, ``Large scale online kernel
  learning,'' \emph{The Journal of Machine Learning Research}, vol.~17, no.~1,
  pp. 1613--1655, 2016.

\bibitem{nallaperuma2019online}
D.~Nallaperuma, R.~Nawaratne, T.~Bandaragoda, A.~Adikari, S.~Nguyen,
  T.~Kempitiya, D.~De~Silva, D.~Alahakoon, and D.~Pothuhera, ``Online
  incremental machine learning platform for big data-driven smart traffic
  management,'' \emph{IEEE Transactions on Intelligent Transportation Systems},
  2019.

\bibitem{friedman2001elements}
J.~Friedman, T.~Hastie, and R.~Tibshirani, \emph{The elements of statistical
  learning}.\hskip 1em plus 0.5em minus 0.4em\relax Springer series in
  statistics New York, 2001, vol.~1, no.~10.

\bibitem{kandel2012enterprise}
S.~Kandel, A.~Paepcke, J.~M. Hellerstein, and J.~Heer, ``Enterprise data
  analysis and visualization: An interview study,'' \emph{IEEE Transactions on
  Visualization and Computer Graphics}, vol.~18, no.~12, pp. 2917--2926, 2012.

\bibitem{kumar2016model}
A.~Kumar, R.~McCann, J.~Naughton, and J.~M. Patel, ``Model selection management
  systems: The next frontier of advanced analytics,'' \emph{ACM SIGMOD Record},
  vol.~44, no.~4, pp. 17--22, 2016.

\bibitem{zhang2019ps2}
Z.~Zhang, B.~Cui, Y.~Shao, L.~Yu, J.~Jiang, and X.~Miao, ``Ps2: Parameter
  server on spark,'' in \emph{Proceedings of the 2019 International Conference
  on Management of Data}, 2019, pp. 376--388.

\bibitem{jiang2018sketchml}
J.~Jiang, F.~Fu, T.~Yang, and B.~Cui, ``Sketchml: Accelerating distributed
  machine learning with data sketches,'' in \emph{Proceedings of the 2018
  International Conference on Management of Data}, 2018, pp. 1269--1284.

\bibitem{huang2018flexps}
Y.~Huang, T.~Jin, Y.~Wu, Z.~Cai, X.~Yan, F.~Yang, J.~Li, Y.~Guo, and J.~Cheng,
  ``Flexps: Flexible parallelism control in parameter server architecture,''
  \emph{Proceedings of the VLDB Endowment}, vol.~11, no.~5, pp. 566--579, 2018.

\bibitem{haralick1992performance}
R.~M. Haralick, ``Performance characterization in computer vision,'' in
  \emph{BMVC92}.\hskip 1em plus 0.5em minus 0.4em\relax Springer, 1992, pp.
  1--8.

\bibitem{forstner1997dagm}
W.~F{\"o}rstner, ``Dagm workshop on performance characteristics and quality of
  computer vision algorithms,'' \emph{Technical University of Brunswick,
  Germany}, 1997.

\bibitem{phillips1999empirical}
P.~J. Phillips and K.~W. Bowyer, ``Empirical evaluation of computer vision
  algorithms,'' \emph{IEEE Transactions on Pattern Analysis and Machine
  Intelligence}, vol.~21, no.~4, pp. 289--290, 1999.

\bibitem{christensen1997special}
H.~Christensen and W.~Foerstner, ``Special issue on performance evaluation,''
  \emph{Machine Vision and Applications}, vol.~9, no.~5, 1997.

\bibitem{nghiem2007etiseo}
A.~T. Nghiem, F.~Bremond, M.~Thonnat, and V.~Valentin, ``Etiseo, performance
  evaluation for video surveillance systems,'' in \emph{Advanced Video and
  Signal Based Surveillance, 2007. AVSS 2007. IEEE Conference on}.\hskip 1em
  plus 0.5em minus 0.4em\relax IEEE, 2007, pp. 476--481.

\bibitem{hauptmann2004successful}
A.~G. Hauptmann and M.~G. Christel, ``Successful approaches in the trec video
  retrieval evaluations,'' in \emph{Proceedings of the 12th annual ACM
  international conference on Multimedia}.\hskip 1em plus 0.5em minus
  0.4em\relax ACM, 2004, pp. 668--675.

\bibitem{branch2006imagery}
H.~O. S.~D. Branch, ``Imagery library for intelligent detection systems
  (i-lids),'' in \emph{2006 IET Conference on Crime and Security}.\hskip 1em
  plus 0.5em minus 0.4em\relax IET, 2006, pp. 445--448.

\bibitem{singh2015survey}
D.~Singh and C.~K. Reddy, ``A survey on platforms for big data analytics,''
  \emph{Journal of big data}, vol.~2, no.~1, p.~8, 2015.

\bibitem{wu2019deep}
D.~Wu, X.~Luo, M.~Shang, Y.~He, G.~Wang, and M.~Zhou, ``A deep latent factor
  model for high-dimensional and sparse matrices in recommender systems,''
  \emph{IEEE Transactions on Systems, Man, and Cybernetics: Systems}, 2019.

\bibitem{mazrekaj2016pricing}
A.~Mazrekaj, I.~Shabani, and B.~Sejdiu, ``Pricing schemes in cloud computing:
  an overview,'' \emph{International Journal of Advanced Computer Science and
  Applications}, vol.~7, no.~2, pp. 80--86, 2016.

\bibitem{khalid2019real}
S.~Khalid, S.~Wu, A.~Alam, and I.~Ullah, ``Real-time feedback query expansion
  technique for supporting scholarly search using citation network analysis,''
  \emph{Journal of Information Science}, p. 0165551519863346, 2019.

\bibitem{kang2019challenges}
D.~Kang, P.~Bailis, and M.~Zaharia, ``Challenges and opportunities in dnn-based
  video analytics: A demonstration of the blazeit video query engine.'' in
  \emph{CIDR}, 2019.

\bibitem{lu2016visflow}
Y.~Lu, A.~Chowdhery, and S.~Kandula, ``Visflow: a relational platform for
  efficient large-scale video analytics,'' in \emph{ACM Symposium on Cloud
  Computing (SoCC)}, 2016.

\bibitem{kangblazeit}
D.~Kang, P.~Bailis, and M.~Zaharia, ``Blazeit: Optimizing declarative
  aggregation and limit queries for neural network-based video analytics.''

\bibitem{ghosh2010guest}
A.~Ghosh and I.~Arce, ``Guest editors' introduction: In cloud computing we
  trust-but should we?'' \emph{IEEE security \& privacy}, vol.~8, no.~6, pp.
  14--16, 2010.

\bibitem{khusro2017social}
S.~Khusro, A.~Alam, and S.~Khalid, ``Social question and answer sites: the
  story so far,'' \emph{Program}, vol.~51, no.~2, pp. 170--192, 2017.

\bibitem{alam2017confluence}
A.~Alam, S.~Khusro, I.~Ullah, and M.~S. Karim, ``Confluence of social network,
  social question and answering community, and user reputation model for
  information seeking and experts generation,'' \emph{Journal of Information
  Science}, vol.~43, no.~2, pp. 260--274, 2017.

\bibitem{crosby2016blockchain}
M.~Crosby, P.~Pattanayak, S.~Verma, V.~Kalyanaraman \emph{et~al.}, ``Blockchain
  technology: Beyond bitcoin,'' \emph{Applied Innovation}, vol.~2, no. 6-10,
  p.~71, 2016.

\end{thebibliography}

\begin{IEEEbiography}[{\includegraphics[width=1in,height=1.25in,clip,keepaspectratio]{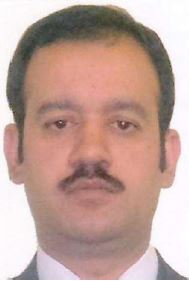}}]{Aftab Alam} has received his master's degree in computer science with a specialization in Web engineering from the Department of Computer Science, University of Peshawar, Peshawar, Pakistan. He is currently pursuing his Ph.D. at the Department of Computer Science and Engineering, Kyung Hee University (Global Campus), South Korea. His research interests in the areas of: Big data analytics, distributed computing, video retrieval, social computing, machine/deep learning, information service engineering, knowledge engineering, knowledge graph and graph mining. 
\end{IEEEbiography}

\begin{IEEEbiography}[{\includegraphics[width=1in,height=1.25in,clip,keepaspectratio]{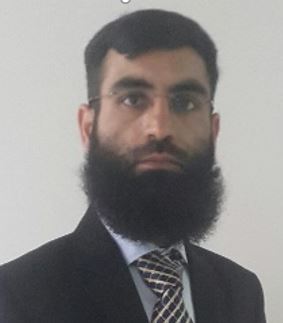}}]{Irfan~Ullah} has received his master's degree in computer science from the National University of Sciences and Technology (NUST), Islamabad, Pakistan. Currently, he is serving as a research assistant at the Department of Computer Science and Engineering, Kyung Hee University (Global Campus), South Korea. His research interests in the areas of: Natural language processing, social computing, crisis informatics, machine/deep learning, OS design and optimization on memory systems, big data analytics, and distributed computing.
\end{IEEEbiography}

\begin{IEEEbiography}[{\includegraphics[width=1in,height=1.25in,clip,keepaspectratio]{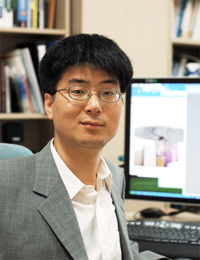}}]{Young-Koo~Lee} has received the B.S., M.S., and Ph.D. degrees in computer science from the Korea Advanced Institute of Science and Technology, Daejeon, South Korea, in 1992, 1994, and 2002. Since 2004, he has been with the Department of Computer Science and Engineering, Kyung Hee University. His researches have been concentrated on data mining, online analytical processing, and big data processing.
\end{IEEEbiography}

\EOD

\end{document}